\documentclass[fleqn,usenatbib]{mnras}
\pdfoutput=1

\usepackage{graphicx}
\usepackage{enumerate}
\usepackage{newtxtext,newtxmath}
\usepackage[T1]{fontenc}
\usepackage{gensymb}
\usepackage{mathtools}
\usepackage{physics}
\usepackage{natbib}
\usepackage{hyperref}
\usepackage{color}
\usepackage{ulem}
\usepackage{soul}
\usepackage{url}
\usepackage{xspace}
\usepackage{multirow}
\usepackage{CJKutf8}
\usepackage[caption=false]{subfig}

\DeclareRobustCommand{\VAN}[3]{#2}
\let\VANthebibliography\thebibliography
\def\thebibliography{\DeclareRobustCommand{\VAN}[3]{##3}\VANthebibliography}

\newcommand{\Hipparcos}{{\sl Hipparcos}}
\newcommand{\Gaia}{{\sl Gaia}}
\newcommand{\HST}{{\sl HST}}

\newcommand{\Msun}{\mbox{$M_{\sun}$}}

\newcommand{\Teff}{\mbox{$T_{\rm eff}$}}
\newcommand{\logg}{\mbox{$\log(g)$}}

\newcommand{\ms}{\hbox{m\,s$^{-1}$}}
\newcommand{\kms}{\mbox{km\,s$^{-1}$}}
\newcommand{\masyr}{\hbox{mas\,yr$^{-1}$}}

\newcommand{\changes}[1]{{#1}}

\title{Dynamical Masses and Ages of Sirius-like Systems}

\author[H.~Zhang et al.]{
Hengyue Zhang  \begin{CJK*}{UTF8}{gbsn} (张\,恒悦) \end{CJK*},$^{1,2}$\thanks{E-mail: hengyue.zhang@physics.ox.ac.uk}
Timothy D.~Brandt,$^{1}$
Rocio Kiman,$^{3}$
Alexander Venner,$^{4}$
\newauthor{
Qier An,$^{1}$
Minghan Chen,$^{1}$
Yiting Li$^{1}$
}
\\
$^{1}$Department of Physics, University of California, Santa Barbara, Santa Barbara, CA 93106, USA\\
$^{2}$Sub-department of Astrophysics, Department of Physics, University of Oxford, Keble Road, Oxford OX1 3RH, UK\\
$^{3}$Kavli Institute for Theoretical Physics, University of California, Santa Barbara, CA 93106, USA\\
$^{4}$Centre for Astrophysics, University of Southern Queensland, Toowoomba, QLD 4350, Australia
}

\date{Accepted XXX. Received YYY; in original form ZZZ}
\pubyear{2023}

\begin{document}
\label{firstpage}
\pagerange{\pageref{firstpage}--\pageref{lastpage}}
\maketitle

\begin{abstract}
We measure precise orbits and dynamical masses and derive age constraints for six confirmed and one candidate Sirius-like systems, including the Hyades member HD 27483. Our orbital analysis incorporates radial velocities, relative astrometry, and {\Hipparcos}-{\Gaia} astrometric accelerations. We constrain the main-sequence lifetime of a white dwarf's progenitor from the remnant's dynamical mass and semi-empirical initial-final mass relations and infer the cooling age from mass and effective temperature. We present new relative astrometry of HD 27483 B from Keck/NIRC2 observations and archival {\HST} data, and obtain the first dynamical mass of ${0.798}_{-0.041}^{+0.10}$ $M_{\sun}$, and an age of ${450}_{-180}^{+570}$ Myr, consistent with previous age estimates of Hyades. We also measure precise dynamical masses for HD 114174 B ($0.591 \pm 0.011$ $M_{\sun}$) and HD 169889 B (${0.526}_{-0.037}^{+0.039}$ $M_{\sun}$), but their age precisions are limited by their uncertain temperatures. For HD 27786 B, the unusually small mass of $0.443 \pm 0.012$ $M_{\sun}$ suggests a history of rapid mass loss, possibly due to binary interaction in its progenitor's AGB phase. The orbits of HD 118475 and HD 136138 from our RV fitting are overall in good agreement with {\Gaia} DR3 astrometric two-body solutions, despite moderate differences in the eccentricity and period of HD 136138. The mass of ${0.580}_{-0.039}^{+0.052}$ $M_{\sun}$ for HD 118475 B and a speckle imaging non-detection confirms that the companion is a white dwarf. Our analysis shows examples of a rich number of precise WD dynamical mass measurements enabled by {\Gaia} DR3 and later releases, which will improve empirical calibrations of the white dwarf initial-final mass relation.
\end{abstract}


\section{Introduction} \label{sec:intro}

While isolated white dwarfs (WDs) simply cool and fade indefinitely, WDs in binary systems can give rise to some of the most dramatic events in the Universe. These include novae and Type Ia supernovae, the source of much of the Universe's iron-peak elements \citep{Iwamoto_1999} and standardizable candles used to measure the expansion history of the Universe \citep[e.g.,][]{Phillips_1993,Riess_1998, Perlmutter_1999}.  The nearest known WD, Sirius B, is itself one component in a binary system \citep{Bond_1862}; it orbits 20\,AU \citep{Bond_2017b} from the early A-type star \citep{Morgan_1953} Sirius A. 
Widely separated Sirius-like systems enable mass measurements based only on Newtonian dynamics and they form the widely-separated analogs of the progenitors of novae and supernovae.  

Stars below $\approx$8\,$M_{\odot}$ will end their lives as WDs, while stars below $\approx$0.8\,$M_\odot$ live longer than a Hubble time on the main sequence (MS), according to stellar evolutionary models \citep[e.g.,][]{MIST_2016}. At least half of the stars in this mass interval are born in binaries \citep{Raghavan_2010}. The Solar neighborhood is rich in the Sirius-like systems left after the more massive star in the binary has evolved off the MS. Before turning into a WD, the star loses mass by expelling its outer shell and exposing its central core during the asymptotic giant phase \changes{(see e.g. \citealp{Hofner_2018} for a review)}. The final mass of the remnant WD is related to the initial mass of its progenitor by the initial-final mass relation (IFMR). Constraining the relation is crucial to understanding the complex physical processes underlying the final stages of stellar evolution. Observations of orbits of Sirius-like systems enable precise measurements of WD masses without requiring constraints on surface gravity and are independent of WD evolutionary models. Combing the mass with a $\Teff$ or luminosity measurement constrains the WD's cooling age, which informs us about the WD's initial mass when compared to the age of the MS companion. These provide direct measurements of the IFMR. Conversely, one can derive the total age of the WD by assuming a particular IFMR. Comparing the WD's age to the MS star's age tests the accuracy of the assumed IFMR.

Fitting dynamical masses and orbits to Sirius-like systems requires either long-term astrometric monitoring like that available for Sirius 
\citep[e.g.,][]{Bond_2017b}, or the combination of different data types to compensate for limited phase coverage. On the one hand, absolute astrometry from the {\Hipparcos} \citep{ESA_1997} and {\Gaia} \citep{Gaia_2016} missions probe the transverse acceleration of the MS star due to the tug of a WD companion. These powerful astrometric measurements have been \changes{compiled} and calibrated to the same reference frame by the Hipparcos-Gaia Catalog of Accelerations \citep[HGCA,][]{Brandt_2018,Brandt_2021}. On the other hand, spectrographs like \changes{the University College London Echelle Spectrograph (UCLES) on the Anglo-Australian Telescope (AAT)} \citep{Diego_1990}, \changes{the High-Resolution Echelle Spectrometer (HIRES) at the Keck observatory}\citep{Vogt_1994}, \changes{the Levy spectrometer for the Automated Planet Finder (APF)}\citep{Vogt_2014}, and \changes{the Hamilton Echelle Spectrometer at the Lick Observatory}\citep{Fischer_2014} provide high-precision radial velocities (RVs) that probe the MS star's acceleration along the line-of-sight. These measurements, together with relative astrometry of a short orbital arc, enable a direct measurement of WD mass \citep[e.g.,][]{Brandt_2019,Bowler_2021,Zeng_2022}.  

In this paper, we combine absolute astrometry from the HGCA with \changes{RVs} and relative astrometry from various sources to measure orbits and masses of six confirmed and one candidate nearby Sirius-like systems, including the Hyades member HD 27483. \changes{Here, we define a "Sirius-like system" as any binary or multiple star system containing at least one WD and at least one non-compact star with spectral type earlier than M0, consistent with the definition in \cite{Holberg_2013}.} For HD 118475 and HD 136138, we compare our results against {\Gaia} DR3 two-body solutions \citep{Gaia_DR3,Holl_2022}. We use the open-source code \textsc{orvara} \citep{Brandt_2021_orvara} to perform the fits and to derive constraints on the WDs' masses and orbits. Then, we infer age constraints on four WDs from their dynamical masses and effective temperatures ($\Teff$). 

We structure the paper as follows. Section \ref{sec:targets} introduces our target selection method and summarizes the properties of the selected targets. In Section \ref{sec:data+obs}, we outline the observational data available for each target and present our new observation for HD 27483 and the corresponding data reduction. We introduce our orbit-fitting method and discuss our results in Section \ref{sec:orbit_fit}. In Section \ref{sec:age}, we perform a Bayesian age analysis on the WDs and compare our results with existing age estimates by other methods. \changes{In Section \ref{sec:discuss}, we discuss whether the progenitors of the WDs could have interacted with their companions. Then, we compute the radii of the WDs from photometry and spectroscopy and compare them to theoretical WD mass-radius relations.} We conclude in Section \ref{sec:conclude}.

\section{Target Selection} \label{sec:targets}

We selected our sample from the list of known Sirius-like systems in \cite{Holberg_2013} and the more recently detected WD + \changes{early non-compact} star binaries in the Montreal White Dwarf Database\footnote{https://www.montrealwhitedwarfdatabase.org/home.html} \citep{Dufour_2017}. Of more than 100 such systems, we first selected 62 targets that have astrometric accelerations in the HGCA \citep{Brandt_2018,Brandt_2021}. Then, we excluded 49 targets with neither high-precision RVs nor at least two epochs of relative astrometry. We choose not to redo the dynamical mass analysis for \changes{six} systems that already have precise dynamical mass measurements: \changes{40 Eri \citep{Mason_2017}, Procyon \citep{Bond_2015}, Sirius \citep{Bond_2017b}, Gl 86 \citep{Brandt_2019}, 12 Psc and HD 159062 \citep{Bowler_2021}}. In addition to the selection criteria, we removed HD 149499 as we found that its $\sim 0.45$ $M_{\rm \sun}$ inner companion \citep{Tokovinin_2019} alone explains the host star's astrometric acceleration and RV trend, which leaves almost no information to constrain the outer WD companion (first discovered by \citealt{Jordan_1997}). Furthermore, we added HD 118475 B, an unconfirmed WD detected by RV in \cite{Kane_2019}. We ended up with seven targets, three of which have both high-precision RVs and relative astrometry, two of which only have relative astrometry (HD 27483 and HD 27786), and two of which (HD 118475 and HD 136138) only have RVs. The host star properties are summarized in Table \ref{tab:host_prop}. We collected the spectral types from \cite{Skiff_2014_catalogue} and parallaxes from {\Gaia} DR3 \citep{Gaia_DR3}. We adopted the most recent isochronological mass for each host star except HD~27483~A, for which we used the total dynamical mass of the two \changes{F6 V} components of the spectroscopic binary, measured by \cite{Konacki_2004}.

\begin{table} 
    \centering
    \caption{Summary of host star properties.}
    \begin{tabular}{ccccc} \hline
        \multirow{2}{1.3em}{HD} & \multirow{2}{1.8em}{SpT} & $\varpi$ & $M$ & \multirow{2}{1.9em}{Ref.} \\
        & & (mas) & ($M_{\sun}$)&\\ \hline \hline
        19019 & \changes{G0 V} & $31.979 \pm 0.029$ & $1.06 \pm 0.06$ & 1 \\ \hline
        27483 & \changes{F6 V+F6 V} & $21.094 \pm 0.032$ & $2.77 \pm 0.26$ & 2 \\ \hline
        27786 & \changes{F4 IV-V} & $23.79 \pm 0.11$ & $1.54 \pm 0.08$ & 3 \\ \hline
        114174 & \changes{G3 IV} & $37.868 \pm 0.024$ & $0.97 \pm 0.04$ & 4 \\ \hline
        118475 & \changes{F9 V} & $29.537 \pm 0.017$ & $1.16 \pm 0.06$ & 3 \\ \hline
        136138 & \changes{G8 IIIa} & $9.011 \pm 0.051$ & $1.84 \pm 0.40$ & 5 \\ \hline
        169889 & \changes{G7 V} & $28.279 \pm 0.026$ & $0.98 \pm 0.05$ & 3 \\ \hline
    \end{tabular}
    \centering
    
    {The spectral types are collected from \citep{Skiff_2014_catalogue} and parallaxes from {\Gaia} DR3 \citep{Gaia_DR3}. The references for stellar masses are: 1. \cite{Landstreet_2020}; 2. \cite{Konacki_2004}; 3. \cite{Kervella_2019}; 4. \cite{Rosenthal_2021}; 5. \cite{Stefanik_2011}}
    \label{tab:host_prop}
\end{table}

\section{Observations and Data} \label{sec:data+obs}
We collect high-precision RVs, relative astrometry, and the {\Hipparcos}-{\Gaia} absolute astrometry to determine each system's orbital parameters and masses. Because a WD companion is fainter than a MS companion with the same mass, it needs to be further from the primary star to be detectable by direct imaging. However, such a companion is often too widely separated to impose significant radial or astrometric acceleration on the primary star. Hence, high-precision RVs and multiple epochs of relative astrometry are not simultaneously available for more than half of our systems. Below we summarize the data that we adopted for each system. \changes{We use J2000 as the astrometric reference frame throughout this work.}

\subsection{Absolute Astrometry} \label{subsec:abs_ast}
Absolute astrometry shows the proper motion anomaly of stars and is thus powerful in constraining the orbits of massive, long-period companions. We use the absolute astrometry in the EDR3 version of the HGCA \citep{Brandt_2021}, which calibrates {\Gaia} EDR3 and {\Hipparcos} astrometry to reveal the proper motion difference between {\Hipparcos}, {\Gaia}, and the {\Hipparcos}-{\Gaia} mean motion. This gives the astrometric acceleration of the host star, which is directly proportional to the companion's mass. Table \ref{tab:ast_acc} lists the difference between the {\Gaia} proper motion ($\mu_{\alpha\star,\delta}^{\rm Gaia}$) and the {\Hipparcos}-{\Gaia} mean motion ($\mu_{\alpha\star,\delta}^{\rm H-G}$) for each target, with the corresponding significance levels of astrometric accelerations. Five of our targets have levels $\gg 5 \sigma$, meaning that they have accelerated significantly due to the gravitational pull of the companions.

\begin{table} 
    \centering
    \caption{Proper motion anomaly ($\mu_{\alpha\star,\delta}^{\rm Gaia}-\mu_{\alpha\star,\delta}^{\rm H-G}$) of the targets as given by the HGCA.}
    \begin{tabular}{cccc} \hline
        HD & $\Delta\mu_{\alpha\star} (\masyr)$ & $\Delta\mu_{\delta}$ (\masyr)& Sig. level\\ \hline \hline
        19019 & $0.045 \pm 0.049$ & $-0.017 \pm 0.040$ & $0.7\sigma$ \\ \hline
        27483 & $0.79 \pm 0.06$ & $2.15 \pm 0.04$ & $52\sigma$ \\ \hline
        27786 & $-0.91 \pm 0.21$ & $-19.42 \pm 0.13$ & $150\sigma$ \\ \hline
        114174 & $0.54 \pm 0.04$ & $-3.55 \pm 0.04$ & $88\sigma$ \\ \hline
        118475 & $28.99 \pm 0.51$ & $18.37 \pm 0.57$ & $64\sigma$ \\ \hline
        136138 & $-0.89 \pm 0.30$ & $-0.97 \pm 0.32$ & $4.1\sigma$ \\ \hline
        169889 & $0.51 \pm 0.04$ & $-1.77 \pm 0.04$ & $48\sigma$ \\ \hline
    \end{tabular}
    \label{tab:ast_acc}
\end{table}

We also checked {\Gaia} DR3 for astrometric non-single stars among our targets. We found two-body fits \citep{Holl_2022} for HD 118475 and HD 136138, which significantly corrected the proper motion and parallax measurements from their one-body fits. \changes{This suggests that the orbital periods of the two systems are comparable to the duration of the {\Gaia} mission. For such systems, the proper motion of the primary star changes considerably between each scan of {\Gaia}, making the {\Gaia} one-body proper motion measurement, obtained by fitting a single proper motion value to data from multiple scans, hard to interpret before the release of {\Gaia} intermediate astrometric data (IAD). The two-body proper motion, on the other hand, refers to the barycenter motion of the binary and contains no information on the orbit between the two stars.} Therefore, we decided not to use \changes{these} proper motions, fitting the two systems only from their RVs and the corrected parallaxes. Section \ref{subsec:orbit_results} compares our best-fitting orbital parameters to the {\Gaia} DR3 two-body solution. Checking the consistency between the two types of solutions helps validate {\Gaia} two-body astrometry. The companion of HD 19019 is detected in {\Gaia} at a separation of $11.7$ arcsec. We use the {\Gaia} proper motions of the companion, calibrated to the same reference frame as the primary using the prescription in \cite{Cantat-Gaudin_2021}, to help constrain the orbit.

\subsection{Radial Velocity and Relative Astrometry}

Our RVs and relative astrometry come from various literature sources, archival data, and, in the case of HD~27483, new data.  In this section, we summarize the data available for each system.

\subsubsection{HD 19019}
The HD~19019 system contains a strongly magnetic WD, its spectroscopic signature detected by \cite{Landstreet_2020} with the \changes{Intermediate-dispersion Spectrograph and Imaging System (ISIS)} on the William Herschel telescope \citep{WHT_1985}. It has 17 RV observations between 2002 and 2014 \citep{Butler_2017} by the HIRES instrument on Keck \citep{Vogt_1994}. The system is in the Washington Double Star Catalog as WDS J03038+0608, first resolved by {\Gaia} DR2 and reported by \cite{Knapp_2019} as a common proper motion pair. The WD companion was also detected in {\Gaia} DR3, which gives a total of two epochs of relative astrometry with precisions better than 0.1 $\rm{mas}$. The extreme precision dominates over the precision of the rest of the data, causing the orbital fit routine (described in Section \ref{sec:orbit_fit}) to ignore the RV and absolute astrometry and struggle to converge. We inflated the uncertainties of {\Gaia} relative astrometry by a factor of 5 to account for potential systematics and to avoid convergence issues.

\subsubsection{HD 27483}
\cite{Boehm-Vitense_1993} discovered a WD companion to the Hyades \changes{F6 V} binary from the \changes{{\sl International Ultraviolet Explorer} ({\sl IUE})} spectrum of the system. The companion was then resolved twice by {\HST} in July 1999 \citep{Barstow_2001} and November 2011 (in WFC3 F218W, unpublished, \changes{{\HST}} proposal 12606, PI Martin Barstow). We derive relative astrometry and photometry from the unpublished data, using the host star's \changes{point spread function (PSF)} as a template to fit the companion's position and magnitude. We apply a least-squares routine, assuming uniform $\sigma_{\rm data}$ on the intensity at each pixel and demanding a reduced $\chi^2$ of unity. We fit for three parameters: an offset in each of two directions, and a contrast.  We estimate the uncertainty of a parameter by fixing all other parameters at their best-fitting values and finding the $\Delta \chi^2=1$ interval of the free parameter. The results are listed in the first row of Table \ref{tab:new_relast}. The two existing epochs of relative astrometry gave only loose constraints on the orbital parameters when we applied our orbit fitting procedure in Section \ref{sec:orbit_fit}.

To obtain additional relative astrometry and photometry, we observed HD~27483
on 2021 November 25 UT with the \changes{second generation of the near-infrared camera (NIRC2)} and the natural guide star adaptive optics system at the Keck II telescope \citep{Wizinowich_2000}. In the $K_s$ filter, we took 21 deep exposures with an integration time of 0.18s per coadd, 150 coadds, and 1024 $\times$ 1024 pixels, obtaining high S/N images for the companion, but with the host star saturated. We then took 11 shallow (0.006s integration time per coadd, 150 coadds, 128 $\times$ 120 pixels) exposures with unsaturated host star PSFs. We used the unsaturated PSFs in shallow exposures as templates to fit both the host star and the companion in deep exposures. For the host star, we masked saturated pixels, fitting the template only to the outer, unsaturated speckles. By comparing the results from different PSF templates, we confirm that the uncertainties of the fit were smaller than 0.1 pixels in both directions. 

\begin{figure*}
    \centering
    \includegraphics[width=0.8\linewidth]{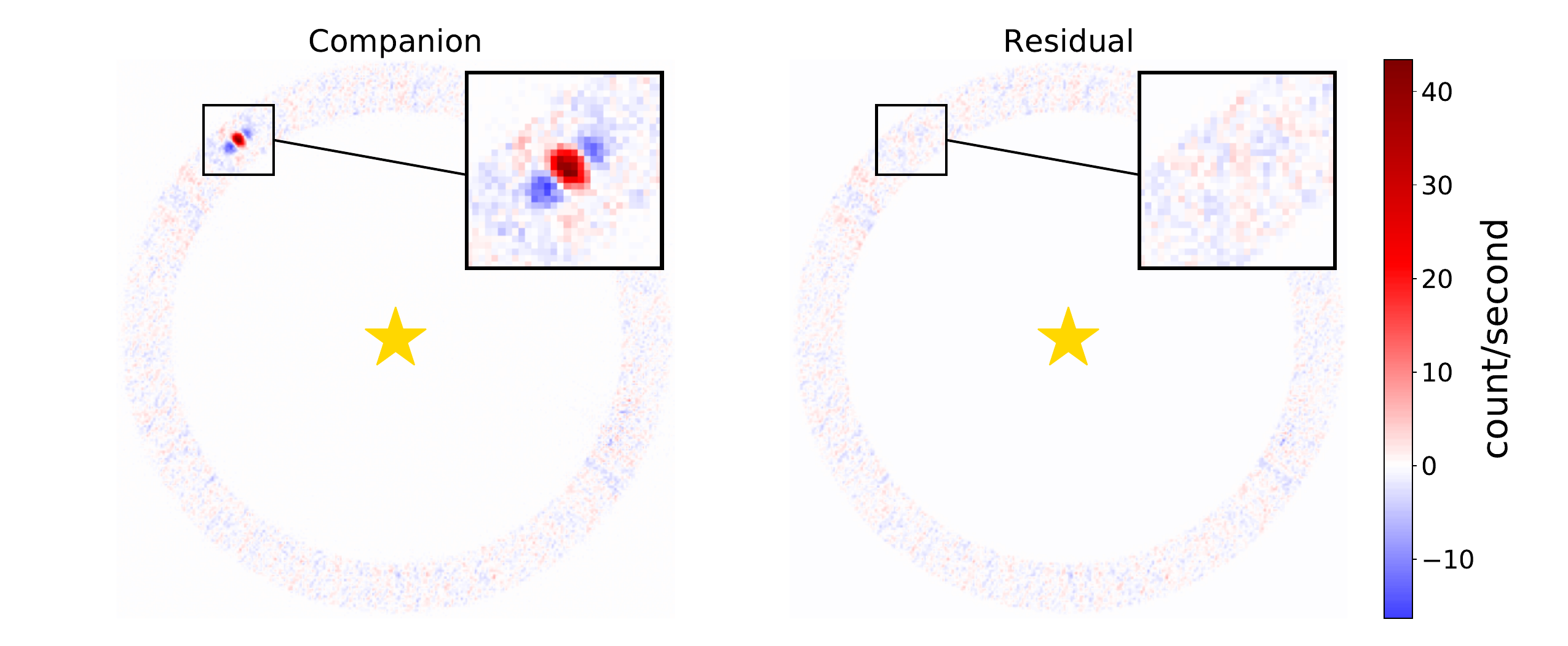}
    \caption{PSF fitting for the NIRC2 observations of HD 27483. The left panel shows the annulus obtained by performing ADI on the data cube. The companion (enlarged in the figure inset) is detected at a separation of $1.08$ arcsec. The right panel shows the residual of subtracting the best-fitting fake companion from the annulus. The host star, represented by a star symbol, is at the center of the annulus.}
    \label{fig:residual_27483}
\end{figure*}

After centering the host star in each frame, we detected the companion by performing angular differential imaging \citep[ADI]{Marois_2006} using the \textsc{vip-hci} \citep{Gonzalez_2017} package. Finally, we applied the negative fake companion method \citep[NEGFC]{Wertz_2017} to fit the companion, subtracting the PSF template from the ADI annulus at different locations and magnitudes until minimizing the root-mean-square residual of the subtraction. We used the \changes{Markov chain Monte Carlo (MCMC)} routine in  \textsc{vip-hci} to perform the fit and estimate the uncertainties. We applied the distortion correction in \cite{Service_2016}, rotating the images by $\theta_{\rm north}=0.262\degree \pm 0.020\degree$ clockwise to align them with the celestial north. The plate scale we adopted is $9.971 \pm 0.004 \pm 0.001$ mas pixel$^{-1}$ \citep{Service_2016}. We added these uncertainties to the uncertainties of the astrometry, in pixels, from the MCMC posteriors. The left panel of Figure \ref{fig:residual_27483} shows the ADI annulus, with the companion clearly revealed at a separation of $1.08$ arcsec, and the right panel shows the residual of the NEGFC subtraction. The derived relative astrometry and photometry are given in the second row of Table \ref{tab:new_relast}.

\begin{table*} 
    \centering
    \caption{New relative astrometry and photometry from this work.}
    \begin{tabular}{ccccccc} \hline
        Target & Date (UT) & Instrument & Sep (arcsec) & PA (deg) & $\Delta m$ & Filter \\ \hline \hline
         HD 27483 & 2011 Nov 26 & WFC3 & $1.1880 \pm 0.0038$ & $25.32 \pm 0.18$ & $3.11 \pm 0.17$ & F218W \\ \hline
         HD 27483 & 2021 Nov 25 & NIRC2 & $1.0832 \pm 0.0016$ & $38.67 \pm 0.07 $ & $7.79 \pm 0.06$ & $K_{s}$ \\ \hline
         HD 27786 Aa-Ab & 2003 Jan 08 & WFPC2 & $0.4120 \pm 0.0070$ & $288.56 \pm 0.97$ & $2.67 \pm 0.22$ & F170W \\ \hline
         Ba-Bb & 2003 Jan 08 & WFPC2 & $0.6200 \pm 0.0096$ & $290.18 \pm 0.65$ & $1.74 \pm 0.22$ & F170W \\ \hline
         Aa-B & 2003 Jan 08 & WFPC2 & $4.2924 \pm 0.0072$ & $14.31 \pm 0.10$ & ... & F170W \\ \hline
        HD 27786 Aa-Ab & 2003 Dec 14 & WFPC2 & $0.4024 \pm 0.0096$ & $280.41 \pm 1.00$ & $2.55 \pm 0.24$ & F170W \\ \hline
         Ba-Bb & 2003 Dec 14 & WFPC2 & $0.6173 \pm 0.0059$ & $289.23 \pm 0.54$ & $1.63 \pm 0.23$ & F170W \\ \hline
         Aa-B & 2003 Dec 14 & WFPC2 & $4.2681 \pm 0.0071$ & $14.11 \pm 0.10$ & ... & F170W \\ \hline
        HD 27786 Aa-Ab & 2012 Feb 28 & WFC3 & $0.2159 \pm 0.0061$ & $191.49 \pm 1.62$ & $4.64 \pm 0.32$ & F218W \\ \hline
         Ba-Bb & 2012 Feb 28 & WFC3 & $0.5272 \pm 0.0041$ & $273.22 \pm 0.45$ & $2.41 \pm 0.17$ & F218W \\ \hline
         Aa-B & 2012 Feb 28 & WFC3 & $4.2347 \pm 0.0026$ & $12.95 \pm 0.04$ & ... & F218W \\ \hline
    \end{tabular}
    \label{tab:new_relast}
\end{table*}

\subsubsection{HD 27786}
HD~27786 (56 Per) is a hierarchical four-body system with astrometric observations \citep[e.g.,][]{Dembowski_1870, Rabe_1953, Kallarakal_1969} dating back to 1847. \cite{Landsman_1996} first discovered an inner WD companion. \cite{Barstow_2001} first resolved all four components, finding that the outer companion is itself two M-dwarfs. From their astrometry and contrast values, we modeled the mass ratio between the two outer M-dwarfs to be $1.475 \pm 0.042$ and derived that their barycenter is at a separation of $4.325$ arcsec and a PA of $14.\!\!\degree95 $ relative to the host star. This allows us to treat the outer binary as a single companion and perform a three-body fit. 

Unpublished observations by \changes{{\HST}} resolved the system twice in the WFPC2 F170W filter in January 2003 (\changes{{\HST}} proposal 9334, PI Howard Bond) and December 2003 (\changes{{\HST}} proposal 9964, PI Howard Bond) and once in the WFC3 F218W filter in February 2012 (\changes{{\HST}} proposal 12606, PI Martin Barstow). We derived relative astrometry and photometry from these unpublished observations using methods similar to that for the HD 27483 \changes{{\HST}} observation. The only difference is that the WD companion is too close to the host star that we cannot use the host star PSF as a template. Instead, we adopted the observation of HIP 66578 on 2000 January 18 (\changes{{\HST}} proposal 8496, PI Stefano Casertano) as the PSF template for WFPC F170W observations and the previously analyzed image of HD 27483 A as the PSF template for the WFC3 F218W observation. We find that adopting a different image as the PSF template results in positional differences of up to 7 mas, comparable to our statistical uncertainties. We add this to our error budget. The results are listed in Table \ref{tab:new_relast}. 

Other epochs of relative astrometry did not resolve all four components and only measured the photocenter of the outer binary, which has a filter-dependent offset from the barycenter. We chose not to include them in the fit.

\subsubsection{HD 114174} We have 66 high-precision RV measurements between 1997 and 2019, all of which are from HIRES as part of the California Legacy Survey \citep{Rosenthal_2021}. Relative astrometry of the system consists of 4 observations by \cite{Crepp_2013} in 2011 (first discovery) and 2012, one from \cite{Matthews_2014} in 2013, one from \cite{Bacchus_2017} in 2014, and 18 observations by \cite{Gratton_2021} between 2014 and 2019 using the Spectro-Polarimetric High-contrast Exoplanet REsearch \citep[SPHERE,][]{Beuzit_2019} instrument at the \changes{Very Large Telescope (VLT)}. 

\subsubsection{HD 118475}
\cite{Kane_2019} acquired RV data of the system using the
UCLES high-resolution spectrograph \citep{Diego_1990} on the \changes{AAT}. The data consist of 11 observations from 2002 to 2014 and reveal a companion with a minimum mass of $0.445 \Msun$. \cite{Kane_2019} found that a $0.445 \Msun$ M-dwarf would be inconsistent by $3.3\sigma$ with the non-detection in their 880 nm direct imaging observations, showing evidence that the companion is a compact object.

\subsubsection{HD 136138}
\cite{de_Medeiros_1999} first identified the system as a spectroscopic binary using RV observations, and the RV orbit was derived by \cite{Massarotti_2008} and \cite{Griffin_2009}. The {\sl IUE} spectrum of the system shows that the companion is a hot ($T_{\rm eff}=30400 \pm 780$) WD \citep{Stefanik_2011}. \cite{Stefanik_2011} acquired 46 additional measurements of the RV between 2003 and 2009 using the \changes{Harvard-Smithsonian Center for Astrophysics Digital Speedometers} \citep{Mayor_1985, Latham_1992}. We adopt all of them in our fit. Previous RVs dating back to 1924 are available in \cite{Griffin_2009}, but we did not include those data because their measurement uncertainties are unavailable. 

\subsubsection{HD 169889}
We have five epochs of relative astrometry published by \cite{Crepp_2018}, who first discovered the WD companion. Four observations were conducted using the NIRC2 instrument on Keck, and the observation on 2016 June 20 UT came from the \changes{Large Binocular Telescope mid-infrared camera (LMIRCam)} \citep{Skrutskie_2010}. Eight precise RV measurements were obtained with HIRES in the California Planet Survey \citep{Howard_2010}, showing a significant RV trend.

\section{Orbit Fitting} \label{sec:orbit_fit}

\subsection{Method}

We use the orbit-fitting code \textsc{orvara} \citep{Brandt_2021_orvara} to infer the orbital parameters of each system. \textsc{orvara} implements a parallel-tempered MCMC with \textsc{ ptemcee} \citep{emcee_2013,Vousden_2016,ptemcee_2021}. 
At each chain step, it uses \textsc{htof} \citep{HTOF_2021} to model the positions and proper motions of the host star relative to the system's barycenter as they would be seen by the {\Hipparcos} and {\Gaia} intermediate data. \textsc{orvara} then computes the likelihood of each sample orbit given the relative astrometry, HGCA absolute astrometry, and RVs.

We run the MCMC with 20 temperatures, 100 walkers per temperature, and at least 100,000 steps per walker. We discard at least the first 25\% and up to the first 75 \% of the chain as the burn-in. Finally, we thin the chain by a factor of 50. We confirm that the chains have converged by reading \textsc{orvara}'s diagnostic plots that show the path of each walker in the parameter space, checking that every walker has reached the same posterior and has fully sampled the posterior. 

We impose Gaussian mass priors on the host stars, adopting masses in Table \ref{tab:host_prop}, but use $1/M$ mass priors for the companions unless otherwise stated in Section \ref{subsec:orbit_results}. A dynamical mass prior is available for HD~27483~A because the star is itself a spectroscopic binary with {\Hipparcos} astrometry \citep{Konacki_2004}. For stars without previous dynamical mass measurements, we use their most recent isochronological masses as priors. We adopt the {\Gaia} DR3 parallaxes as our parallax priors. We assume the standard geometric prior for the inclination and log-uniform priors for the semimajor axis and the RV jitter. We use uniform priors for all other parameters.

\subsection{Results} \label{subsec:orbit_results}
\begin{figure*}
\captionsetup[subfigure]{labelformat=empty}
    \centering
    \subfloat[]{
    \includegraphics[height=4.4cm,width=0.32\linewidth]{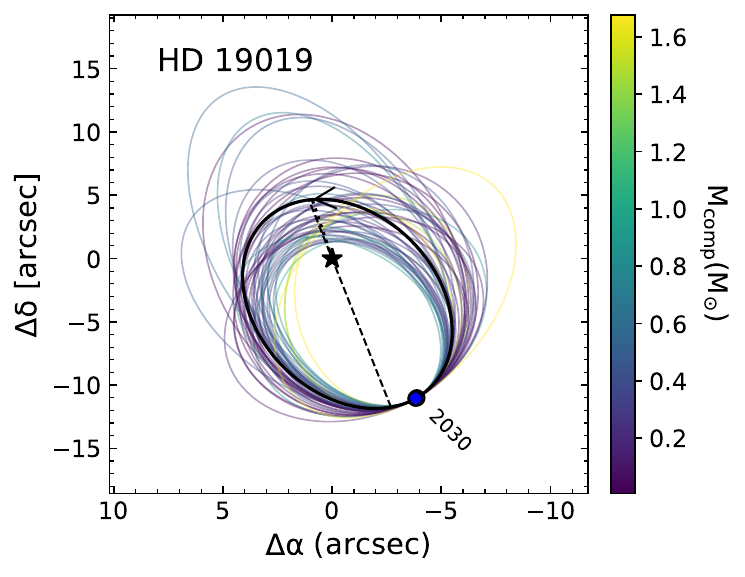}}
    \subfloat[]{
    \includegraphics[height=4.4cm,width=0.32\linewidth]{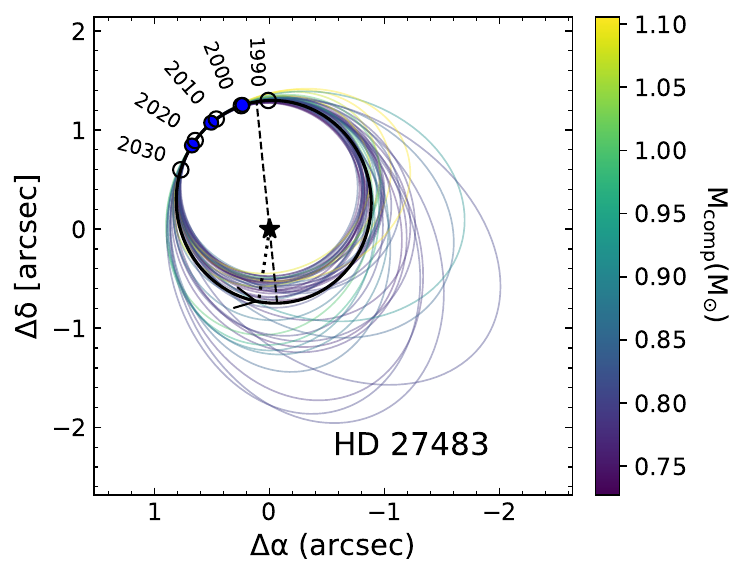}} \\
    \subfloat[]{
    \includegraphics[height=4.4cm,width=0.32\linewidth]{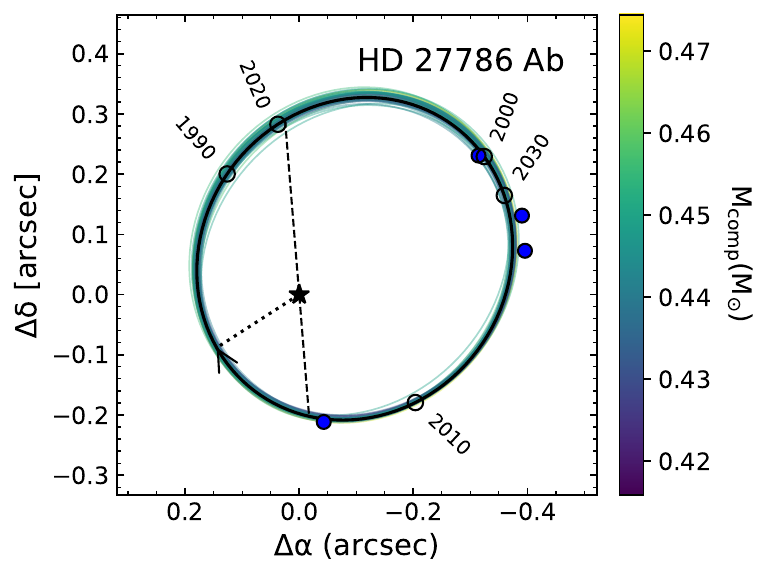}}
    \subfloat[]{
    \includegraphics[height=4.4cm,width=0.32\linewidth]{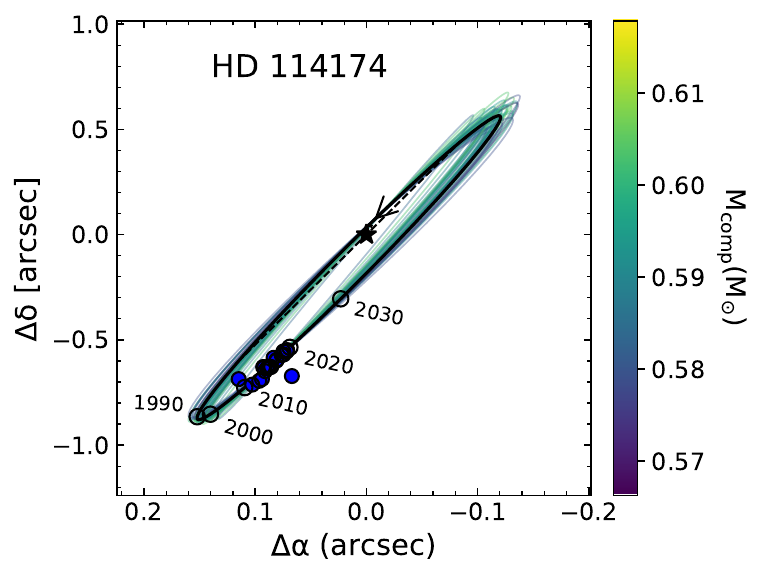}} \\
    \subfloat[]{
    \includegraphics[height=4.4cm,width=0.32\linewidth]{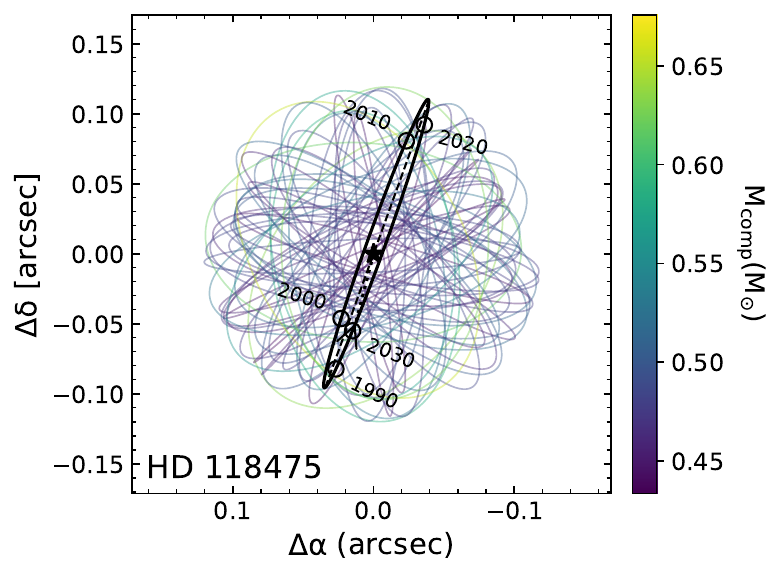}}
    \subfloat[]{
    \includegraphics[height=4.4cm,width=0.32\linewidth]{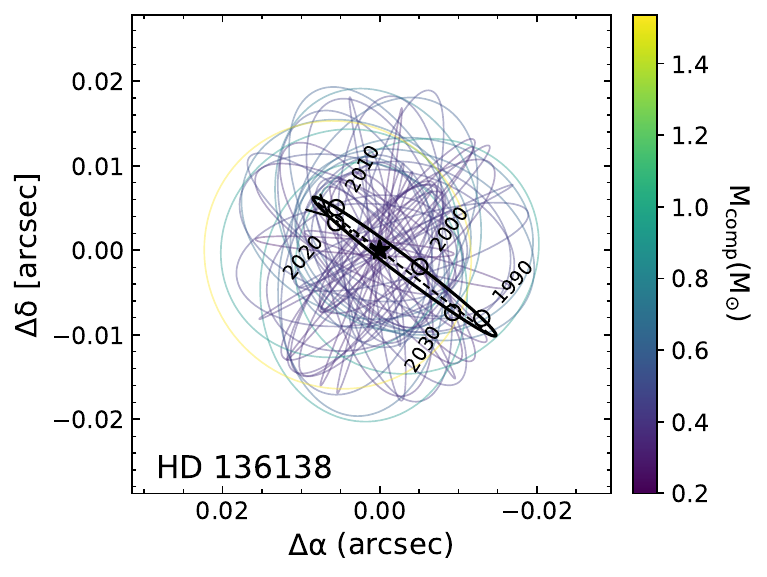}} 
    \subfloat[]{
    \includegraphics[height=4.4cm,width=0.32\linewidth]{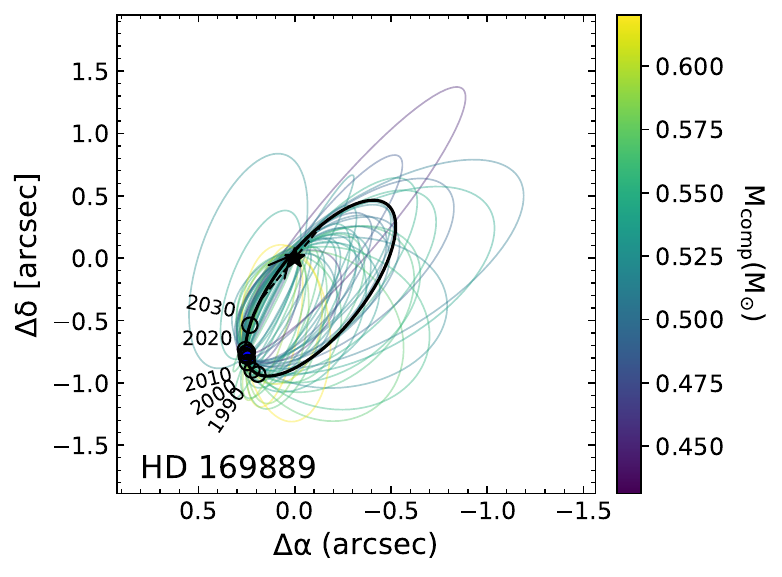}} \\
    \caption{Relative astrometric orbits of companions. The thick black line indicates the maximum likelihood orbit. The thin lines, color-coded by the companion mass, are 50 orbits drawn randomly from the posterior distribution. The dotted line connects the primary star to the periastron. The dashed line is the intersection between the orbital plane and the sky plane. The blue dots show the relative astrometry measurements, with error bars typically smaller than the size of the symbols. The black circles give the predicted positions of the companion at several past and future times along the orbit.}
    \label{fig:orbits}
\end{figure*}

Our MCMC analysis improves the orbital parameters and gives unprecedented dynamical mass constraints for all systems except HD 118475 and HD 136138, for which we have imprecise minimum masses due to not using {\Hipparcos}-{\Gaia} proper motions. Figure \ref{fig:orbits} shows the best-fitting relative astrometric orbits of the systems. \changes{Figure \ref{fig:corner_19019} is the corner plot of the key orbital parameters of HD 19019. Figure \ref{fig:RV_19019} gives the best-fit RV orbit of HD 19019 and the corresponding residuals. Figure \ref{fig:relAst_19019} shows the fit to relative and absolute astrometry of HD 19019 and the corresponding residual. Similar plots for other systems are available as supplementary material in the online version of the paper. Tables \ref{tab:post_19019} to \ref{tab:post_169889} list the priors and posteriors of fitted and derived parameters of all seven systems.} We discuss the details of each system below.

\subsubsection{HD 19019}
The dynamical mass constraint on the WD is loose because of the system's long orbital period and insignificant astrometric acceleration. We place a uniform prior instead of a $1/M$ prior on the companion because the $1/M$ prior would dominate over the weak dynamical mass constraint and give a mass posterior very close to zero. The companion mass posterior of ${0.32}_{-0.23}^{+0.37}$ $M_{\sun}$ is surprisingly in tension with (2-$\sigma$ away from) the spectroscopic mass of $1.12 \pm 0.15$ $M_{\sun}$ in \cite{Landstreet_2020} and does not rule out $M_{\rm comp}=0$. 
The left panel of Figure \ref{fig:RV_19019} shows that the RV data cover only a tiny fraction of the orbit with an insignificant RV trend. Our best-fitting RV jitter of ${12.2}_{-2.0}^{+2.6}$ \ms is within the error bars of the jitter of $13.36\pm 2.86$ \ms derived by \cite{Luhn_2020} that assumed no companions, suggesting that the companion does not induce a measurable RV signal. The proper motion of the $G=6.8$ star in {\Gaia} DR4 will likely be $\sim4$ times more precise than that in {\Gaia} DR3 \citep{Gaia_DR3}. The uncertainties will be better than $0.01$ \masyr and comparable to the star's proper motion change over the baseline of the {\Gaia} mission, allowing substantial improvement in the system's orbital parameters.

\subsubsection{HD 27483}
The host star, HD 27483 A, is itself a tight binary of roughly equal masses and magnitudes, with a period of 3.06 d \citep{Konacki_2004}. We treat the host binary as a single star because the photocenter of its two components almost coincides with the barycenter. 

The WD companion, HD 27483 B, has a long history of spectroscopic observations. \cite{Boehm-Vitense_1993} derived a mass of $\sim 0.6$ $M_{\sun}$ from the WD's \changes{{\sl IUE}} spectrum. \cite{Burleigh_1998} re-analyzed the spectrum and got $M_{\rm WD}\approx 0.94$ $M_{\sun}$. \cite{Joyce_2018} analyzed two \changes{{\HST}} spectra of the WD and got $0.748 \pm 0.072$ $M_{\sun}$ and $0.711 \pm 0.137$ $M_{\sun}$, respectively.

With a $1/M$ prior, our dynamical mass posterior of ${0.798}_{-0.041}^{+0.10}$ $M_{\sun}$ agrees reasonably with and has comparable precision to the spectroscopic masses in \cite{Joyce_2018}. Compared to \cite{Burleigh_1998}, the smaller WD mass implies a longer progenitor lifetime and would bring the total age of the WD to closer agreement with that of the Hyades cluster. We perform a detailed analysis of the WD's age in Section \ref{subsec:age_27483}. The WD's orbital inclination of ${30\degree}_{-15\degree}^{+13\degree}$ is reasonably aligned with the inclination of the host binary's orbit, $45.1\degree \pm 1.7\degree$ \citep{Konacki_2004}. Using a uniform companion mass prior instead of the log-flat prior does not change the MCMC posterior noticeably.

\subsubsection{HD 27786}

While HD~27786 is a quadruple system, we treat the tight outer M-dwarf binary as a single unit and perform a three-body fit with the primary and the inner WD companion. We place a uniform mass prior on the outer binary. The three-body fit to the system has 16 separate parameters and is therefore difficult to converge. Hence, we adopt the orbital parameters for the outer companion in \cite{Tokovinin_2016} as the starting condition for the MCMC to speed up convergence.

Our dynamical mass of $M_{\rm WD}=0.443 \pm 0.012$ $M_{\sun}$ places the companion near the low-mass end in the known sample of WDs. \cite{Landsman_1996} fitted possible spectroscopic masses of the WD for a grid of assumed distances. On their grid, the closest point to the {\Gaia} DR3 distance of $42.0 \pm 0.2$ pc was at 43 pc, for which they derived a mass of $0.435$ $M_{\sun}$. The mass is within 1-$\sigma$ of our dynamical mass.

If we assume a standard evolutionary track, such a small dynamical mass implies a $< 1$ $M_{\sun}$ progenitor, which does not have enough time to evolve off the MS within the primary star's age of ${2.11}_{-0.79}^{+1.26}$ Gyr \citep{David_2015}. One plausible explanation is that the WD had a $\sim 2$ $M_{\sun}$ progenitor but underwent faster-than-usual mass loss in its \changes{asymtotic giant branch (AGB)} phase due to interactions with the primary star, such as common envelope evolution \citep[for a review, see][]{Ivanova_2013}. This scenario is not uncommon in compact binary systems with periods of days \citep{Kilic_2007,Brown_2011} but is not known to happen for wider WD companions. Future spectroscopic observations of the system may provide evidence for or against such interaction.

Because the inner WD companion contributes $> 99\%$ of the astrometric acceleration, it is unrealistic to tightly constrain the mass of the outer binary companion. The dynamical mass of ${1.36}_{-0.41}^{+0.51}$ $M_{\sun}$ agrees loosely with the total photometric mass of the two companions, $0.84$ $M_{\sun}$ \citep{Tokovinin_2016}. Our result of $P=1400 \pm 200$ yr and $a=4.4 \pm 0.4$ arcsec is marginally consistent with $P = 1945 \pm 535$ yr and $a = 7.5 \pm 2.7$ arcsec from \cite{Izmailov_2019}.

\subsubsection{HD 114174}

Literature spectroscopic masses of the WD companion are discrepant, depending on assumptions on core composition and the specific WD-evolution model applied. \cite{Matthews_2014} reported a mass of $0.54 \pm 0.01$ $M_{\sun}$ using the theoretical models in \cite{Tremblay_2011} for a pure-H atmosphere and a C/O core. \cite{Bacchus_2017}, however, added new \changes{spectroscopy} and got $M=0.336 \pm 0.014$ $M_{\sun}$ for low-mass He-core evolutionary models \citep{Althaus_2001} and $M=1.198 \pm 0.006$ $M_{\sun}$ for high-mass C/O core evolutionary models \citep{Fontaine_2001}. Recent observations with SPHERE \citep{Gratton_2021} gave $M=0.75 \pm 0.03$ $M_{\sun}$, derived from pure-H C/O core models in \cite{Bergeron_1995}. The only previous measurement of the companion's dynamical mass was from \cite{Crepp_2013}, who inferred a minimum mass of $0.260 \pm 0.010$ $M_{\sun}$ from the host star's RV trend.

 Our mass of $M=0.591 \pm 0.011$ $M_{\sun}$ is the first precise dynamical mass for the WD companion and agrees with the dynamical mass lower limit from \cite{Crepp_2013}. Comparing to the WD's spectroscopic masses, it is the closest ($\approx 3\sigma$) to the mass of $0.54 \pm 0.01$ $M_{\sun}$ by \cite{Matthews_2014}, suggesting that the WD has a C/O core. Our orbital period of $108.2_{-4.8}^{+5.2}$ yr and inclination of $88.86\degree \pm 0.21\degree$ agree reasonably with $P\approx124$ yr and $i=88.11\degree \pm 0.11\degree$ published by \cite{Gratton_2021}. Our semimajor axis of $26.25_{-0.66}^{+0.71}$ AU and eccentricity of $0.690 \pm 0.017$ are slightly smaller than $a=30.11 \pm 0.03$ AU and $e=0.89$ by \cite{Gratton_2021} because those authors assumed a companion mass higher than the dynamical mass we found. Other orbital parameters of the system are summarized in Table \ref{tab:post_114174}.

\subsubsection{HD 118475}
Because we fitted only to the RV data, there are no informative constraints on the orbital inclination. Instead of a precise dynamical mass, we obtain a minimum mass of $M_{\rm sec}\sin{i}={0.461}_{-0.015}^{+0.017}$ $M_{\sun}$, agreeing with $M_{\rm sec}\sin{i}=0.445 \pm 0.025$ from \cite{Kane_2019}, using the same RVs but a different fitting procedure. Other parameters ($a={3.76}_{-0.06}^{+0.07}$ AU, $P=2070.47 \pm 0.16$ d, $e=0.128 \pm 0.001$, $\omega=237.8\degree \pm 0.2\degree$) are nearly identical to the values in \cite{Kane_2019} ($a=3.69 \pm 0.11$ AU, $P={2070.47}_{-0.2}^{+0.19}$ d, $e=0.128 \pm 0.001$, $\omega=237.7\degree \pm 0.3\degree$). The complete set of orbital parameters is listed in Table \ref{tab:post_118475}.

In {\Gaia} DR3, HD 118475 has a two-body solution, given in terms of $P$, $e$, time of periastron $T_0$, and the Thiele-Innes constants $A$, $B$, $F$, $G$ in units of mas (multiplied by the semi-major axis of the photocenter motion $a_0$). We first convert the Thiele-Innes constants into physical orbit parameters with the following equations:
\begin{align} \label{eq:orbit_para}
    a_0&=\sqrt{p+\sqrt{p^2-q^2}}  \\
    i&=\cos^{-1}{\frac{q}{a_{0}^2}}  \\ 
    \omega&=\frac{r+s}{2}  \\
    \Omega&=\frac{r-s}{2}
\end{align}
with $p$, $q$, $r$, $s$ defined to be

\begin{align}
    p &\equiv \frac{1}{2}\left(A^2+B^2+F^2+G^2\right) \\
    q &\equiv AG-BF \\
    r &\equiv \tan^{-1}\left(\frac{F-B}{G+A}\right) \\
    s &\equiv \tan^{-1}\left(\frac{F+B}{G-A}\right)
\end{align}

Because the primary star is much brighter than the companion, the system's photocenter almost coincides with the primary. Therefore, the semi-major axis of the photocenter motion is approximately that of the primary motion, which is related to the total semi-major axis $a$ by:
\begin{equation} \label{eq:semi-major}
    a=\frac{M_{\rm tot}}{M_{\rm sec}} a_0
\end{equation}
We also have Kepler's Third Law:
\begin{equation} \label{eq:Kepler_third}
    P^2 = \frac{a^3}{M_{\rm tot}}
\end{equation}
where $P$ is in years, $a$ is in AU, and $M_{\rm tot}=M_{\rm pri}+M_{\rm sec}$ is in solar masses.

Combining Equations \eqref{eq:semi-major} and \eqref{eq:Kepler_third}, we get
\begin{equation} \label{eq:solve_mass}
    M_{\rm sec}^3-(M_{\rm pri}+M_{\rm sec})^2 \frac{a_0^3}{P^2}=0
\end{equation}
from which we solve for the companion mass $M_{\rm sec}$, given a primary mass of $M_{\rm pri}=1.16 \pm 0.06$ $M_{\sun}$ \citep{Kervella_2019}.

The last two columns of Table \ref{tab:post_118475} compare the orbital elements from our RV fit to those from the {\Gaia} DR3 astrometric solution. Both solutions tightly constrain $M_{\rm sec}$, $a$, $P$, $e$, and the argument of periastron $\omega$, and they agree remarkably well. Once the {\Gaia} IAD are published, a joint analysis of {\Hipparcos} IAD, {\Gaia} IAD, and RV will place tight constraints on all orbital parameters. Our analysis is an example of verifying a {\Gaia} orbital solution against orbital parameters derived by other means. The same procedure could be applied to a large sample of binary systems with known orbits to validate the {\Gaia} two-body solutions.

Our $M_{\rm sec}\sin{i}$ of ${0.461}_{-0.015}^{+0.017}$ $M_{\sun}$ and the {\Gaia} DR3 inclination of ${53.\!\!\degree1}_{-5.\!\!\degree5}^{+4.\!\!\degree8}$ give a dynamical mass of ${0.580}_{-0.039}^{+0.052}$ $M_{\sun}$. The precise dynamical mass is much greater than the minimum mass of $\approx 0.445$ $M_{\sun}$ in \cite{Kane_2019}, implying that if the companion were a MS star, it would have a magnitude more inconsistent with the non-detection in the direct imaging observations by \cite{Kane_2019} than previously derived. Using \textsc{orvara}, we confirm that the separation between the primary star and the companion was $\approx0.09$ arcsec at the observation epoch, so we refer to Figure 3 of \cite{Kane_2019} as the non-detection significance curve. According to evolutionary tracks \changes{in MESA Isochrones and Stellar Tracks (MIST)} \citep{MIST_2016}, a 0.58 $M_{\sun}$ MS star would be only $\approx3.87$ mag fainter, near 880 nm, than the 1.16 $M_{\sun}$ primary at the current age ($\sim4.1$ Gyr) and metallicity ($\rm{[Fe/H]}=0.10$) of the system \citep{Valenti_2005}. Hence, we can rule out a MS companion at a significance level of more than $8\sigma$. The dynamical mass also eliminates the possibility that the companion is a neutron star or a black hole. Therefore, we conclude that HD 118475 B is a WD.

\subsubsection{HD 136138}
\cite{Stefanik_2011} inferred a WD mass of $M_{\rm WD}=0.79 \pm 0.09$ $M_{\sun}$ from the {\sl IUE} spectrum of the system \citep{Landsman_1996}, using C/O-core cooling models in \cite{Wood_1995}. The authors also obtained a dynamical mass of $0.59 \pm 0.12$ $M_{\sun}$ from a joint fit of {\Hipparcos} IAD and RV. Analyzing only the RVs, we get a minimum dynamical mass of $M\sin{i}={0.389}_{-0.080}^{+0.081}$ $M_{\sun}$, agreeing well with the dynamical mass and $i=42.9\degree \pm 6.7\degree$ in \cite{Stefanik_2011} ($M\sin{i}={0.395}^{+0.099}_{-0.091}$ $M_{\sun}$) but only marginally with the spectroscopic mass. With a semimajor axis of only $a={1.65}_{-0.12}^{+0.10}$ AU, it is possible that the progenitor of the WD once transferred some of its mass to the giant primary. \changes{We investigate this possibility in Section \ref{subsec:pre_interact}.} Our period of $509.6 \pm 1.2$ d, eccentricity of $e={0.336}_{-0.014}^{+0.015}$, and RV semi-amplitude of ${6.22}_{-0.11}^{+0.14}$ \kms all agree reasonably with $P=506.45 \pm 0.18$ d, $e=0.3353 \pm 0.0056$, and $K=6.340 \pm 0.044$ \kms in \cite{Stefanik_2011}.

HD 136138 also has a two-body solution in {\Gaia} DR3. We derive the corresponding orbital elements with Equations \eqref{eq:orbit_para} to \eqref{eq:solve_mass} and list the results in the last column of Table \ref{tab:post_136138}. The {\Gaia} solution gives a semi-major axis and $M\sin{i}$ almost identical to our RV solution, but it has a slightly larger ($1.7\sigma$) eccentricity, shorter (by $1.7\sigma$) period, and larger (by $3.2\sigma$) argument of periastron, all in moderate tension with our values.

\subsubsection{HD 169889}
Our dynamical mass of ${0.526}_{-0.037}^{+0.039}$ $M_{\sun}$ is the first precise mass of the WD, consistent with the dynamical mass lower limit of $0.369 \pm 0.010$ $M_{\sun}$ from \cite{Crepp_2018}. The eccentricity of $e={0.896}_{-0.088}^{+0.064}$ is the largest in our sample. A large eccentricity is not \changes{uncommon} because the eccentricity distribution of wide MS binaries with separations smaller than 100 AU is nearly uniform \citep{Hwang_2022}, and the isotropic and adiabatic mass loss of a star during the AGB phase does not modify the eccentricity of its orbit \citep{Dosopoulou_2016} \changes{unless it once tidally interacted with a close companion. We discuss the possibility of such close tidal interactions in Section \ref{subsec:pre_interact}.} Our fit to relative separation has a $\chi^2$ of 32 for only five observations, suggesting that the relative astrometry in \cite{Crepp_2018} may have underestimated uncertainties. Inflating the uncertainties to get a reduced $\chi^2$ of 1 does not affect the results significantly. The WD only has two broadband photometry measurements in the $H$-band and the $L'$-band, which are insufficient to precisely measure its $\Teff$, $\logg$, or photometric mass \citep[\changes{see}][\changes{and our attempt in Section \ref{subsec:MRR}}]{Crepp_2018}. Additional photometry or spectroscopy will be necessary to characterize the fundamental properties of the WD.

\section{WD Age Inference} \label{sec:age}
A dynamical mass comes directly from the solution to the Kepler problem, independent of stellar models and their theoretical uncertainties. Hence, it is a reliable starting point to constrain the fundamental parameters of a WD or test WD evolutionary models. A Sirius-like binary system is an especially ideal testing ground because the age of the system is often known by measuring the activity or the rotation of the MS star. The dynamical mass of the WD, closely related to the WD's cooling age and progenitor lifetime, allows for an independent measurement of the system's age. In this section, we outline our methods to infer the MS lifetimes and the cooling ages of the WDs and discuss our results for each system.

\subsection{MS Lifetime}
The mass of a WD implies the mass of its MS progenitor, and the progenitor mass determines the progenitor's lifetime before evolving into a WD. We adopt the empirical IFMR in \cite{El-Badry_2018} to compute the likelihood of a WD's progenitor mass from its dynamical mass posterior in the MCMC chain. Then, we multiply the distribution with the initial-mass function (IMF) in \cite{Chabrier_2003} (corrected for binaries) to obtain the posterior distribution of the initial mass. Finally, we perform cubic spline interpolation on MIST evolutionary tracks \citep{MIST_2016} to find the MS lifetime distribution from the initial mass posterior, assuming that the WD companion has the same metallicity as the MS primary. We remove any probability density beyond the age of the universe \citep[13.8 Gyr]{Planck_Collab_2020} and re-normalize the distribution.

Our analysis is subject to potential systematics in our choice of IFMR, IMF, and stellar evolutionary model. We repeat the above procedures with the \cite{Cummings_2018} IFMR, the \cite{Kroupa_2001} IMF, and evolutionary tracks \changes{from the PAdova and TRieste Stellar Evolution Code \citep[PARSEC,][]{Bressan_2012}}, confirming that our conclusions in Section \ref{subsec:age_results} do not change significantly.

\subsection{Cooling Age}
The cooling age of a WD relates primarily to the starting condition of cooling (decided by its mass) and by how much it has cooled (its present effective temperature). A dynamical mass alone does not constrain the cooling age, but when paired with a {\Teff} measurement or multi-band photometry, it tightens existing constraints. We perform multi-dimensional linear interpolation on the latest generation of Montr{\'e}al cooling sequences\footnote{http://www.astro.umontreal.ca/~bergeron/CoolingModels} \citep{Bedard_2020} to compute the cooling age from dynamical mass and other parameters\changes{, assuming a pure-hydrogen (DA) atmosphere and a C/O core}. The synthetic photometry in the Montr{\'e}al cooling sequences is computed using prescriptions in \cite{Holberg_2006} and relies on models of \cite{Blouin_2018}, \cite{Tremblay_2011}, and \cite{Bedard_2020} for low, intermediate, and high-temperature DA WDs, respectively. Our interpolation method is similar to that in \cite{Kiman_2022}.

\subsection{Results} \label{subsec:age_results}
We place informative constraints on the ages of all WDs in this work except HD 27786 Ab, HD 118475 B, and HD 136138 B. The unusually low mass of HD 27786 Ab is smaller than the applicable range of the \cite{El-Badry_2018} IFMR and the \cite{Cummings_2018} IFMR and is not explainable by standard single-star evolutionary tracks. One needs to consider binary interaction to infer the age of the WD. HD 118475 B and HD 136138 B only have dynamical mass lower limits. The corresponding upper limits on age are over the age of the universe. We present our results for the rest of the systems below.

\begin{figure*}
\includegraphics[height=4.6cm,width=0.32\linewidth]{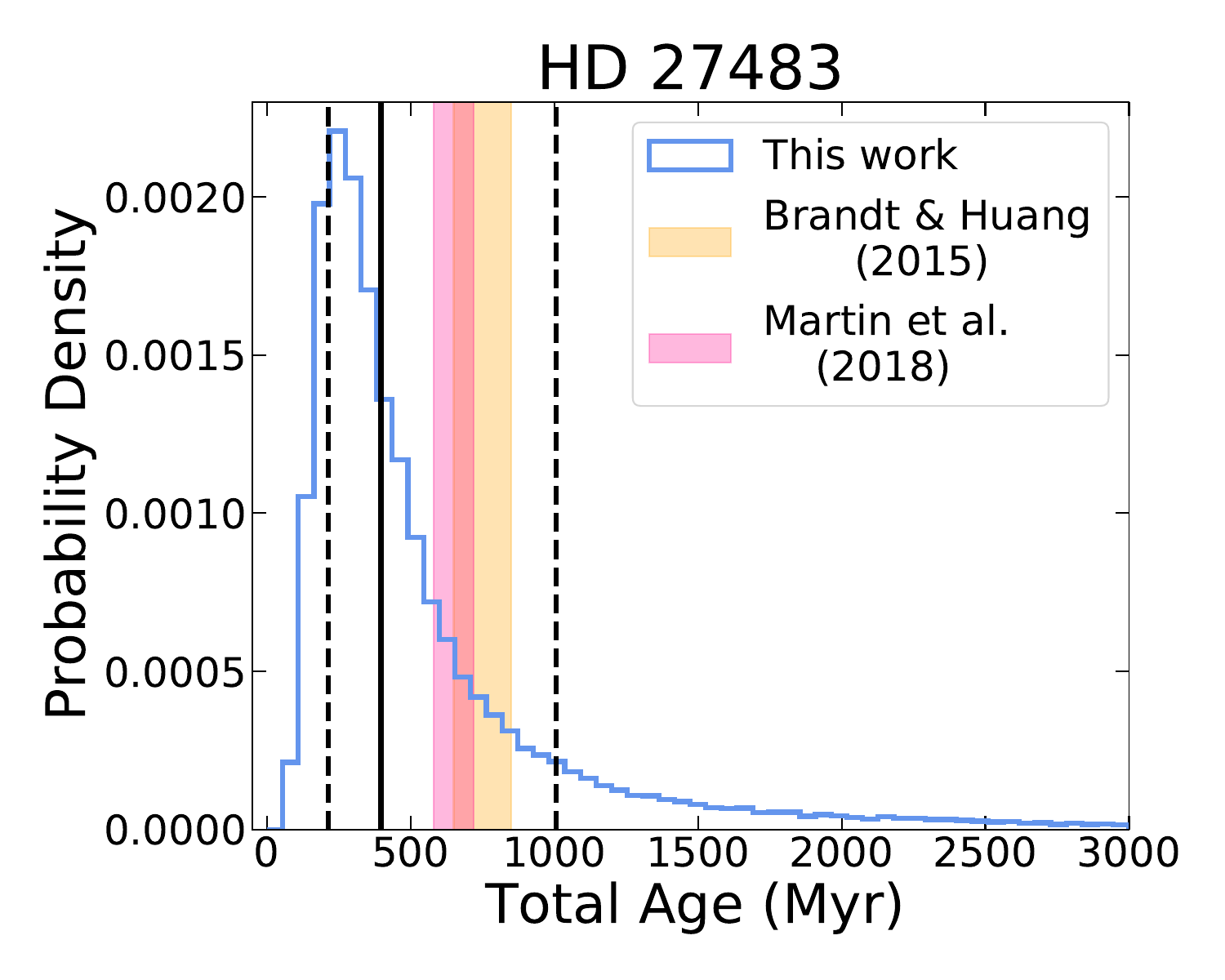}
\includegraphics[height=4.6cm,width=0.32\linewidth]{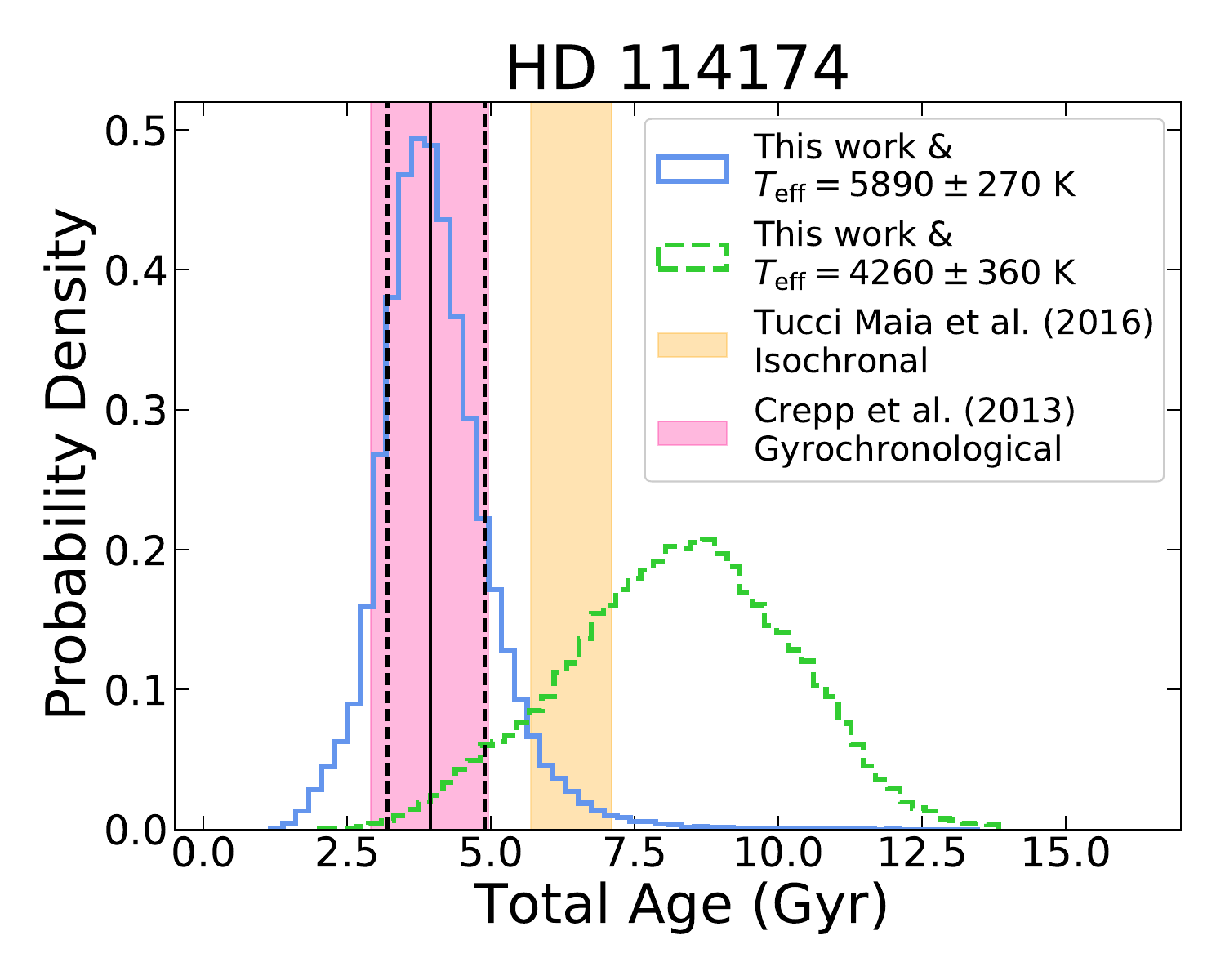}
\includegraphics[height=4.6cm,width=0.32\linewidth]{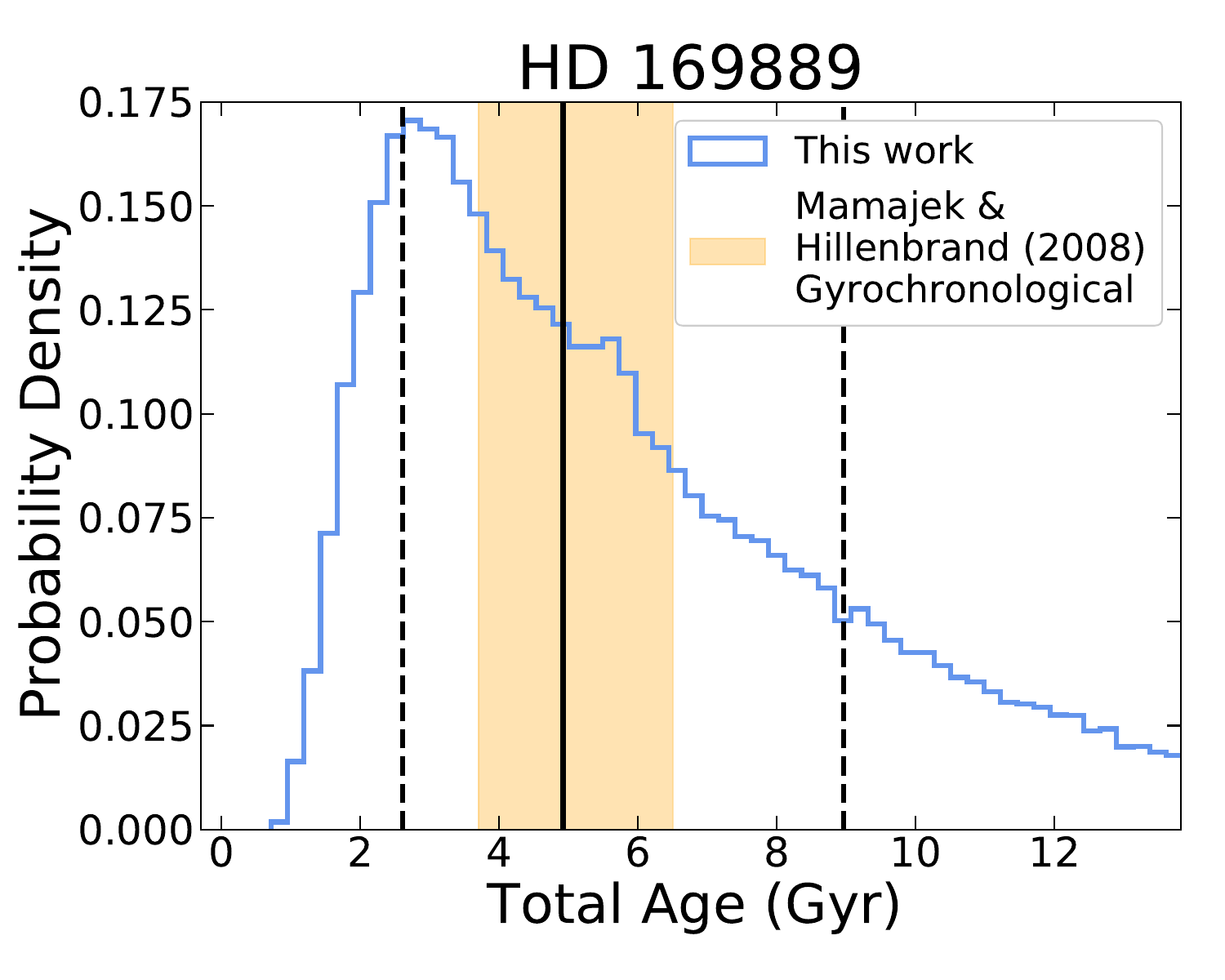}
    \caption{Comparison of the WD age posteriors (plotted as histograms) derived from our dynamical masses to the $1\sigma$ intervals (represented by shaded regions) of previous age measurements. In each panel, the vertical solid black line marks the 50th percentile of the distribution, while the two dashed back lines correspond to the 16th and the 84th percentiles. Left panel (HD 27483): our result does not resolve the age tension between \citet{Brandt_2015} (gold region, the age of the Hyades cluster derived from rotating isochrones) and \citet{Martin_2018} (pink region, the lithium depletion boundary age of the Hyades cluster). Middle panel (HD 114174): two discrepant $\Teff$ give different ages. The age derived from $\Teff=5890 \pm 270$ K \citep[solid blue histogram,][]{Gratton_2021} agrees nicely with the primary's gyrochronological age (pink region, \citealp{Crepp_2013}), but the result from $\Teff=4260 \pm 360$ K \citep[dashed green histogram,][]{Matthews_2014} agrees better with the primary's isochronal age (orange region, \citealp{Tucci_Maia_2016}). Right panel (HD 169889): our age constraint is loose but matches well with the primary's gyrochronological age (orange region, \citealp{Mamajek_2008}).}
    \label{fig:age_posterior}
\end{figure*}

\subsubsection{HD 19019}
The precision of our dynamical mass is insufficient to place good constraints on the MS lifetime of HD 19019 B. A considerable fraction of the mass posterior is below the applicable range of any IFMRs. If we naively apply the \cite{El-Badry_2018} IFMR, we would obtain a progenitor mass of $M_{\rm pro}={2.8}_{-1.3}^{+2.3}$ $M_{\sun}$ and a maximum likelihood MS lifetime of $\sim 0.5$ Gyr (assuming $[\rm{Fe}/\rm{H}]=-0.15$ as in \citealp{Arentsen_2019}), but we cannot rule out any lifetime within the age of the universe. If we instead adopt the spectroscopic mass of $1.12 \pm 0.15$ $M_{\sun}$ \citep{Landstreet_2020}, we would get a progenitor mass of $M_{\rm pro}={6.3}_{-1.7}^{+1.9}$ $M_{\sun}$ and a MS lifetime of ${110}_{-60}^{+320}$ Myr.

Because the spectroscopic analysis in \cite{Landstreet_2020} gives a relatively hot {\Teff} of $18200 \pm 3000$ K, but our dynamical analysis favors small masses, we get a young cooling age of ${51}_{-44}^{+157}$ Myr. If adopting $\logg=8.85 \pm 0.15$ in \cite{Landstreet_2020} instead of our dynamical mass, we obtain a much older cooling age of ${530}_{-220}^{+380}$ Myr. Given that the age of the MS primary is $\sim 3.0$ Gyr \citep{Landstreet_2020}, if the WD had a mass of $\sim 0.54$ $M_{\sun}$, it would have a cooling age of $\sim 0.1$ Myr and a MS lifetime of $\sim 2.9$ Myr for a $\sim 1.4$ $M_{\sun}$ progenitor, giving a consistent total age. This mass is within the 1-$\sigma$ interval of our dynamical mass.

\subsubsection{HD 27483} \label{subsec:age_27483}
HD 27483 is a member of the Hyades cluster. For the cluster, different age determination methods give values in moderate tension.

\cite{Perryman_1998} determine an age of $625 \pm 50$~Myr from non-rotating stellar models.
\cite{Brandt_2015} infer an age of $750 \pm 100$ Myr from rotational stellar models from \cite{Ekstrom+Georgy+Eggenberger+etal_2012}. \cite{Gossage+Conroy+Dotter+etal_2018} infer a somewhat younger age of $\approx$680~Myr from models with a different treatment of stellar rotation \citep{MESA_rotation}.  \cite{Martin_2018} report a younger age of $650 \pm 70$ Myr by determining the lithium depletion boundary of two members. \cite{DeGennaro_2009} obtained an age of $648 \pm 45$ Myr by fitting the WD portion of the color-magnitude diagram to stellar evolution models. Here, we present an independent age estimate for the Hyades by inferring the age of HD 27483 B.

The ${0.798}_{-0.041}^{+0.10}$ $M_{\sun}$ WD has a ${3.59} _{-0.78}^{+1.43}$ $M_{\sun}$ progenitor, which lived a lifetime of ${350}_{-180}^{+570}$ Myr, assuming a metallicity of $[\rm{Fe}/\rm{H}]=0.04$ \citep{Bochanski_2018}. The imprecision originates mainly from the uncertainty in the empirical IFMR. We neglect uncertainties in the MS lifetime from, e.g., the effects of stellar rotation \citep{Brandt_2015}; these would be $\lesssim$20\% of the age. Using a theoretical IFMR, like the MIST IFMR described in \cite{MIST_2016}, would reduce statistical uncertainties but overlook the systematics in theoretical modeling.

The photometry of a WD constrains its effective temperature, mass and hence the cooling age. To find the magnitudes of HD 27483 B from the $\Delta m$ measurements in Table \ref{tab:new_relast}, we need precise magnitudes of the host star in the two filters. We adopt $K_{\rm pri}=5.062 \pm 0.018$ from \cite{Cutri_2003} as the $K$-band apparent magnitude of the primary. However, we are unaware of any measurement of the primary's F218W magnitude. Hence, we apply the \textsc{species} package \citep{Stolker_2020} to compute the magnitude from synthetic photometry. We modeled the spectrum of the primary with the BT-NextGen model \citep{Allard_2012}, adopting $\Teff=6549 \pm 80$ K, $\logg=4.04$ from \cite{Casagrande_2011} and $R={2.02}^{+0.10}_{-0.11}$ $R_{\sun}$ and $\rm{[Fe/H]}={0.04}^{+0.15}_{-0.13}$ from \cite{Bochanski_2018}. Integrating the model spectrum with the F218W filter, we obtain an F218W absolute magnitude of ${5.29}_{-0.24}^{+0.26}$ for HD 27483 A. Finally, we add the host star magnitudes to the contrast values in Table \ref{tab:new_relast} and get $M_{K}=11.47 \pm 0.06$ and $M_{\rm F218W}=8.40 \pm 0.30$ for the WD companion.

Comparing the WD absolute magnitudes to the Montr{\'e}al cooling sequences, we get a cooling age of ${61}_{-17}^{+21}$ Myr, an effective temperature of $21000 \pm 3000$ K, and a photometric mass of ${0.669}_{-0.066}^{+0.079}$ $M_{\sun}$. Our $\Teff$ agrees well with the spectroscopic $\Teff$ of $20790 \pm 187$ K \citep{Joyce_2018} but has a much larger uncertainty. The photometric mass is in mild tension with ($1.4\sigma$ below) our dynamical mass. The observed magnitudes depend mainly on the WD's $\Teff$ and size, while the size depends on the WD's mass via the mass-radius relation \citep[e.g.,][]{Joyce_2018, Romero_2019, Chandra_2020}. The Montr{\'e}al cooling sequences match the photometry of HD 27483 B at a radius larger than that expected for a 0.8\,$M_\odot$ WD, suggesting a lower mass. The lower mass would, in turn, imply a less massive progenitor and a longer MS lifetime (and older Hyades age). We note that with $\Teff\approx21000$ K, the WD has its emission peak at about 1400 {\AA}, much bluer than both photometric bands. Additional photometry at shorter wavelengths and a more precise dynamical mass will hopefully resolve the discrepancy.

Combining all of our constraints, we perform a joint analysis of photometry and dynamical mass, treating the dynamical mass as a prior when fitting the cooling sequences to the photometry. We get a mass posterior of ${0.763}_{-0.026} ^{+0.034}$ $M_{\sun}$ and a cooling age of ${45}_{-9}^{+12}$ Myr. The cooling time is slightly smaller than that from photometry alone due to the mass tension.

Adding the cooling age from the joint analysis to the MS lifetime, we get a total age of ${400}_{-180}^{+570}$ Myr. The left panel of Figure \ref{fig:age_posterior} compares our age posterior (blue histogram) to the $1\sigma$ intervals of previous age estimates (light orange region for \cite{Brandt_2015} and light pink region for \cite{Martin_2018}). Our current analysis does not have the precision to resolve the age tension, mainly due to the uncertainties in the semi-empirical IFMR. Reducing the dynamical mass uncertainty would improve the cooling age but have negligible effects on the precision of the MS lifetime. On the contrary, if one could reduce the uncertainty of the IFMR by a factor of 4, they would constrain the total age to 10\% ($\approx 70$ Myr), comparable to the precision of other age estimates of Hyades.

\subsubsection{HD 114174} \label{subsec:age_114174}
The \cite{El-Badry_2018} IFMR gives a progenitor mass of ${1.90}_{-0.27}^{+0.30}$ $M_{\sun}$ for the $0.591 \pm 0.011$ $M_{\sun}$ WD. The \cite{Cummings_2018} IFMR suggests a smaller progenitor of $1.28 \pm 0.47$ $M_{\sun}$. We adopt the former mass because \cite{El-Badry_2018} has more low-mass ($<0.6$ $M_{\sun}$) WDs in their calibration sample. Comparing the progenitor mass to MIST tracks, we get a MS lifetime of ${1.46}_{-0.48}^{+0.84}$ Gyr (for a metallicity of $[\rm{Fe}/\rm{H}]= 0.056$ as in \citealp{Casali_2020}).

Previous works on the WD reported disagreeing effective temperatures leading to discrepant cooling ages. Our dynamical mass and $\Teff = 4260 \pm 360$ K from \cite{Matthews_2014} combing spectroscopy and photometry suggests a cooling age of $t_{\rm cool}={8.3}_{-2.1}^{+1.9}$ Gyr, but assuming $\Teff = 5890 \pm 270$ K from \cite{Gratton_2021} using the SPHERE \changes{integral-field spectrograph} instead gives $t_{\rm cool}={2.40}_{-0.41}^{+0.52}$ Gyr. Other $\Teff$ measurements are too imprecise to constrain the cooling age. Summing the MS lifetime and the cooling age, the former $\Teff$ corresponds to a total age of ${8.3}_{-2.1}^{+1.9}$ Gyr, while the latter gives a much younger age of ${3.93}_{-0.74}^{+0.98}$ Gyr. The middle panel of Figure \ref{fig:age_posterior} compares the total age posterior of the WD, for different $\Teff$, to the $1\sigma$ intervals of existing age measurements for the primary. The total age from our dynamical mass and the \cite{Gratton_2021} $\Teff$ agrees remarkably with the primary's gyrochronological age by \cite{Crepp_2013} but only marginally with the isochronal age from \cite{Tucci_Maia_2016}. On the contrary, the much older age from the \cite{Matthews_2014} $\Teff$ is consistent with \cite{Tucci_Maia_2016} but in moderate tension with \cite{Crepp_2013}. Additional spectroscopy will be necessary to resolve the $\Teff$ discrepancy and confirm the age of the WD.

\subsubsection{HD 169889} \label{subsec:age_169889}
The relatively light (${0.526}_{-0.037}^{+0.039}$ $M_{\sun}$) WD has a ${1.33}_{-0.26}^{+0.38}$ $M_{\sun}$ progenitor, with a MS lifetime of ${4.50}_{-2.38}^{+3.90}$ Gyr (assuming $[\rm{Fe}/\rm{H}]=-0.14$ as in \citealp{Brewer_2016}). Due to the uncertainty of the IFMR, our progenitor mass is not precise enough to rule out low-mass progenitors that lived up to the age of the universe.

\cite{Crepp_2018} could not determine the WD's precise $\Teff$ from their $H$-band and $L'$-band photometry, proposing two different temperatures consistent with observation. The cooler temperature, $\Teff \approx 2150$ K, plus our dynamical mass, would bring the cooling age to $\approx 12$ Gyr, much longer than the primary's gyrochronological age of ${5.2}_{-1.5}^{+1.3}$ Gyr \citep{Mamajek_2008}. The maximum likelihood total age would be even beyond the universe's age. A WD this cool and old has never been detected. Hence, we adopt the hotter temperature, $\Teff \approx 10000$ K, assuming a conservative uncertainty of 1000 K. We find a young cooling age of ${0.48}_{-0.15}^{+0.20}$ Gyr. The WD's total age is, thus, ${5.0}_{-2.4}^{+3.9}$ Gyr, in good agreement with the primary's gyrochronological age. The right panel of Figure \ref{fig:age_posterior} displays our age posterior on top of a shaded region representing the $1\sigma$ interval of the gyrochronological age. We cut the tail of the distribution at the universe's age. The age constraint could be considerably improved if one obtains a more precise IFMR in the low-mass range, ruling out light progenitors with extremely long lifetimes.

\changes{\section{Discussion} \label{sec:discuss}
\subsection{Orbital Evolution and Possible Previous Interactions in Sirius-like Systems}\label{subsec:pre_interact}}
\changes{A long-standing puzzle for many Sirius-like systems, including Sirius itself, is the lack of evidence of previous interactions despite the proximity of the binary during the WD progenitor's AGB phase \citep[see e.g.,][]{Oomen_2018}. For example, the progenitor of Sirius B was only $\sim$ 1.5-1.6 AU from Sirius A at the periastron, which is smaller than its radius in the AGB phase. However, the spectrum of Sirius shows no evidence of a common-envelope event, nor did any component of Sirius deviate significantly from single-star evolutionary tracks \citep{Bond_2017b}. In addition, it is expected that when a star fills its Roche lobe substantially, its orbit tidally circularizes on a timescale shorter than that of stellar evolution \citep[e.g.,][]{BM_2008}, but Sirius B's orbit remains eccentric \citep[$e=0.59$,][]{Bond_2017b}.}

\changes{To investigate if our Sirius-like systems present similar puzzles, we derive the separations between the WD progenitors and their companions using their present-day orbital parameters. We assume that the mass loss is adiabatic (on a timescale much longer than the orbital period) so that \citep[see derivations in, e.g.,][]{Dosopoulou_2016}:}
\changes{
\begin{align}
M_{\rm tot} \hspace{0.05cm} a&=\rm constant \\
e&=\rm constant
\end{align}}
\changes{Hence, we can compute the semi-major axis of the progenitor's orbit, $a_{\rm pro}$, from the current semi-major axis, the WD's dynamical mass, the progenitor mass given by the IFMR, and the mass of the WD's companion. Then, we take the progenitor's orbital eccentricity to equal the present eccentricity, assuming that the orbit did not tidally circularize. Table \ref{tab:prog_orbit} lists the eccentricity, the progenitor semi-major axis, the progenitor mass, and the mass ratio between the progenitor and its companion (now the primary star), $q_{\rm pro}=M_{\rm pro}/M_{\rm pri}$, for five WDs. We take the present-day masses of HD 118475 B and HD 136138 B to be the masses from our $M\sin{i}$ and the {\Gaia} DR3 inclinations. HD 19019 B and HD 27786 Ab are excluded because their dynamical masses are too small for existing IFMRs to give reliable progenitor masses.}

\changes{The approximate radius of the Roche lobe of a star a separation $A$ from its companion is given by \citep{Eggleton_1983}:
\begin{equation}
\frac{R_{\rm Roche}}{A}=\frac{0.49q^{2/3}}{0.6q^{2/3}+\ln{\left(1+q^{1/3}\right)}}
\end{equation}
Here, we take $A$ to be the separation between the WD progenitor and its companion at the periastron, $A=a_{\rm pro}(1-e)$. This gives the progenitor's minimum Roche lobe radius along its orbit, $R_{\rm Roche, min}$. We list the value of $R_{\rm Roche, min}$ for each WD in the last column of Table \ref{tab:prog_orbit}.} 

\changes{
HD 114174 B, HD 118475 B, HD 136138 B and HD 169889 B have minimum Roche lobe radii comparable to or even smaller than the typical radius of a $\sim2$ $M_{\sun}$ AGB star ($\sim$ 1-2 AU), suggesting that they likely once filled their Roche lobes and transferred some of their masses to their companions. Yet, we are unaware of any observations of these systems that reported obvious signs of mass transfer. Moreover, the eccentricities of the systems are still high, contradicting the expectation that such close interactions would tidally circularize their orbits. These puzzles are very similar to those of Sirius, suggesting that the physics of interacting stars on the AGB phase is still poorly understood. Several mechanisms have been proposed to explain the high eccentricities of Sirius-like systems with close progenitors, such as eccentricity pumping due to phase-dependent mass loss \citep{BM_2008} and interactions with an unseen third component \citep{Perets_2012}, but more work is needed to solve the puzzle for each system.}

\begin{table*}
    \centering
    \changes{\caption{Masses and orbital parameters of the progenitors.}
    \label{tab:prog_orbit}}
    \begin{tabular}{cccccc} \hline
      \multirow{2}{1.3em}{HD} & $M_{\rm pro}$ & \multirow{2}{1.3em}{$q_{\rm pro}$} & $a_{\rm pro}$ & \multirow{2}{0.6em}{$e$} & $R_{\rm Roche, min}$ \\ 
      & ($M_{\sun}$) & & (AU) & & (AU) \\ \hline \hline
      27483 B & ${3.59}_{-0.78}^{+1.43}$ & ${1.30}_{-0.31}^{+0.55}$ & ${28.2}_{-5.5}^{+7.7}$ & ${0.342}_{-0.19}^{+0.094}$ & ${7.6}_{-1.6}^{+2.4}$ \\ \hline 
      114174 B & ${1.90}_{-0.27}^{+0.30}$ & ${1.96}_{-0.28}^{+0.32}$ & ${14.3}_{-1.4}^{+ 1.5}$ & $0.690 \pm 0.017$ & $2.18 \pm 0.14$\\ \hline
      118475 B & ${1.76}_{-0.43}^{+0.60}$ & ${1.52}_{-0.38}^{+0.52}$ & $2.24 \pm 0.34$ & $0.1278 \pm 0.0008$ & ${0.834}_{-0.029}^{+0.025}$ \\ \hline
      136138 B & ${1.69}_{-0.52}^{+0.87}$ & ${0.94}_{-0.33}^{+0.55}$ & ${1.11}_{-0.22}^{+0.20}$ & ${0.336}_{-0.014}^{+0.015}$ & ${0.235}_{-0.039}^{+0.032}$ \\ \hline
      169889 B & ${1.90}_{-0.27}^{+0.30}$ & ${1.36}_{-0.28}^{+0.40}$ & ${26.8}_{-7.3}^{+10.4}$ & ${0.896}_{-0.088}^{+0.064}$ & ${1.2}_{-0.7}^{+ 1.1}$ \\ \hline
    \end{tabular}
\end{table*}

\changes{
\subsection{WD Mass-Radius Relation}
\label{subsec:MRR}}

\changes{
Theoretical models \citep[e.g.,][]{Chandrasekhar_1931,Fontaine_2001} predict a mass-radius relation (MRR) of WDs at any given effective temperature. The MRR is crucial to our understanding of WDs and useful in inferring WD masses from photometry or spectroscopy when dynamical measurements are unavailable. In this section, we present additional data points to test the theoretical MRR by computing the radii of our WDs and comparing them to the dynamical masses we derived in Section \ref{sec:orbit_fit}.}

\changes{We derive the radius of a WD by interpolating Montr{\'e}al cooling sequences to compute the radius from multi-band photometry. The interpolation method is similar to that used in finding the cooling ages. We successfully constrain the radii of HD 19019 B, HD 27483 B, HD 114174 B, and HD 169889 B with their available photometry. For HD 27786 Ab and HD 136138 B, we adopt the spectroscopic radius measurements in the literature, rescaled to their {\Gaia} DR3 distances.
Finally, as there are currently no photometric or spectroscopic observations of HD 118475 B, we postpone its radius measurement to future studies.}

\begin{figure}
    \centering
    \includegraphics[width=\linewidth]{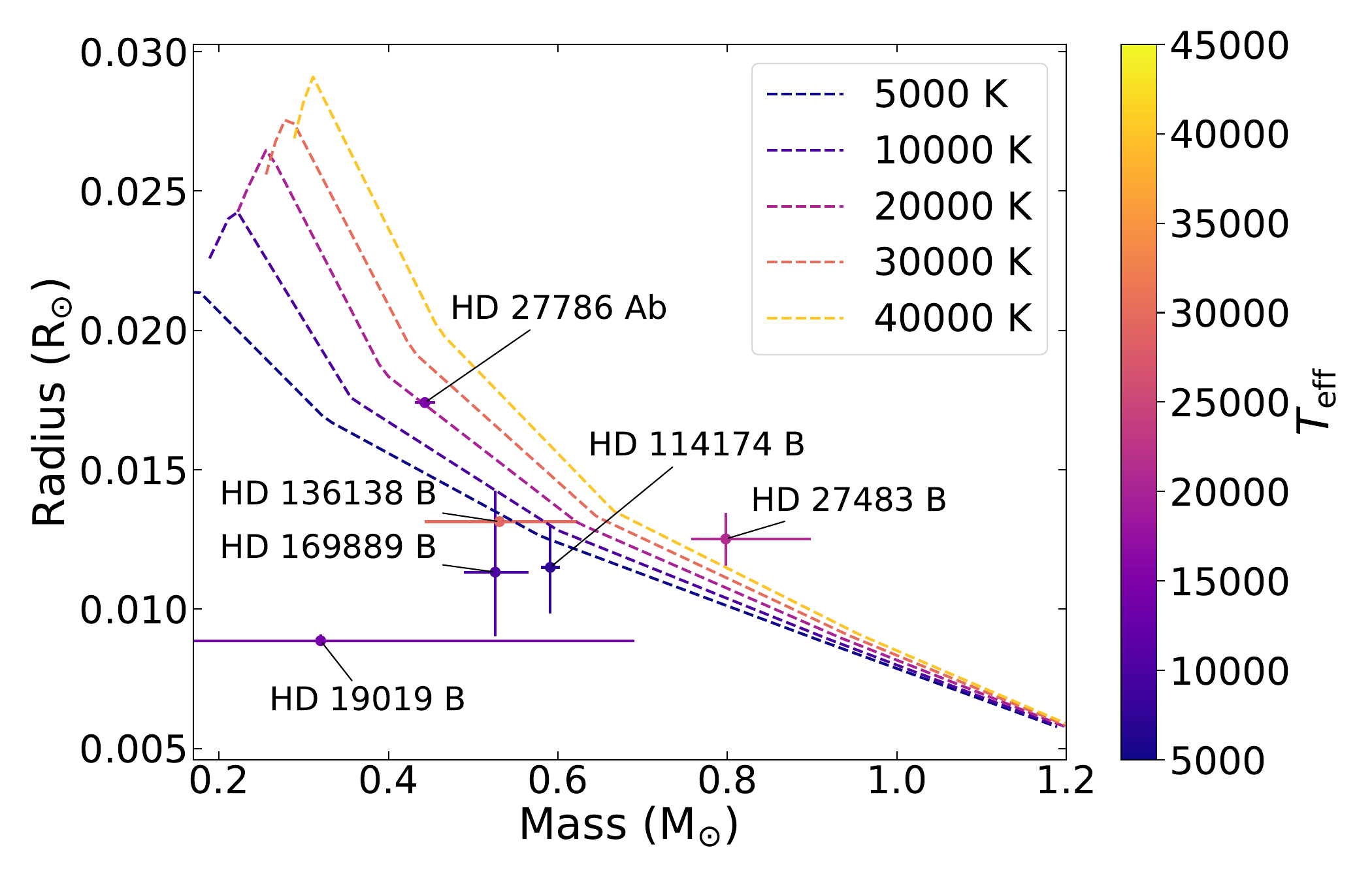}
    \changes{\caption{Masses and radii of six of our WDs (filled circles with 1-$\sigma$ error bars) compared to the theoretical MRRs given by Montr{\'e}al cooling models (dashed lines). All data points and model curves are colored according to their effective temperatures. Our WDs slightly deviate from but are statistically compatible with (< 2-$\sigma$ from) the theoretical MRRs at their effective temperatures.}
    \label{fig:M-R relation}}
\end{figure}

\changes{
Table \ref{tab:WDproperties} lists the radii, dynamical masses, and effective temperatures of the WDs. We take the mass of HD 136138 B to be the mass from our $M \sin{i}$ and the {\Gaia} DR3 inclination. Figure \ref{fig:M-R relation} compares the positions of our WDs on the mass-radius diagram to the theoretical MRR given by Montr{\'e}al cooling models at different effective temperatures. We note that this is equivalent to comparing a WD's dynamical mass to its photometric mass, as theoretical models compute the photometric mass directly from the photometric radius and the mass-radius relation. Our WDs slightly deviate from but are statistically compatible with (< 2-$\sigma$ from) the theoretical MRRs at their effective temperatures. Below, we outline the photometry we adopted and elaborate on the results.}

\changes{\subsubsection{HD 19019 B}
Because the WD was only resolved by {\Gaia}, we adopt the {\Gaia} DR3 photometry of $M_{\rm G_{\rm bp}}=12.30 \pm 0.04$ and $M_{\rm G_{\rm rp}}=12.36 \pm 0.07$, which give $\Teff=14100 \pm 200$ K and $R={0.887}_{-0.019}^{+0.023} \times 10^{-2}$ $R_{\odot}$. The results agree marginally with $\Teff=18200 \pm 3000$ K and $R={0.656}_{-0.112}^{+0.134} \times 10^{-2}$ $R_{\odot}$ from the spectroscopy in \cite{Landstreet_2020}. Because of the remarkable precision of {\Gaia} photometry, our statistical uncertainties are likely smaller than the systematic uncertainties associated with our model assumptions (e.g., pure-hydrogen atmosphere). Therefore, the $R$ and $\Teff$ uncertainties listed in Table \ref{tab:WDproperties} should be considered lower bounds of the actual uncertainties, so it is unclear how much HD 19019 B agrees with the theoretical MRR.}

\changes{\subsubsection{HD 27483 B}
From the $K$-band and the F218W magnitudes in Section \ref{subsec:age_27483}, we get $\Teff=21000 \pm 3000$ K and $R= (1.252 \pm 0.095) \times 10^{-2}$ $R_{\odot}$. The radius agrees remarkably with the spectroscopic radius of $R=(1.235 \pm 0.018) \times 10^{-2}$ $R_{\odot}$ \citep{Joyce_2018}. The photometry of the WD suggests a radius larger than that expected from the MRR, consistent with our finding in Section \ref{subsec:age_27483} that the WD's photometric mass is $1.4\sigma$ below its dynamical mass.}

\changes{\subsubsection{HD 27786 Ab}
Although we have measured the WD's contrast with its host star in HST WFPC2 F170W and WFC3 F218W (Table \ref{tab:new_relast}), we are unaware of any measurements of the host star's magnitudes in the two filters. We may estimate the host magnitudes from synthetic photometry, but the accuracy will be limited by the assumptions of our model spectrum. Also, both filters are very close to the peak of a black-body emission ($\approx 2000$ {\AA} for $\Teff\approx 14500$ as in \citealt{Landsman_1996}), making it unlikely to constrain the WD's $\Teff$ and radius simultaneously. Hence, additional observations are required to reliably determine the WD's photometric radius.}

\changes{\cite{Landsman_1996} fitted the WD's radius and effective temperature from its {\sl IUE} spectrum and obtained different values for a grid of different assumed distances. Interpolating the grid linearly to the {\Gaia} DR3 distance of $42.03 \pm 0.19$ pc gives $\Teff=14650 \pm 30$ K and $R = (1.741 \pm 0.015) \times 10^{-2}$ $R_{\sun}$. The uncertainties reported here reflect only the parallax uncertainties and do not incorporate uncertainties in the original spectroscopic fitting, which are not provided in \cite{Landsman_1996}. Therefore, the actual error bars of $R$ and $\Teff$ are likely much larger. As shown in Figure \ref{fig:M-R relation}, the WD has a radius compatible with a C/O core despite its usually low mass.}

\changes{\subsubsection{HD 114174 B}
We use $M_{y}=14.09 \pm 0.05$, $M_{J}=13.55 \pm 0.06$, $M_{H}=13.44 \pm 0.03$, and $M_{K_s}=13.11 \pm 0.02$, measured by \cite{Gratton_2021} using SPHERE data, plus $M_{L^\prime}=13.18 \pm 0.16$ from \cite{Matthews_2014} using NIRC2. When we fit Montr{\'e}al cooling models to these measurements, the posterior distributions of $\Teff$ and radius are bimodal, with one peak giving $\Teff=5100$ K and $R=1.71 \times 10^{-2}$ $R_{\sun}$ and the other giving $\Teff=9500$ K and $R=0.75 \times 10^{-2}$ $R_{\sun}$. This is likely related to the $\Teff$ discrepancy discussed in Section \ref{subsec:age_114174}. We attribute this to potentially underestimated systematic uncertainties from the calibration and data reduction of SPHERE observations, assuming that the statistical uncertainties reported by \cite{Gratton_2021} are much smaller than the actual uncertainties. Given this assumption, we inflate the errors of all SPHERE photometry by 0.14 dex to achieve a reduced $\chi^2$ of unity. Doing so leads to $\Teff={7200}_{-1000}^{+1400}$ K and $R={1.15}_{-0.17}^{+0.15}$, which are consistent with the WD's dynamical mass within 1-$\sigma$. Additional observations will be necessary to validate our assumption and resolve the discrepancy.}

\changes{\subsubsection{HD 136138 B}
Due to the lack of photometry, we adopt the $R$ and $\Teff$ obtained by \cite{Stefanik_2011} by fitting the WD's {\sl IUE} spectrum. Similar to our method for HD 27786 Ab, we interpolate the spectroscopic fit results to the {\Gaia} DR3 distance of $110.97 \pm 0.63$ pc and get $\Teff=29510 \pm 50$ and $R = (1.314 \pm 0.011) \times 10^{-2}$ $R_{\odot}$. Again, these uncertainty values do not incorporate uncertainties of the original spectroscopic fitting and should only be considered lower bounds of the actual uncertainties. At this radius and $\Teff$, the WD's dynamical mass is slightly lower than the mass expected from the MRR.}

\changes{\subsubsection{HD 169889 B}
$M_H=13.59 \pm 0.16$ and $M_{L^{\prime}}=13.32 \pm 0.08$ \citep{Crepp_2018} give $R={1.13}_{-0.23}^{+0.29} \times 10^{-2}$ $R_{\odot}$ and a $\Teff$ anywhere between $\sim4000$ K and $\sim12000$ K. Our results are consistent with those in \cite{Crepp_2018}. As in Section \ref{subsec:age_169889}, we assume $\Teff \approx 10000 \pm 1000$ K, noting that the uncertainty is a conservative estimate after ruling out cooler temperatures incompatible with the age of the host star. The uncertainties are large because both photometric measurements lie on the Rayleigh-Jeans tail of the WD's spectrum, causing a degeneracy between $\Teff$ and $R$. Assuming $\Teff=10000$ K, additional photometry bluer than 14000 {\AA} will help break the degeneracy and determine how much the WD agrees with the theoretical MRR.}

\begin{table*}
    \centering
    \changes{\caption{Radii, effective temperatures, and dynamical masses of the WDs.}\label{tab:WDproperties}}
    \begin{tabular}{ccccc} \hline
       \multirow{2}{1.3em}{HD} & $R$ & $\Teff$ & \multirow{2}{4em}{Reference} & $M$ \\ 
       & ($0.01 R_{\sun}$) & (K) & & ($M_{\sun}$) \\ \hline \hline
       19019 B & ${0.887}_{-0.019}^{+0.023}$ & $14100 \pm 200$ & This work & ${0.32}_{-0.23}^{+0.37}$ \\ \hline
       27483 B & $1.252 \pm 0.095$ & $21000 \pm 3000$ & This work & ${0.798}_{-0.041}^{+0.10}$ \\ \hline
       27786 Ab & $1.741 \pm 0.015^*$ & $14650 \pm 30^*$ & \cite{Landsman_1996} & $0.443 \pm 0.012$ \\ \hline
       114174 B & ${1.15}_{-0.17}^{+0.15}$ & ${7200}_{-1000}^{+1400}$ & This work & $0.591 \pm 0.011$ \\ \hline
       136138 B & $1.314 \pm 0.011^*$ & $29510 \pm 50^*$ & \cite{Stefanik_2011} & ${0.531}_{-0.088}^{+0.092}$ \\ \hline
       168889 B & ${1.13}_{-0.23}^{+0.29}$ & $10000 \pm 1000$ & This work & ${0.526}_{-0.037}^{+0.039}$ \\ \hline
    \end{tabular}
    \centering
    
    \changes{{Note: the $R$ and $\Teff$ of HD 27786 Ab and HD 136138 B are computed by interpolating the spectroscopic fit results in the listed references to the {\Gaia} DR3 distances. The uncertainties here (marked with "*") reflect only the parallax uncertainties and do not incorporate uncertainties of the spectroscopic fitting, which are likely much larger but are unavailable in the original references.}}
\end{table*}

\section{Conclusions} \label{sec:conclude}
In this paper, we have derived the masses and orbits for six confirmed and one candidate Sirius-like systems from a joint analysis of high-precision RVs, relative astrometry, and {\Hipparcos}-{\Gaia} proper motion anomaly. From the dynamical masses of the WDs, we have constrained the ages of four systems and compared them to age estimates by other means. \changes{We have also discussed the possibilities of previous interactions between the WDs and their companions. Finally, we have derived the radii of the WDs and compared them to theoretical MRRs.} We summarize our main results below:

\begin{enumerate}
    \item For HD 27483 B, our analysis of unpublished Hubble data and new observations with NIRC2 introduce two additional epochs of relative astrometry and photometry, allowing us to obtain the first-ever dynamical mass of the WD (${0.798}_{-0.041}^{+0.10}$ $M_{\sun}$). The $\sim$10\%-precision mass leads to a total age of ${400}_{-180}^{+570}$ Myr, consistent with previous age estimates of the Hyades cluster but not sufficiently precise to resolve the age discrepancy between various methods. Additional relative astrometry would improve the mass precision considerably, but the MS lifetime precision is limited by the calibration uncertainty of semi-empirical IFMRs. Constraints on the cooling age could be improved by additional photometry or adopting a $\Teff$ from high-resolution spectroscopy.

    \item Our three-body fit on HD 27786 results in an unusually low mass ($0.443 \pm 0.012$ $M_{\sun}$) for the WD HD 27786 Ab, suggesting the possibility of common envelope evolution with the primary star during its progenitor's AGB phase.
     
    \item We obtain the first precise dynamical masses of HD 114174 B ($0.591 \pm 0.011$ $M_{\sun}$) and HD 169889 B (${0.526}_{-0.037}^{+0.039}$ $M_{\sun}$). The ages of both WDs depend strongly on the assumed $\Teff$, showing the need for high-resolution spectroscopy to resolve the $\Teff$ uncertainty and improve the age constraints.

    \item For HD 118475, the orbital elements we derived from RVs agree remarkably with the {\Gaia} DR3 two-body solution from intermediate astrometry. Whereas for HD 136138, the best-fitting eccentricity and period from the RVs are in moderate tension with that from {\Gaia}. Such a comparison, applicable to a large sample of systems, provides validations of {\Gaia}'s two-body solutions.

    \item With $M\sin{i}={0.461}_{-0.015}^{+0.017}$ $M_{\sun}$ and a {\Gaia} DR3 inclination of ${53.\!\!\degree1}_{-5.\!\!\degree5}^{+4.\!\!\degree8}$, HD 118475 B has a dynamical mass of ${0.580}_{-0.039}^{+0.052}$ $M_{\sun}$. By comparing the magnitude of a $0.58$ $M_{\sun}$ MS star to the non-detection significance curve in \cite{Kane_2019}, we rule out the hypothesis that the companion is a MS star at $>8\sigma$ level, confirming that it is a WD.

    \changes{\item The progenitors of HD 114174 B, HD 118475 B, HD 136138 B, and HD 169889 B may have filled their Roche lobes during the AGB phase. Yet, their orbits did not tidally circularize, presenting a puzzle similar to that of Sirius.}

    \changes{\item The masses, radii, and effective temperatures of our WDs are compatible with the theoretical MRR from Montr{\'e}al cooling models. In particular, HD 27786 B has a radius consistent with a C/O core despite its usually low mass.}
\end{enumerate}

\changes{The precisions of such dynamical measurements of white dwarf masses can reach $\approx$1\% for the very best stars. As shown in Equation (8) of \cite{Brandt_2019}, this level of mass precision requires $\lesssim$1\% precision each on separation, RV acceleration, and proper motion acceleration. Achieving this on all three measurements will be challenging for all but the best and closest targets, and the limiting factor will be different for each target. For example, HD 27483 and HD 27786 have multiple epochs of relative astrometry but no RVs. HD 118475 and HD 136138 have high-quality RV data over multiple orbital periods, but their orbital periods are too short for the {\Gaia} proper motions to be helpful, nor do they have relative astrometry, preventing us from constraining their WD masses. The orbit fitting of HD 19019 will benefit the most from proper motion measurements in a future {\Gaia} release precise enough to detect its weak astrometric acceleration, while improving the orbit fitting of HD 169889 requires better relative astrometry. Despite the challenges, continued RV and astrometric monitoring and exceptional data quality from future {\Gaia} data releases will steadily grow the sample of white dwarfs with very precise masses. }

For WDs with better than $\sim 15\%$ dynamical masses, such as HD 27483 B, HD 114174 B, and HD 169889 B, the dominant source of age uncertainty (assuming a good $\Teff$) is the scatter in semi-empirical IFMRs. Theoretically modeled IFMRs, in comparison, are free from calibration uncertainties but replace them with systematic uncertainties of specific mass-loss models \citep[e.g.,][]{Reimers_1975, Bloecker_1995}. Hence, we choose not to use them to disentangle our analysis from systematics. From another perspective, our dynamical masses are themselves measurements of the IFMR. With a dynamical final mass, one can infer an initial mass, independent of mass-loss models, from the difference between the MS primary's age and the WD's cooling age. With help from precise astrometric solutions in {\Gaia} DR3 and later releases, repeating our analysis on additional Sirius-like systems would increase the sample size for a future re-calibration of the WD IFMR.

\section*{ACKNOWLEDGEMENTS}

\changes{We thank the anonymous referee for their thoughtful comments.} This research has made use of the SIMBAD database \citep{Wenger_2000}, operated at CDS, Strasbourg, France. This research has made use of the SVO Filter Profile Service \citep{SVO_2020} supported by the Spanish MINECO through grant AYA2017-84089. 
This work has made use of data from the European Space Agency (ESA) mission {\it Gaia} (\url{https://www.cosmos.esa.int/gaia}), processed by the {\it Gaia} Data Processing and Analysis Consortium (DPAC, \url{https://www.cosmos.esa.int/web/gaia/dpac/consortium}). Funding for the DPAC has been provided by national institutions, in particular the institutions participating in the {\it Gaia} Multilateral Agreement. \changes{This research has made use of the Keck Observatory Archive (KOA), which is operated by the W. M. Keck Observatory and the NASA Exoplanet Science Institute (NExScI), under contract with the National Aeronautics and Space Administration. This research has made use of observations made with the NASA/ESA Hubble Space Telescope, and obtained from the Hubble Legacy Archive, which is a collaboration between the Space Telescope Science Institute (STScI/NASA), the Space Telescope European Coordinating Facility (ST-ECF/ESA) and the Canadian Astronomy Data Centre (CADC/NRC/CSA).}

\section*{Data Availability}
The observations used in this article are available in the Keck Observatory Archive (\url{https://koa.ipac.caltech.edu}), the Hubble Legacy Archive (\url{https://hla.stsci.edu}), and the references listed in Section \ref{sec:data+obs}.

\bibliographystyle{mnras}
\bibliography{refs}

\begin{thebibliography}{}
\makeatletter
\relax
\def\mn@urlcharsother{\let\do\@makeother \do\$\do\&\do\#\do\^\do\_\do\%\do\~}
\def\mn@doi{\begingroup\mn@urlcharsother \@ifnextchar [ {\mn@doi@}
  {\mn@doi@[]}}
\def\mn@doi@[#1]#2{\def\@tempa{#1}\ifx\@tempa\@empty \href
  {http://dx.doi.org/#2} {doi:#2}\else \href {http://dx.doi.org/#2} {#1}\fi
  \endgroup}
\def\mn@eprint#1#2{\mn@eprint@#1:#2::\@nil}
\def\mn@eprint@arXiv#1{\href {http://arxiv.org/abs/#1} {{\tt arXiv:#1}}}
\def\mn@eprint@dblp#1{\href {http://dblp.uni-trier.de/rec/bibtex/#1.xml}
  {dblp:#1}}
\def\mn@eprint@#1:#2:#3:#4\@nil{\def\@tempa {#1}\def\@tempb {#2}\def\@tempc
  {#3}\ifx \@tempc \@empty \let \@tempc \@tempb \let \@tempb \@tempa \fi \ifx
  \@tempb \@empty \def\@tempb {arXiv}\fi \@ifundefined
  {mn@eprint@\@tempb}{\@tempb:\@tempc}{\expandafter \expandafter \csname
  mn@eprint@\@tempb\endcsname \expandafter{\@tempc}}}

\bibitem[\protect\citeauthoryear{{Allard}, {Homeier}  \& {Freytag}}{{Allard}
  et~al.}{2012}]{Allard_2012}
{Allard} F.,  {Homeier} D.,   {Freytag} B.,  2012, \mn@doi [Philosophical
  Transactions of the Royal Society of London Series A]
  {10.1098/rsta.2011.0269}, \href
  {https://ui.adsabs.harvard.edu/abs/2012RSPTA.370.2765A} {370, 2765}

\bibitem[\protect\citeauthoryear{{Althaus}, {Serenelli}  \&
  {Benvenuto}}{{Althaus} et~al.}{2001}]{Althaus_2001}
{Althaus} L.~G.,  {Serenelli} A.~M.,   {Benvenuto} O.~G.,  2001, \mn@doi
  [\mnras] {10.1046/j.1365-8711.2001.04324.x}, \href
  {https://ui.adsabs.harvard.edu/abs/2001MNRAS.324..617A} {324, 617}

\bibitem[\protect\citeauthoryear{{Arentsen} et~al.,}{{Arentsen}
  et~al.}{2019}]{Arentsen_2019}
{Arentsen} A.,  et~al., 2019, \mn@doi [\aap] {10.1051/0004-6361/201834273},
  \href {https://ui.adsabs.harvard.edu/abs/2019A&A...627A.138A} {627, A138}

\bibitem[\protect\citeauthoryear{{Bacchus} et~al.,}{{Bacchus}
  et~al.}{2017}]{Bacchus_2017}
{Bacchus} E.,  et~al., 2017, \mn@doi [\mnras] {10.1093/mnras/stx1171}, \href
  {https://ui.adsabs.harvard.edu/abs/2017MNRAS.469.4796B} {469, 4796}

\bibitem[\protect\citeauthoryear{{Barstow}, {Bond}, {Burleigh}  \&
  {Holberg}}{{Barstow} et~al.}{2001}]{Barstow_2001}
{Barstow} M.~A.,  {Bond} H.~E.,  {Burleigh} M.~R.,   {Holberg} J.~B.,  2001,
  \mn@doi [\mnras] {10.1046/j.1365-8711.2001.04203.x}, \href
  {https://ui.adsabs.harvard.edu/abs/2001MNRAS.322..891B} {322, 891}

\bibitem[\protect\citeauthoryear{{B{\'e}dard}, {Bergeron}, {Brassard}  \&
  {Fontaine}}{{B{\'e}dard} et~al.}{2020}]{Bedard_2020}
{B{\'e}dard} A.,  {Bergeron} P.,  {Brassard} P.,   {Fontaine} G.,  2020,
  \mn@doi [\apj] {10.3847/1538-4357/abafbe}, \href
  {https://ui.adsabs.harvard.edu/abs/2020ApJ...901...93B} {901, 93}

\bibitem[\protect\citeauthoryear{{Bergeron}, {Wesemael}  \&
  {Beauchamp}}{{Bergeron} et~al.}{1995}]{Bergeron_1995}
{Bergeron} P.,  {Wesemael} F.,   {Beauchamp} A.,  1995, \mn@doi [\pasp]
  {10.1086/133661}, \href
  {https://ui.adsabs.harvard.edu/abs/1995PASP..107.1047B} {107, 1047}

\bibitem[\protect\citeauthoryear{{Beuzit} et~al.,}{{Beuzit}
  et~al.}{2019}]{Beuzit_2019}
{Beuzit} J.~L.,  et~al., 2019, \mn@doi [\aap] {10.1051/0004-6361/201935251},
  \href {https://ui.adsabs.harvard.edu/abs/2019A&A...631A.155B} {631, A155}

\bibitem[\protect\citeauthoryear{{Bloecker}}{{Bloecker}}{1995}]{Bloecker_1995}
{Bloecker} T.,  1995, \aap, \href
  {https://ui.adsabs.harvard.edu/abs/1995A&A...297..727B} {297, 727}

\bibitem[\protect\citeauthoryear{{Blouin}, {Dufour}  \& {Allard}}{{Blouin}
  et~al.}{2018}]{Blouin_2018}
{Blouin} S.,  {Dufour} P.,   {Allard} N.~F.,  2018, \mn@doi [\apj]
  {10.3847/1538-4357/aad4a9}, \href
  {https://ui.adsabs.harvard.edu/abs/2018ApJ...863..184B} {863, 184}

\bibitem[\protect\citeauthoryear{{Bochanski}, {Faherty}, {Gagn{\'e}}, {Nelson},
  {Coker}, {Smithka}, {Desir}  \& {Vasquez}}{{Bochanski}
  et~al.}{2018}]{Bochanski_2018}
{Bochanski} J.~J.,  {Faherty} J.~K.,  {Gagn{\'e}} J.,  {Nelson} O.,  {Coker}
  K.,  {Smithka} I.,  {Desir} D.,   {Vasquez} C.,  2018, \mn@doi [\aj]
  {10.3847/1538-3881/aaaebe}, \href
  {https://ui.adsabs.harvard.edu/abs/2018AJ....155..149B} {155, 149}

\bibitem[\protect\citeauthoryear{{Boehm-Vitense}}{{Boehm-Vitense}}{1993}]{Boehm-Vitense_1993}
{Boehm-Vitense} E.,  1993, \mn@doi [\aj] {10.1086/116709}, \href
  {https://ui.adsabs.harvard.edu/abs/1993AJ....106.1113B} {106, 1113}

\bibitem[\protect\citeauthoryear{{Boksenberg}}{{Boksenberg}}{1985}]{WHT_1985}
{Boksenberg} A.,  1985, \mn@doi [Vistas in Astronomy]
  {10.1016/0083-6656(85)90074-1}, \href
  {https://ui.adsabs.harvard.edu/abs/1985VA.....28..531B} {28, 531}

\bibitem[\protect\citeauthoryear{{Bona{\v{c}}i{\'c} Marinovi{\'c}}, {Glebbeek}
  \& {Pols}}{{Bona{\v{c}}i{\'c} Marinovi{\'c}} et~al.}{2008}]{BM_2008}
{Bona{\v{c}}i{\'c} Marinovi{\'c}} A.~A.,  {Glebbeek} E.,   {Pols} O.~R.,  2008,
  \mn@doi [\aap] {10.1051/0004-6361:20078297}, \href
  {https://ui.adsabs.harvard.edu/abs/2008A&A...480..797B} {480, 797}

\bibitem[\protect\citeauthoryear{{Bond}}{{Bond}}{1862}]{Bond_1862}
{Bond} G.,  1862, Astronomische Nachrichten, \href
  {https://ui.adsabs.harvard.edu/abs/1862AN.....57..131B} {57, 131}

\bibitem[\protect\citeauthoryear{{Bond} et~al.,}{{Bond}
  et~al.}{2015}]{Bond_2015}
{Bond} H.~E.,  et~al., 2015, \mn@doi [\apj] {10.1088/0004-637X/813/2/106},
  \href {https://ui.adsabs.harvard.edu/abs/2015ApJ...813..106B} {813, 106}

\bibitem[\protect\citeauthoryear{{Bond} et~al.,}{{Bond}
  et~al.}{2017}]{Bond_2017b}
{Bond} H.~E.,  et~al., 2017, \mn@doi [\apj] {10.3847/1538-4357/aa6af8}, \href
  {https://ui.adsabs.harvard.edu/abs/2017ApJ...840...70B} {840, 70}

\bibitem[\protect\citeauthoryear{{Bowler} et~al.,}{{Bowler}
  et~al.}{2021}]{Bowler_2021}
{Bowler} B.~P.,  et~al., 2021, \mn@doi [\aj] {10.3847/1538-3881/abd243}, \href
  {https://ui.adsabs.harvard.edu/abs/2021AJ....161..106B} {161, 106}

\bibitem[\protect\citeauthoryear{{Brandt}}{{Brandt}}{2018}]{Brandt_2018}
{Brandt} T.~D.,  2018, \mn@doi [\apjs] {10.3847/1538-4365/aaec06}, \href
  {https://ui.adsabs.harvard.edu/abs/2018ApJS..239...31B} {239, 31}

\bibitem[\protect\citeauthoryear{{Brandt}}{{Brandt}}{2021}]{Brandt_2021}
{Brandt} T.~D.,  2021, \mn@doi [\apjs] {10.3847/1538-4365/abf93c}, \href
  {https://ui.adsabs.harvard.edu/abs/2021ApJS..254...42B} {254, 42}

\bibitem[\protect\citeauthoryear{{Brandt} \& {Huang}}{{Brandt} \&
  {Huang}}{2015}]{Brandt_2015}
{Brandt} T.~D.,  {Huang} C.~X.,  2015, \mn@doi [\apj]
  {10.1088/0004-637X/807/1/58}, \href
  {https://ui.adsabs.harvard.edu/abs/2015ApJ...807...58B} {807, 58}

\bibitem[\protect\citeauthoryear{{Brandt}, {Dupuy}  \& {Bowler}}{{Brandt}
  et~al.}{2019}]{Brandt_2019}
{Brandt} T.~D.,  {Dupuy} T.~J.,   {Bowler} B.~P.,  2019, \mn@doi [\aj]
  {10.3847/1538-3881/ab04a8}, \href
  {https://ui.adsabs.harvard.edu/abs/2019AJ....158..140B} {158, 140}

\bibitem[\protect\citeauthoryear{Brandt, Argafal  \& trace andreason}{Brandt
  et~al.}{2021a}]{HTOF_2021}
Brandt G.~M.,  Argafal  trace andreason 2021a, gmbrandt/HTOF: 0.4.2,
  \mn@doi{10.5281/zenodo.5035801}, \url
  {https://doi.org/10.5281/zenodo.5035801}

\bibitem[\protect\citeauthoryear{{Brandt}, {Dupuy}, {Li}, {Brandt}, {Zeng},
  {Michalik}, {Bardalez Gagliuffi}  \& {Raposo-Pulido}}{{Brandt}
  et~al.}{2021b}]{Brandt_2021_orvara}
{Brandt} T.~D.,  {Dupuy} T.~J.,  {Li} Y.,  {Brandt} G.~M.,  {Zeng} Y.,
  {Michalik} D.,  {Bardalez Gagliuffi} D.~C.,   {Raposo-Pulido} V.,  2021b,
  \mn@doi [\aj] {10.3847/1538-3881/ac042e}, \href
  {https://ui.adsabs.harvard.edu/abs/2021AJ....162..186B} {162, 186}

\bibitem[\protect\citeauthoryear{{Bressan}, {Marigo}, {Girardi}, {Salasnich},
  {Dal Cero}, {Rubele}  \& {Nanni}}{{Bressan} et~al.}{2012}]{Bressan_2012}
{Bressan} A.,  {Marigo} P.,  {Girardi} L.,  {Salasnich} B.,  {Dal Cero} C.,
  {Rubele} S.,   {Nanni} A.,  2012, \mn@doi [\mnras]
  {10.1111/j.1365-2966.2012.21948.x}, \href
  {https://ui.adsabs.harvard.edu/abs/2012MNRAS.427..127B} {427, 127}

\bibitem[\protect\citeauthoryear{{Brewer}, {Fischer}, {Valenti}  \&
  {Piskunov}}{{Brewer} et~al.}{2016}]{Brewer_2016}
{Brewer} J.~M.,  {Fischer} D.~A.,  {Valenti} J.~A.,   {Piskunov} N.,  2016,
  \mn@doi [\apjs] {10.3847/0067-0049/225/2/32}, \href
  {https://ui.adsabs.harvard.edu/abs/2016ApJS..225...32B} {225, 32}

\bibitem[\protect\citeauthoryear{{Brown}, {Kilic}, {Brown}  \&
  {Kenyon}}{{Brown} et~al.}{2011}]{Brown_2011}
{Brown} J.~M.,  {Kilic} M.,  {Brown} W.~R.,   {Kenyon} S.~J.,  2011, \mn@doi
  [\apj] {10.1088/0004-637X/730/2/67}, \href
  {https://ui.adsabs.harvard.edu/abs/2011ApJ...730...67B} {730, 67}

\bibitem[\protect\citeauthoryear{{Burleigh}, {Barstow}  \&
  {Holberg}}{{Burleigh} et~al.}{1998}]{Burleigh_1998}
{Burleigh} M.~R.,  {Barstow} M.~A.,   {Holberg} J.~B.,  1998, \mn@doi [\mnras]
  {10.1046/j.1365-8711.1998.01914.x}, \href
  {https://ui.adsabs.harvard.edu/abs/1998MNRAS.300..511B} {300, 511}

\bibitem[\protect\citeauthoryear{{Butler} et~al.,}{{Butler}
  et~al.}{2017}]{Butler_2017}
{Butler} R.~P.,  et~al., 2017, \mn@doi [\aj] {10.3847/1538-3881/aa66ca}, \href
  {https://ui.adsabs.harvard.edu/abs/2017AJ....153..208B} {153, 208}

\bibitem[\protect\citeauthoryear{{Cantat-Gaudin} \& {Brandt}}{{Cantat-Gaudin}
  \& {Brandt}}{2021}]{Cantat-Gaudin_2021}
{Cantat-Gaudin} T.,  {Brandt} T.~D.,  2021, \mn@doi [\aap]
  {10.1051/0004-6361/202140807}, \href
  {https://ui.adsabs.harvard.edu/abs/2021A&A...649A.124C} {649, A124}

\bibitem[\protect\citeauthoryear{{Casagrande}, {Sch{\"o}nrich}, {Asplund},
  {Cassisi}, {Ram{\'\i}rez}, {Mel{\'e}ndez}, {Bensby}  \&
  {Feltzing}}{{Casagrande} et~al.}{2011}]{Casagrande_2011}
{Casagrande} L.,  {Sch{\"o}nrich} R.,  {Asplund} M.,  {Cassisi} S.,
  {Ram{\'\i}rez} I.,  {Mel{\'e}ndez} J.,  {Bensby} T.,   {Feltzing} S.,  2011,
  \mn@doi [\aap] {10.1051/0004-6361/201016276}, \href
  {https://ui.adsabs.harvard.edu/abs/2011A&A...530A.138C} {530, A138}

\bibitem[\protect\citeauthoryear{{Casali} et~al.,}{{Casali}
  et~al.}{2020}]{Casali_2020}
{Casali} G.,  et~al., 2020, \mn@doi [\aap] {10.1051/0004-6361/202038055}, \href
  {https://ui.adsabs.harvard.edu/abs/2020A&A...639A.127C} {639, A127}

\bibitem[\protect\citeauthoryear{{Chabrier}}{{Chabrier}}{2003}]{Chabrier_2003}
{Chabrier} G.,  2003, \mn@doi [\pasp] {10.1086/376392}, \href
  {https://ui.adsabs.harvard.edu/abs/2003PASP..115..763C} {115, 763}

\bibitem[\protect\citeauthoryear{{Chandra}, {Hwang}, {Zakamska}  \&
  {Cheng}}{{Chandra} et~al.}{2020}]{Chandra_2020}
{Chandra} V.,  {Hwang} H.-C.,  {Zakamska} N.~L.,   {Cheng} S.,  2020, \mn@doi
  [\apj] {10.3847/1538-4357/aba8a2}, \href
  {https://ui.adsabs.harvard.edu/abs/2020ApJ...899..146C} {899, 146}

\bibitem[\protect\citeauthoryear{{Chandrasekhar}}{{Chandrasekhar}}{1931}]{Chandrasekhar_1931}
{Chandrasekhar} S.,  1931, \mn@doi [\apj] {10.1086/143324}, \href
  {https://ui.adsabs.harvard.edu/abs/1931ApJ....74...81C} {74, 81}

\bibitem[\protect\citeauthoryear{{Choi}, {Dotter}, {Conroy}, {Cantiello},
  {Paxton}  \& {Johnson}}{{Choi} et~al.}{2016}]{MIST_2016}
{Choi} J.,  {Dotter} A.,  {Conroy} C.,  {Cantiello} M.,  {Paxton} B.,
  {Johnson} B.~D.,  2016, \mn@doi [\apj] {10.3847/0004-637X/823/2/102}, \href
  {https://ui.adsabs.harvard.edu/abs/2016ApJ...823..102C} {823, 102}

\bibitem[\protect\citeauthoryear{{Crepp}, {Johnson}, {Howard}, {Marcy},
  {Gianninas}, {Kilic}  \& {Wright}}{{Crepp} et~al.}{2013}]{Crepp_2013}
{Crepp} J.~R.,  {Johnson} J.~A.,  {Howard} A.~W.,  {Marcy} G.~W.,  {Gianninas}
  A.,  {Kilic} M.,   {Wright} J.~T.,  2013, \mn@doi [\apj]
  {10.1088/0004-637X/774/1/1}, \href
  {https://ui.adsabs.harvard.edu/abs/2013ApJ...774....1C} {774, 1}

\bibitem[\protect\citeauthoryear{{Crepp} et~al.,}{{Crepp}
  et~al.}{2018}]{Crepp_2018}
{Crepp} J.~R.,  et~al., 2018, \mn@doi [\apj] {10.3847/1538-4357/aad381}, \href
  {https://ui.adsabs.harvard.edu/abs/2018ApJ...864...42C} {864, 42}

\bibitem[\protect\citeauthoryear{{Cummings}, {Kalirai}, {Tremblay},
  {Ramirez-Ruiz}  \& {Choi}}{{Cummings} et~al.}{2018}]{Cummings_2018}
{Cummings} J.~D.,  {Kalirai} J.~S.,  {Tremblay} P.~E.,  {Ramirez-Ruiz} E.,
  {Choi} J.,  2018, \mn@doi [\apj] {10.3847/1538-4357/aadfd6}, \href
  {https://ui.adsabs.harvard.edu/abs/2018ApJ...866...21C} {866, 21}

\bibitem[\protect\citeauthoryear{{Cutri} et~al.,}{{Cutri}
  et~al.}{2003}]{Cutri_2003}
{Cutri} R.~M.,  et~al., 2003, VizieR Online Data Catalog, \href
  {https://ui.adsabs.harvard.edu/abs/2003yCat.2246....0C} {p. II/246}

\bibitem[\protect\citeauthoryear{{David} \& {Hillenbrand}}{{David} \&
  {Hillenbrand}}{2015}]{David_2015}
{David} T.~J.,  {Hillenbrand} L.~A.,  2015, \mn@doi [\apj]
  {10.1088/0004-637X/804/2/146}, \href
  {https://ui.adsabs.harvard.edu/abs/2015ApJ...804..146D} {804, 146}

\bibitem[\protect\citeauthoryear{{De Gennaro}, {von Hippel}, {Jefferys},
  {Stein}, {van Dyk}  \& {Jeffery}}{{De Gennaro} et~al.}{2009}]{DeGennaro_2009}
{De Gennaro} S.,  {von Hippel} T.,  {Jefferys} W.~H.,  {Stein} N.,  {van Dyk}
  D.,   {Jeffery} E.,  2009, \mn@doi [\apj] {10.1088/0004-637X/696/1/12}, \href
  {https://ui.adsabs.harvard.edu/abs/2009ApJ...696...12D} {696, 12}

\bibitem[\protect\citeauthoryear{{Dembowski}}{{Dembowski}}{1870}]{Dembowski_1870}
{Dembowski} H.,  1870, \mn@doi [Astronomische Nachrichten]
  {10.1002/asna.18700760602}, \href
  {https://ui.adsabs.harvard.edu/abs/1870AN.....76...81D} {76, 81}

\bibitem[\protect\citeauthoryear{{Diego}, {Charalambous}, {Fish}  \&
  {Walker}}{{Diego} et~al.}{1990}]{Diego_1990}
{Diego} F.,  {Charalambous} A.,  {Fish} A.~C.,   {Walker} D.~D.,  1990, in
  {Crawford} D.~L.,  ed.,  Society of Photo-Optical Instrumentation Engineers
  (SPIE) Conference Series Vol. 1235, Instrumentation in Astronomy VII. pp
  562--576, \mn@doi{10.1117/12.19119}

\bibitem[\protect\citeauthoryear{{Dosopoulou} \& {Kalogera}}{{Dosopoulou} \&
  {Kalogera}}{2016}]{Dosopoulou_2016}
{Dosopoulou} F.,  {Kalogera} V.,  2016, \mn@doi [\apj]
  {10.3847/0004-637X/825/1/71}, \href
  {https://ui.adsabs.harvard.edu/abs/2016ApJ...825...71D} {825, 71}

\bibitem[\protect\citeauthoryear{{Dufour}, {Blouin}, {Coutu},
  {Fortin-Archambault}, {Thibeault}, {Bergeron}  \& {Fontaine}}{{Dufour}
  et~al.}{2017}]{Dufour_2017}
{Dufour} P.,  {Blouin} S.,  {Coutu} S.,  {Fortin-Archambault} M.,  {Thibeault}
  C.,  {Bergeron} P.,   {Fontaine} G.,  2017, in {Tremblay} P.~E.,  {Gaensicke}
  B.,   {Marsh} T.,  eds,  Astronomical Society of the Pacific Conference
  Series Vol. 509, 20th European White Dwarf Workshop. p.~3 (\mn@eprint {arXiv}
  {1610.00986})

\bibitem[\protect\citeauthoryear{ESA}{ESA}{1997}]{ESA_1997}
ESA ed. 1997, {The HIPPARCOS and TYCHO catalogues.}  ESA Special Publication
  Vol. 1200

\bibitem[\protect\citeauthoryear{{Eggleton}}{{Eggleton}}{1983}]{Eggleton_1983}
{Eggleton} P.~P.,  1983, \mn@doi [\apj] {10.1086/160960}, \href
  {https://ui.adsabs.harvard.edu/abs/1983ApJ...268..368E} {268, 368}

\bibitem[\protect\citeauthoryear{{Ekstr{\"o}m} et~al.,}{{Ekstr{\"o}m}
  et~al.}{2012}]{Ekstrom+Georgy+Eggenberger+etal_2012}
{Ekstr{\"o}m} S.,  et~al., 2012, \mn@doi [\aap] {10.1051/0004-6361/201117751},
  \href {https://ui.adsabs.harvard.edu/abs/2012A&A...537A.146E} {537, A146}

\bibitem[\protect\citeauthoryear{{El-Badry}, {Rix}  \& {Weisz}}{{El-Badry}
  et~al.}{2018}]{El-Badry_2018}
{El-Badry} K.,  {Rix} H.-W.,   {Weisz} D.~R.,  2018, \mn@doi [\apjl]
  {10.3847/2041-8213/aaca9c}, \href
  {https://ui.adsabs.harvard.edu/abs/2018ApJ...860L..17E} {860, L17}

\bibitem[\protect\citeauthoryear{{Fischer}, {Marcy}  \& {Spronck}}{{Fischer}
  et~al.}{2014}]{Fischer_2014}
{Fischer} D.~A.,  {Marcy} G.~W.,   {Spronck} J. F.~P.,  2014, \mn@doi [\apjs]
  {10.1088/0067-0049/210/1/5}, \href
  {https://ui.adsabs.harvard.edu/abs/2014ApJS..210....5F} {210, 5}

\bibitem[\protect\citeauthoryear{{Fontaine}, {Brassard}  \&
  {Bergeron}}{{Fontaine} et~al.}{2001}]{Fontaine_2001}
{Fontaine} G.,  {Brassard} P.,   {Bergeron} P.,  2001, \mn@doi [\pasp]
  {10.1086/319535}, \href
  {https://ui.adsabs.harvard.edu/abs/2001PASP..113..409F} {113, 409}

\bibitem[\protect\citeauthoryear{{Foreman-Mackey}, {Hogg}, {Lang}  \&
  {Goodman}}{{Foreman-Mackey} et~al.}{2013}]{emcee_2013}
{Foreman-Mackey} D.,  {Hogg} D.~W.,  {Lang} D.,   {Goodman} J.,  2013, \mn@doi
  [\pasp] {10.1086/670067}, \href
  {https://ui.adsabs.harvard.edu/abs/2013PASP..125..306F} {125, 306}

\bibitem[\protect\citeauthoryear{{Gaia Collaboration} et~al.,}{{Gaia
  Collaboration} et~al.}{2016}]{Gaia_2016}
{Gaia Collaboration} et~al., 2016, \mn@doi [\aap]
  {10.1051/0004-6361/201629272}, \href
  {https://ui.adsabs.harvard.edu/abs/2016A&A...595A...1G} {595, A1}

\bibitem[\protect\citeauthoryear{{Gaia Collaboration} et~al.,}{{Gaia
  Collaboration} et~al.}{2022}]{Gaia_DR3}
{Gaia Collaboration} et~al., 2022, arXiv e-prints, \href
  {https://ui.adsabs.harvard.edu/abs/2022arXiv220800211G} {p. arXiv:2208.00211}

\bibitem[\protect\citeauthoryear{{Gomez Gonzalez} et~al.,}{{Gomez Gonzalez}
  et~al.}{2017}]{Gonzalez_2017}
{Gomez Gonzalez} C.~A.,  et~al., 2017, \mn@doi [\aj]
  {10.3847/1538-3881/aa73d7}, \href
  {https://ui.adsabs.harvard.edu/abs/2017AJ....154....7G} {154, 7}

\bibitem[\protect\citeauthoryear{{Gossage}, {Conroy}, {Dotter}, {Choi},
  {Rosenfield}, {Cargile}  \& {Dolphin}}{{Gossage}
  et~al.}{2018}]{Gossage+Conroy+Dotter+etal_2018}
{Gossage} S.,  {Conroy} C.,  {Dotter} A.,  {Choi} J.,  {Rosenfield} P.,
  {Cargile} P.,   {Dolphin} A.,  2018, \mn@doi [\apj]
  {10.3847/1538-4357/aad0a0}, \href
  {https://ui.adsabs.harvard.edu/abs/2018ApJ...863...67G} {863, 67}

\bibitem[\protect\citeauthoryear{{Gratton} et~al.,}{{Gratton}
  et~al.}{2021}]{Gratton_2021}
{Gratton} R.,  et~al., 2021, \mn@doi [\aap] {10.1051/0004-6361/202039601},
  \href {https://ui.adsabs.harvard.edu/abs/2021A&A...646A..61G} {646, A61}

\bibitem[\protect\citeauthoryear{{Griffin}}{{Griffin}}{2009}]{Griffin_2009}
{Griffin} R.~F.,  2009, The Observatory, \href
  {https://ui.adsabs.harvard.edu/abs/2009Obs...129....6G} {129, 6}

\bibitem[\protect\citeauthoryear{{H{\"o}fner} \& {Olofsson}}{{H{\"o}fner} \&
  {Olofsson}}{2018}]{Hofner_2018}
{H{\"o}fner} S.,  {Olofsson} H.,  2018, \mn@doi [\aapr]
  {10.1007/s00159-017-0106-5}, \href
  {https://ui.adsabs.harvard.edu/abs/2018A&ARv..26....1H} {26, 1}

\bibitem[\protect\citeauthoryear{{Holberg} \& {Bergeron}}{{Holberg} \&
  {Bergeron}}{2006}]{Holberg_2006}
{Holberg} J.~B.,  {Bergeron} P.,  2006, \mn@doi [\aj] {10.1086/505938}, \href
  {https://ui.adsabs.harvard.edu/abs/2006AJ....132.1221H} {132, 1221}

\bibitem[\protect\citeauthoryear{{Holberg}, {Oswalt}, {Sion}, {Barstow}  \&
  {Burleigh}}{{Holberg} et~al.}{2013}]{Holberg_2013}
{Holberg} J.~B.,  {Oswalt} T.~D.,  {Sion} E.~M.,  {Barstow} M.~A.,   {Burleigh}
  M.~R.,  2013, \mn@doi [\mnras] {10.1093/mnras/stt1433}, \href
  {https://ui.adsabs.harvard.edu/abs/2013MNRAS.435.2077H} {435, 2077}

\bibitem[\protect\citeauthoryear{{Holl} et~al.,}{{Holl}
  et~al.}{2022}]{Holl_2022}
{Holl} B.,  et~al., 2022, arXiv e-prints, \href
  {https://ui.adsabs.harvard.edu/abs/2022arXiv220605439H} {p. arXiv:2206.05439}

\bibitem[\protect\citeauthoryear{{Howard} et~al.,}{{Howard}
  et~al.}{2010}]{Howard_2010}
{Howard} A.~W.,  et~al., 2010, \mn@doi [\apj] {10.1088/0004-637X/721/2/1467},
  \href {https://ui.adsabs.harvard.edu/abs/2010ApJ...721.1467H} {721, 1467}

\bibitem[\protect\citeauthoryear{{Hwang}, {Ting}  \& {Zakamska}}{{Hwang}
  et~al.}{2022}]{Hwang_2022}
{Hwang} H.-C.,  {Ting} Y.-S.,   {Zakamska} N.~L.,  2022, \mn@doi [\mnras]
  {10.1093/mnras/stac675}, \href
  {https://ui.adsabs.harvard.edu/abs/2022MNRAS.512.3383H} {512, 3383}

\bibitem[\protect\citeauthoryear{{Ivanova} et~al.,}{{Ivanova}
  et~al.}{2013}]{Ivanova_2013}
{Ivanova} N.,  et~al., 2013, \mn@doi [\aapr] {10.1007/s00159-013-0059-2}, \href
  {https://ui.adsabs.harvard.edu/abs/2013A&ARv..21...59I} {21, 59}

\bibitem[\protect\citeauthoryear{{Iwamoto}, {Brachwitz}, {Nomoto}, {Kishimoto},
  {Umeda}, {Hix}  \& {Thielemann}}{{Iwamoto} et~al.}{1999}]{Iwamoto_1999}
{Iwamoto} K.,  {Brachwitz} F.,  {Nomoto} K.,  {Kishimoto} N.,  {Umeda} H.,
  {Hix} W.~R.,   {Thielemann} F.-K.,  1999, \mn@doi [\apjs] {10.1086/313278},
  \href {https://ui.adsabs.harvard.edu/abs/1999ApJS..125..439I} {125, 439}

\bibitem[\protect\citeauthoryear{{Izmailov}}{{Izmailov}}{2019}]{Izmailov_2019}
{Izmailov} I.~S.,  2019, \mn@doi [Astronomy Letters]
  {10.1134/S106377371901002X}, \href
  {https://ui.adsabs.harvard.edu/abs/2019AstL...45...30I} {45, 30}

\bibitem[\protect\citeauthoryear{{Jordan}, {Napiwotzki}, {Koester}  \&
  {Rauch}}{{Jordan} et~al.}{1997}]{Jordan_1997}
{Jordan} S.,  {Napiwotzki} R.,  {Koester} D.,   {Rauch} T.,  1997, \aap, \href
  {https://ui.adsabs.harvard.edu/abs/1997A&A...318..461J} {318, 461}

\bibitem[\protect\citeauthoryear{{Joyce}, {Barstow}, {Casewell}, {Burleigh},
  {Holberg}  \& {Bond}}{{Joyce} et~al.}{2018}]{Joyce_2018}
{Joyce} S.~R.~G.,  {Barstow} M.~A.,  {Casewell} S.~L.,  {Burleigh} M.~R.,
  {Holberg} J.~B.,   {Bond} H.~E.,  2018, \mn@doi [\mnras]
  {10.1093/mnras/sty1425}, \href
  {https://ui.adsabs.harvard.edu/abs/2018MNRAS.479.1612J} {479, 1612}

\bibitem[\protect\citeauthoryear{{Kallarakal}, {Lindenblad}, {Josties},
  {Riddle}, {Miranian}, {Mintz}  \& {Klugh}}{{Kallarakal}
  et~al.}{1969}]{Kallarakal_1969}
{Kallarakal} V.~V.,  {Lindenblad} I.~W.,  {Josties} F.~J.,  {Riddle} R.~K.,
  {Miranian} M.,  {Mintz} B.~F.,   {Klugh} A.~P.,  1969, Publications of the
  U.S. Naval Observatory Second Series, \href
  {https://ui.adsabs.harvard.edu/abs/1969PUSNO..18.....K} {18}

\bibitem[\protect\citeauthoryear{{Kane}, {Dalba}, {Horner}, {Li}, {Wittenmyer},
  {Horch}, {Howell}  \& {Everett}}{{Kane} et~al.}{2019}]{Kane_2019}
{Kane} S.~R.,  {Dalba} P.~A.,  {Horner} J.,  {Li} Z.,  {Wittenmyer} R.~A.,
  {Horch} E.~P.,  {Howell} S.~B.,   {Everett} M.~E.,  2019, \mn@doi [\apj]
  {10.3847/1538-4357/ab0e74}, \href
  {https://ui.adsabs.harvard.edu/abs/2019ApJ...875...74K} {875, 74}

\bibitem[\protect\citeauthoryear{{Kervella}, {Arenou}, {Mignard}  \&
  {Th{\'e}venin}}{{Kervella} et~al.}{2019}]{Kervella_2019}
{Kervella} P.,  {Arenou} F.,  {Mignard} F.,   {Th{\'e}venin} F.,  2019, \mn@doi
  [\aap] {10.1051/0004-6361/201834371}, \href
  {https://ui.adsabs.harvard.edu/abs/2019A&A...623A..72K} {623, A72}

\bibitem[\protect\citeauthoryear{{Kilic}, {Allende Prieto}, {Brown}  \&
  {Koester}}{{Kilic} et~al.}{2007}]{Kilic_2007}
{Kilic} M.,  {Allende Prieto} C.,  {Brown} W.~R.,   {Koester} D.,  2007,
  \mn@doi [\apj] {10.1086/514327}, \href
  {https://ui.adsabs.harvard.edu/abs/2007ApJ...660.1451K} {660, 1451}

\bibitem[\protect\citeauthoryear{{Kiman}, {Xu}, {Faherty}, {Gagn{\'e}},
  {Angus}, {Brandt}, {Casewell}  \& {Cruz}}{{Kiman} et~al.}{2022}]{Kiman_2022}
{Kiman} R.,  {Xu} S.,  {Faherty} J.~K.,  {Gagn{\'e}} J.,  {Angus} R.,  {Brandt}
  T.~D.,  {Casewell} S.~L.,   {Cruz} K.~L.,  2022, \mn@doi [\aj]
  {10.3847/1538-3881/ac7788}, \href
  {https://ui.adsabs.harvard.edu/abs/2022AJ....164...62K} {164, 62}

\bibitem[\protect\citeauthoryear{{Knapp} \& {Nanson}}{{Knapp} \&
  {Nanson}}{2019}]{Knapp_2019}
{Knapp} W.,  {Nanson} J.,  2019, Journal of Double Star Observations, \href
  {https://ui.adsabs.harvard.edu/abs/2019JDSO...15...42K} {15, 42}

\bibitem[\protect\citeauthoryear{{Konacki} \& {Lane}}{{Konacki} \&
  {Lane}}{2004}]{Konacki_2004}
{Konacki} M.,  {Lane} B.~F.,  2004, \mn@doi [\apj] {10.1086/421037}, \href
  {https://ui.adsabs.harvard.edu/abs/2004ApJ...610..443K} {610, 443}

\bibitem[\protect\citeauthoryear{{Kroupa}}{{Kroupa}}{2001}]{Kroupa_2001}
{Kroupa} P.,  2001, \mn@doi [\mnras] {10.1046/j.1365-8711.2001.04022.x}, \href
  {https://ui.adsabs.harvard.edu/abs/2001MNRAS.322..231K} {322, 231}

\bibitem[\protect\citeauthoryear{{Landsman}, {Simon}  \& {Bergeron}}{{Landsman}
  et~al.}{1996}]{Landsman_1996}
{Landsman} W.,  {Simon} T.,   {Bergeron} P.,  1996, \mn@doi [\pasp]
  {10.1086/133718}, \href
  {https://ui.adsabs.harvard.edu/abs/1996PASP..108..250L} {108, 250}

\bibitem[\protect\citeauthoryear{{Landstreet} \& {Bagnulo}}{{Landstreet} \&
  {Bagnulo}}{2020}]{Landstreet_2020}
{Landstreet} J.~D.,  {Bagnulo} S.,  2020, \mn@doi [\aap]
  {10.1051/0004-6361/201937301}, \href
  {https://ui.adsabs.harvard.edu/abs/2020A&A...634L..10L} {634, L10}

\bibitem[\protect\citeauthoryear{{Latham}}{{Latham}}{1992}]{Latham_1992}
{Latham} D.~W.,  1992, in {McAlister} H.~A.,  {Hartkopf} W.~I.,  eds,
  Astronomical Society of the Pacific Conference Series Vol. 32, IAU Colloq.
  135: Complementary Approaches to Double and Multiple Star Research. p.~110

\bibitem[\protect\citeauthoryear{{Luhn}, {Wright}, {Howard}  \&
  {Isaacson}}{{Luhn} et~al.}{2020}]{Luhn_2020}
{Luhn} J.~K.,  {Wright} J.~T.,  {Howard} A.~W.,   {Isaacson} H.,  2020, \mn@doi
  [\aj] {10.3847/1538-3881/ab855a}, \href
  {https://ui.adsabs.harvard.edu/abs/2020AJ....159..235L} {159, 235}

\bibitem[\protect\citeauthoryear{{Mamajek} \& {Hillenbrand}}{{Mamajek} \&
  {Hillenbrand}}{2008}]{Mamajek_2008}
{Mamajek} E.~E.,  {Hillenbrand} L.~A.,  2008, \mn@doi [\apj] {10.1086/591785},
  \href {https://ui.adsabs.harvard.edu/abs/2008ApJ...687.1264M} {687, 1264}

\bibitem[\protect\citeauthoryear{{Marois}, {Lafreni{\`e}re}, {Doyon},
  {Macintosh}  \& {Nadeau}}{{Marois} et~al.}{2006}]{Marois_2006}
{Marois} C.,  {Lafreni{\`e}re} D.,  {Doyon} R.,  {Macintosh} B.,   {Nadeau} D.,
   2006, \mn@doi [\apj] {10.1086/500401}, \href
  {https://ui.adsabs.harvard.edu/abs/2006ApJ...641..556M} {641, 556}

\bibitem[\protect\citeauthoryear{{Mart{\'\i}n}, {Lodieu}, {Pavlenko}  \&
  {B{\'e}jar}}{{Mart{\'\i}n} et~al.}{2018}]{Martin_2018}
{Mart{\'\i}n} E.~L.,  {Lodieu} N.,  {Pavlenko} Y.,   {B{\'e}jar} V. J.~S.,
  2018, \mn@doi [\apj] {10.3847/1538-4357/aaaeb8}, \href
  {https://ui.adsabs.harvard.edu/abs/2018ApJ...856...40M} {856, 40}

\bibitem[\protect\citeauthoryear{{Mason}, {Hartkopf}  \& {Miles}}{{Mason}
  et~al.}{2017}]{Mason_2017}
{Mason} B.~D.,  {Hartkopf} W.~I.,   {Miles} K.~N.,  2017, \mn@doi [\aj]
  {10.3847/1538-3881/aa803e}, \href
  {https://ui.adsabs.harvard.edu/abs/2017AJ....154..200M} {154, 200}

\bibitem[\protect\citeauthoryear{{Massarotti}, {Latham}, {Stefanik}  \&
  {Fogel}}{{Massarotti} et~al.}{2008}]{Massarotti_2008}
{Massarotti} A.,  {Latham} D.~W.,  {Stefanik} R.~P.,   {Fogel} J.,  2008,
  \mn@doi [\aj] {10.1088/0004-6256/135/1/209}, \href
  {https://ui.adsabs.harvard.edu/abs/2008AJ....135..209M} {135, 209}

\bibitem[\protect\citeauthoryear{{Matthews} et~al.,}{{Matthews}
  et~al.}{2014}]{Matthews_2014}
{Matthews} C.~T.,  et~al., 2014, \mn@doi [\apjl] {10.1088/2041-8205/783/2/L25},
  \href {https://ui.adsabs.harvard.edu/abs/2014ApJ...783L..25M} {783, L25}

\bibitem[\protect\citeauthoryear{{Mayor}}{{Mayor}}{1985}]{Mayor_1985}
{Mayor} M.,  1985, in {Philip} A.~G.~D.,  {Latham} D.~W.,  eds, Stellar Radial
  Velocities. pp 21--34

\bibitem[\protect\citeauthoryear{{Morgan}, {Harris}  \& {Johnson}}{{Morgan}
  et~al.}{1953}]{Morgan_1953}
{Morgan} W.~W.,  {Harris} D.~L.,   {Johnson} H.~L.,  1953, \mn@doi [\apj]
  {10.1086/145729}, \href
  {https://ui.adsabs.harvard.edu/abs/1953ApJ...118...92M} {118, 92}

\bibitem[\protect\citeauthoryear{{Oomen}, {Van Winckel}, {Pols}, {Nelemans},
  {Escorza}, {Manick}, {Kamath}  \& {Waelkens}}{{Oomen}
  et~al.}{2018}]{Oomen_2018}
{Oomen} G.-M.,  {Van Winckel} H.,  {Pols} O.,  {Nelemans} G.,  {Escorza} A.,
  {Manick} R.,  {Kamath} D.,   {Waelkens} C.,  2018, \mn@doi [\aap]
  {10.1051/0004-6361/201833816}, \href
  {https://ui.adsabs.harvard.edu/abs/2018A&A...620A..85O} {620, A85}

\bibitem[\protect\citeauthoryear{{Paxton} et~al.,}{{Paxton}
  et~al.}{2013}]{MESA_rotation}
{Paxton} B.,  et~al., 2013, \mn@doi [\apjs] {10.1088/0067-0049/208/1/4}, \href
  {https://ui.adsabs.harvard.edu/abs/2013ApJS..208....4P} {208, 4}

\bibitem[\protect\citeauthoryear{{Perets} \& {Kratter}}{{Perets} \&
  {Kratter}}{2012}]{Perets_2012}
{Perets} H.~B.,  {Kratter} K.~M.,  2012, \mn@doi [\apj]
  {10.1088/0004-637X/760/2/99}, \href
  {https://ui.adsabs.harvard.edu/abs/2012ApJ...760...99P} {760, 99}

\bibitem[\protect\citeauthoryear{{Perlmutter} et~al.,}{{Perlmutter}
  et~al.}{1999}]{Perlmutter_1999}
{Perlmutter} S.,  et~al., 1999, \mn@doi [\apj] {10.1086/307221}, \href
  {https://ui.adsabs.harvard.edu/abs/1999ApJ...517..565P} {517, 565}

\bibitem[\protect\citeauthoryear{{Perryman} et~al.,}{{Perryman}
  et~al.}{1998}]{Perryman_1998}
{Perryman} M.~A.~C.,  et~al., 1998, \aap, \href
  {https://ui.adsabs.harvard.edu/abs/1998A&A...331...81P} {331, 81}

\bibitem[\protect\citeauthoryear{{Phillips}}{{Phillips}}{1993}]{Phillips_1993}
{Phillips} M.~M.,  1993, \mn@doi [\apjl] {10.1086/186970}, \href
  {https://ui.adsabs.harvard.edu/abs/1993ApJ...413L.105P} {413, L105}

\bibitem[\protect\citeauthoryear{{Planck Collaboration} et~al.,}{{Planck
  Collaboration} et~al.}{2020}]{Planck_Collab_2020}
{Planck Collaboration} et~al., 2020, \mn@doi [\aap]
  {10.1051/0004-6361/201833910}, \href
  {https://ui.adsabs.harvard.edu/abs/2020A&A...641A...6P} {641, A6}

\bibitem[\protect\citeauthoryear{{Rabe}}{{Rabe}}{1953}]{Rabe_1953}
{Rabe} W.,  1953, {Mikrometermessungen von Doppelsternen in den Jahren 1932 bis
  1946.}

\bibitem[\protect\citeauthoryear{{Raghavan} et~al.,}{{Raghavan}
  et~al.}{2010}]{Raghavan_2010}
{Raghavan} D.,  et~al., 2010, \mn@doi [\apjs] {10.1088/0067-0049/190/1/1},
  \href {https://ui.adsabs.harvard.edu/abs/2010ApJS..190....1R} {190, 1}

\bibitem[\protect\citeauthoryear{{Reimers}}{{Reimers}}{1975}]{Reimers_1975}
{Reimers} D.,  1975, Memoires of the Societe Royale des Sciences de Liege,
  \href {https://ui.adsabs.harvard.edu/abs/1975MSRSL...8..369R} {8, 369}

\bibitem[\protect\citeauthoryear{{Riess} et~al.,}{{Riess}
  et~al.}{1998}]{Riess_1998}
{Riess} A.~G.,  et~al., 1998, \mn@doi [\aj] {10.1086/300499}, \href
  {https://ui.adsabs.harvard.edu/abs/1998AJ....116.1009R} {116, 1009}

\bibitem[\protect\citeauthoryear{{Rodrigo} \& {Solano}}{{Rodrigo} \&
  {Solano}}{2020}]{SVO_2020}
{Rodrigo} C.,  {Solano} E.,  2020, in Contributions to the XIV.0 Scientific
  Meeting (virtual) of the Spanish Astronomical Society. p.~182

\bibitem[\protect\citeauthoryear{{Romero}, {Kepler}, {Joyce}, {Lauffer}  \&
  {C{\'o}rsico}}{{Romero} et~al.}{2019}]{Romero_2019}
{Romero} A.~D.,  {Kepler} S.~O.,  {Joyce} S.~R.~G.,  {Lauffer} G.~R.,
  {C{\'o}rsico} A.~H.,  2019, \mn@doi [\mnras] {10.1093/mnras/stz160}, \href
  {https://ui.adsabs.harvard.edu/abs/2019MNRAS.484.2711R} {484, 2711}

\bibitem[\protect\citeauthoryear{{Rosenthal} et~al.,}{{Rosenthal}
  et~al.}{2021}]{Rosenthal_2021}
{Rosenthal} L.~J.,  et~al., 2021, \mn@doi [\apjs] {10.3847/1538-4365/abe23c},
  \href {https://ui.adsabs.harvard.edu/abs/2021ApJS..255....8R} {255, 8}

\bibitem[\protect\citeauthoryear{{Service}, {Lu}, {Campbell}, {Sitarski},
  {Ghez}  \& {Anderson}}{{Service} et~al.}{2016}]{Service_2016}
{Service} M.,  {Lu} J.~R.,  {Campbell} R.,  {Sitarski} B.~N.,  {Ghez} A.~M.,
  {Anderson} J.,  2016, \mn@doi [\pasp] {10.1088/1538-3873/128/967/095004},
  \href {https://ui.adsabs.harvard.edu/abs/2016PASP..128i5004S} {128, 095004}

\bibitem[\protect\citeauthoryear{{Skiff}}{{Skiff}}{2014}]{Skiff_2014_catalogue}
{Skiff} B.~A.,  2014, VizieR Online Data Catalog, \href
  {https://ui.adsabs.harvard.edu/abs/2014yCat....1.2023S} {p.~B/mk}

\bibitem[\protect\citeauthoryear{{Skrutskie} et~al.,}{{Skrutskie}
  et~al.}{2010}]{Skrutskie_2010}
{Skrutskie} M.~F.,  et~al., 2010, in {McLean} I.~S.,  {Ramsay} S.~K.,
  {Takami} H.,  eds,  Society of Photo-Optical Instrumentation Engineers (SPIE)
  Conference Series Vol. 7735, Ground-based and Airborne Instrumentation for
  Astronomy III. p. 77353H, \mn@doi{10.1117/12.857724}

\bibitem[\protect\citeauthoryear{{Stefanik}, {Torres}, {Latham}, {Landsman},
  {Craig}  \& {Murrett}}{{Stefanik} et~al.}{2011}]{Stefanik_2011}
{Stefanik} R.~P.,  {Torres} G.,  {Latham} D.~W.,  {Landsman} W.,  {Craig} N.,
  {Murrett} J.,  2011, \mn@doi [\aj] {10.1088/0004-6256/141/5/144}, \href
  {https://ui.adsabs.harvard.edu/abs/2011AJ....141..144S} {141, 144}

\bibitem[\protect\citeauthoryear{{Stolker} et~al.,}{{Stolker}
  et~al.}{2020}]{Stolker_2020}
{Stolker} T.,  et~al., 2020, \mn@doi [\aap] {10.1051/0004-6361/201937159},
  \href {https://ui.adsabs.harvard.edu/abs/2020A&A...635A.182S} {635, A182}

\bibitem[\protect\citeauthoryear{{Tokovinin} \& {Kiyaeva}}{{Tokovinin} \&
  {Kiyaeva}}{2016}]{Tokovinin_2016}
{Tokovinin} A.,  {Kiyaeva} O.,  2016, \mn@doi [\mnras] {10.1093/mnras/stv2825},
  \href {https://ui.adsabs.harvard.edu/abs/2016MNRAS.456.2070T} {456, 2070}

\bibitem[\protect\citeauthoryear{{Tokovinin}, {Mason}, {Mendez}, {Horch}  \&
  {Brice{\~n}o}}{{Tokovinin} et~al.}{2019}]{Tokovinin_2019}
{Tokovinin} A.,  {Mason} B.~D.,  {Mendez} R.~A.,  {Horch} E.~P.,
  {Brice{\~n}o} C.,  2019, \mn@doi [\aj] {10.3847/1538-3881/ab24e4}, \href
  {https://ui.adsabs.harvard.edu/abs/2019AJ....158...48T} {158, 48}

\bibitem[\protect\citeauthoryear{{Tremblay}, {Bergeron}  \&
  {Gianninas}}{{Tremblay} et~al.}{2011}]{Tremblay_2011}
{Tremblay} P.~E.,  {Bergeron} P.,   {Gianninas} A.,  2011, \mn@doi [\apj]
  {10.1088/0004-637X/730/2/128}, \href
  {https://ui.adsabs.harvard.edu/abs/2011ApJ...730..128T} {730, 128}

\bibitem[\protect\citeauthoryear{{Tucci Maia}, {Ram{\'\i}rez}, {Mel{\'e}ndez},
  {Bedell}, {Bean}  \& {Asplund}}{{Tucci Maia} et~al.}{2016}]{Tucci_Maia_2016}
{Tucci Maia} M.,  {Ram{\'\i}rez} I.,  {Mel{\'e}ndez} J.,  {Bedell} M.,  {Bean}
  J.~L.,   {Asplund} M.,  2016, \mn@doi [\aap] {10.1051/0004-6361/201527848},
  \href {https://ui.adsabs.harvard.edu/abs/2016A&A...590A..32T} {590, A32}

\bibitem[\protect\citeauthoryear{{Valenti} \& {Fischer}}{{Valenti} \&
  {Fischer}}{2005}]{Valenti_2005}
{Valenti} J.~A.,  {Fischer} D.~A.,  2005, \mn@doi [\apjs] {10.1086/430500},
  \href {https://ui.adsabs.harvard.edu/abs/2005ApJS..159..141V} {159, 141}

\bibitem[\protect\citeauthoryear{{Vogt} et~al.,}{{Vogt}
  et~al.}{1994}]{Vogt_1994}
{Vogt} S.~S.,  et~al., 1994, in {Crawford} D.~L.,  {Craine} E.~R.,  eds,
  Society of Photo-Optical Instrumentation Engineers (SPIE) Conference Series
  Vol. 2198, Instrumentation in Astronomy VIII. p.~362,
  \mn@doi{10.1117/12.176725}

\bibitem[\protect\citeauthoryear{{Vogt} et~al.,}{{Vogt}
  et~al.}{2014}]{Vogt_2014}
{Vogt} S.~S.,  et~al., 2014, \mn@doi [\pasp] {10.1086/676120}, \href
  {https://ui.adsabs.harvard.edu/abs/2014PASP..126..359V} {126, 359}

\bibitem[\protect\citeauthoryear{{Vousden}, {Farr}  \& {Mandel}}{{Vousden}
  et~al.}{2016}]{Vousden_2016}
{Vousden} W.~D.,  {Farr} W.~M.,   {Mandel} I.,  2016, \mn@doi [\mnras]
  {10.1093/mnras/stv2422}, \href
  {https://ui.adsabs.harvard.edu/abs/2016MNRAS.455.1919V} {455, 1919}

\bibitem[\protect\citeauthoryear{{Vousden}, {Farr}  \& {Mandel}}{{Vousden}
  et~al.}{2021}]{ptemcee_2021}
{Vousden} W.,  {Farr} W.~M.,   {Mandel} I.,  2021, {ptemcee: A
  parallel-tempered version of emcee} (\mn@eprint {ascl} {2101.006})

\bibitem[\protect\citeauthoryear{{Wenger} et~al.,}{{Wenger}
  et~al.}{2000}]{Wenger_2000}
{Wenger} M.,  et~al., 2000, \mn@doi [\aaps] {10.1051/aas:2000332}, \href
  {https://ui.adsabs.harvard.edu/abs/2000A&AS..143....9W} {143, 9}

\bibitem[\protect\citeauthoryear{{Wertz}, {Absil}, {G{\'o}mez Gonz{\'a}lez},
  {Milli}, {Girard}, {Mawet}  \& {Pueyo}}{{Wertz} et~al.}{2017}]{Wertz_2017}
{Wertz} O.,  {Absil} O.,  {G{\'o}mez Gonz{\'a}lez} C.~A.,  {Milli} J.,
  {Girard} J.~H.,  {Mawet} D.,   {Pueyo} L.,  2017, \mn@doi [\aap]
  {10.1051/0004-6361/201628730}, \href
  {https://ui.adsabs.harvard.edu/abs/2017A&A...598A..83W} {598, A83}

\bibitem[\protect\citeauthoryear{{Wizinowich} et~al.,}{{Wizinowich}
  et~al.}{2000}]{Wizinowich_2000}
{Wizinowich} P.,  et~al., 2000, \mn@doi [\pasp] {10.1086/316543}, \href
  {https://ui.adsabs.harvard.edu/abs/2000PASP..112..315W} {112, 315}

\bibitem[\protect\citeauthoryear{{Wood}}{{Wood}}{1995}]{Wood_1995}
{Wood} M.~A.,  1995, in {Koester} D.,  {Werner} K.,  eds, , Vol.~443, White
  Dwarfs.
p.~41, \mn@doi{10.1007/3-540-59157-5_171}

\bibitem[\protect\citeauthoryear{{Zeng} et~al.,}{{Zeng}
  et~al.}{2022}]{Zeng_2022}
{Zeng} Y.,  et~al., 2022, \mn@doi [\aj] {10.3847/1538-3881/ac8ff7}, \href
  {https://ui.adsabs.harvard.edu/abs/2022AJ....164..188Z} {164, 188}

\bibitem[\protect\citeauthoryear{{de Medeiros} \& {Mayor}}{{de Medeiros} \&
  {Mayor}}{1999}]{de_Medeiros_1999}
{de Medeiros} J.~R.,  {Mayor} M.,  1999, \mn@doi [\aaps] {10.1051/aas:1999401},
  \href {https://ui.adsabs.harvard.edu/abs/1999A&AS..139..433D} {139, 433}

\makeatother
\end{thebibliography}
\bsp	
\label{lastpage}

\appendix

\changes{\section{Figures and tables of fit results}}

\begin{figure*}
    \centering
    \includegraphics[width=\linewidth]{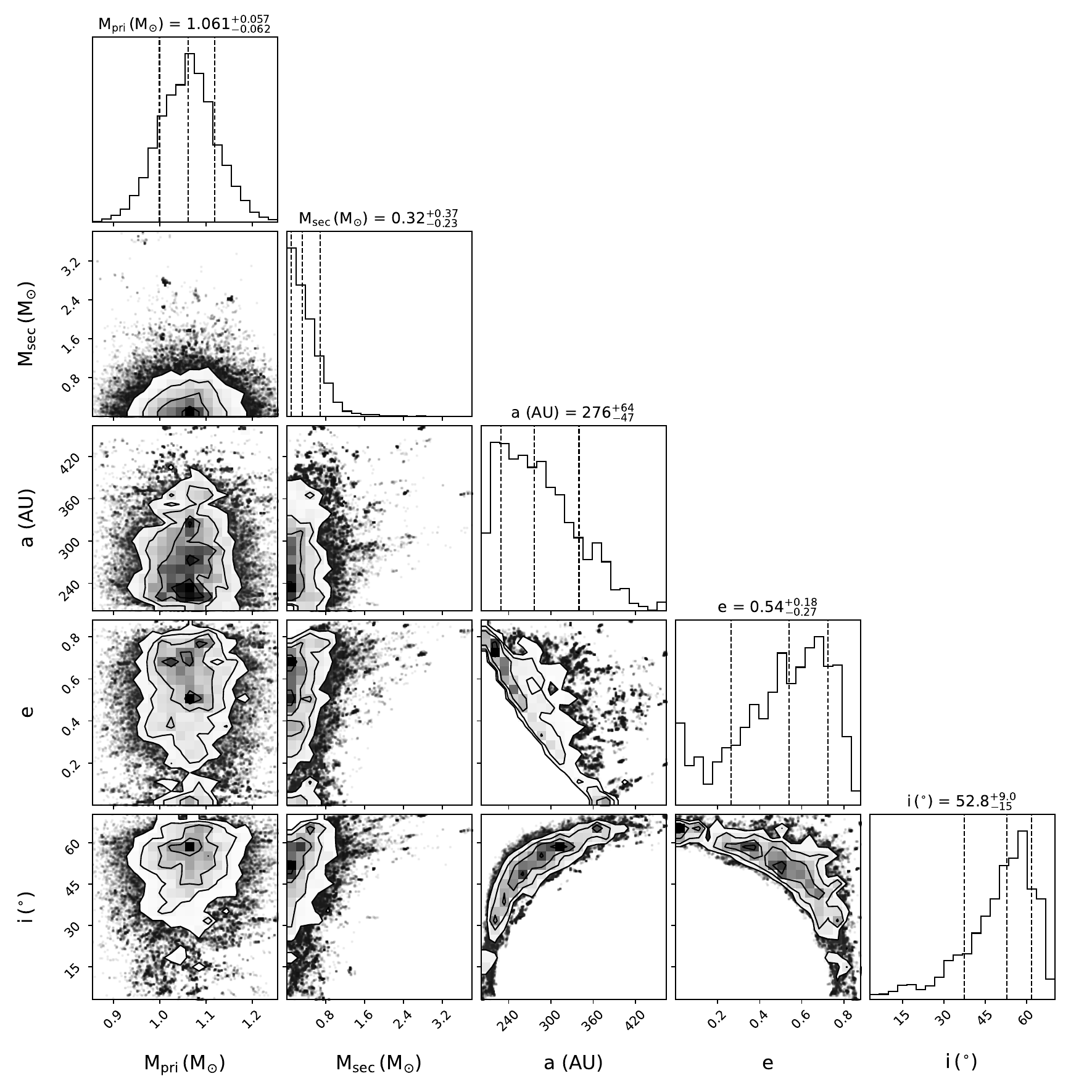}
    \caption{Best-fitting orbital parameters for the HD 19019 system and their posterior distributions from \textsc{orvara}. The selected parameters include the primary mass (in solar masses) $M_{\rm pri}$, the secondary mass (in Jupiter masses) $M_{\rm sec}$, the semi-major axis (in A.U.) $a$, the eccentricity $e$, and the inclination (in degrees) $i$. The contours on the 2-d joint posterior distributions give the 1-$\sigma$, 2-$\sigma$, and 3-$\sigma$ levels. The vertical dashed lines on the 1-d marginalized distributions indicate the 16\% and 84\% quantiles.}
    \label{fig:corner_19019}
\end{figure*}

\clearpage

\begin{figure*}
\captionsetup[subfigure]{labelformat=empty}
    \centering
    \subfloat[]{
    \includegraphics[height=8cm,width=0.43\linewidth]{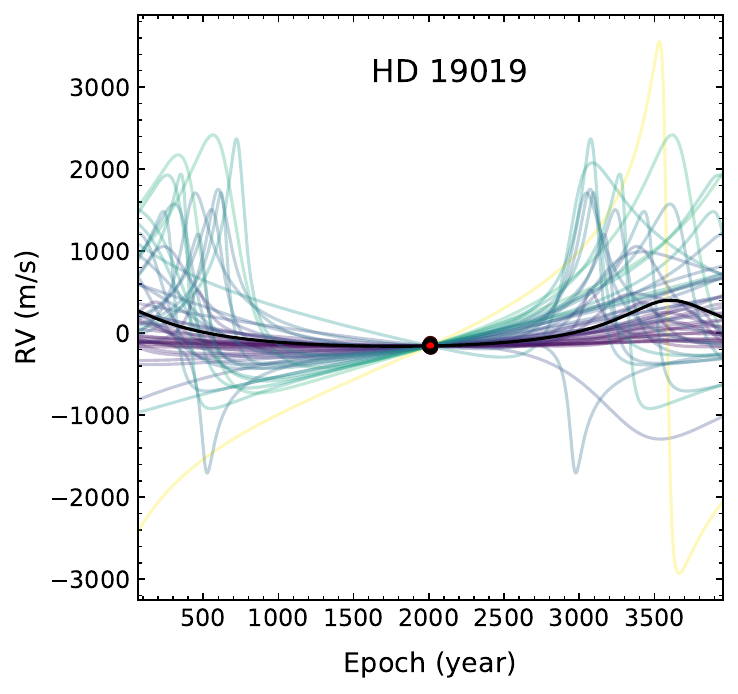}
    }
    \subfloat[]{
    \includegraphics[height=8cm,width=0.53\linewidth]{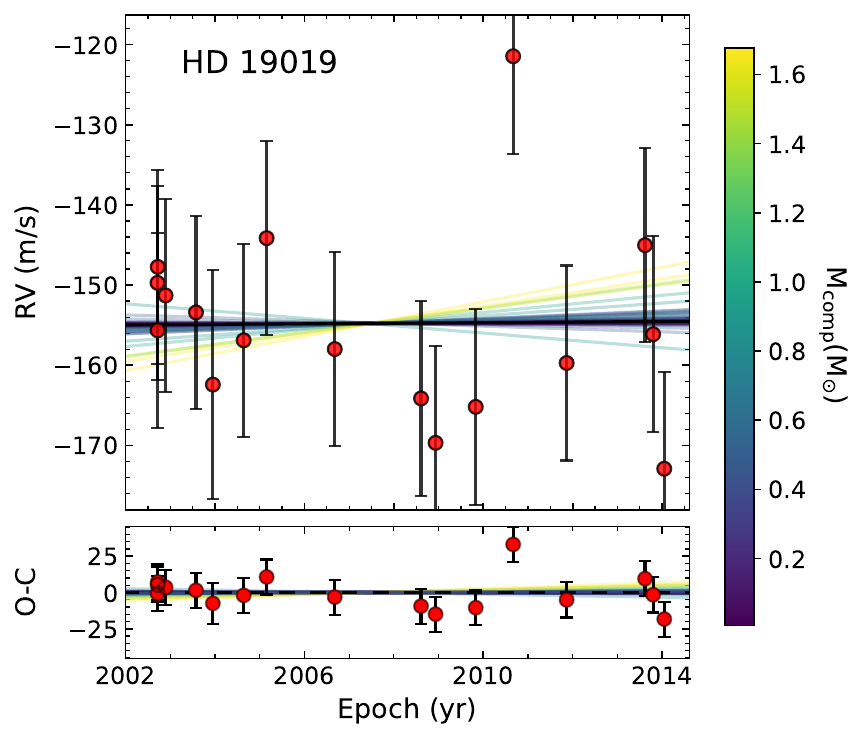}
    }
    \caption{Left panel: RV orbits of HD 19019 over the entire best-fitting orbital period. Right panel: RV orbits of HD 19019 over the observational time frame (top) and the residuals after subtracting the best-fitting orbit (bottom). For both panels, an RV of zero corresponds to the barycentric velocity of the system. The thick black line indicates the maximum likelihood RV orbit. The thin lines, color-coded by the companion mass, are 50 other orbits drawn randomly from the posterior distribution. The colored dots are the RV measurements, with a different color indicating a separate telescope. Most color bars are too small to be visible on the left panel.}
    \label{fig:RV_19019}
\end{figure*}

\begin{table*} 
    \centering
    \begin{tabular}{c@{\hskip 3.5cm}c@{\hskip 3.5cm}c} \hline \hline
         Parameter & Prior Distribution & Posteriors $\pm 1\sigma$ \\ \hline
         \multicolumn{3}{c}{Fitted parameters} \\ \hline
         Primary mass ($M_{\odot}$) & $1.06 \pm 0.06$ &        ${1.061}_{-0.062}^{+0.057}$ \\
         Companion mass ($M_{\odot}$) & Uniform &   ${0.32}_{-0.23}^{+0.37}$ \\
         Parallax (mas) & $31.979 \pm 0.029$ & $31.979 \pm 0.028$ \\
         Semimajor axis $a$ (AU) & $1/a$ (log-flat) &  ${276}_{-47}^{+64}$ \\
         Inclination $i$ ($\degree$) & $\sin{i}$ &  ${52.8}_{-15}^{+9.0}$ \\
         $\sqrt{e}\sin{\omega}$ & Uniform &  ${0.18}_{-0.43}^{+0.42}$ \\
         $\sqrt{e}\cos{\omega}$ & Uniform & ${0.53}_{-0.26}^{+0.24}$ \\
         Mean longitude at $t_{\rm ref}=2455197.50$ JD ($\degree$) & Uniform &  ${207}_{-11}^{+22}$ \\
         PA of the ascending node $\Omega$ ($\degree$) & Uniform &   ${23}_{-14}^{+326}$ \\
         RV jitter $\sigma$ (\ms) & Log-flat over [0,1000 \ms] &  ${12.2}_{-2.0}^{+2.6}$ \\ \hline
         \multicolumn{3}{c}{Derived parameters} \\ \hline
         Orbital period (yr) & ... &  ${3888}_{-956}^{+1304}$ \\
         Semimajor axis (mas) & ... &  ${8830}_{-1518}^{+2031}$ \\
         Eccentricity $e$ & ... &   ${0.54}_{-0.27}^{+0.18}$ \\
         Argument of periastron $\omega$ $(\degree)$ & ... &  ${59}_{-40}^{+281}$ \\
         Time of periastron $T_0$ (JD) & ... & ${3033863}_{-164261}^{+373450}$ \\
         Mass ratio & ... &   ${0.30}_{-0.22}^{+0.35}$ \\ \hline
    \end{tabular}
    \caption{Posteriors of the HD 19019 system.}
    \label{tab:post_19019}
\end{table*}

\clearpage

\begin{figure*}
\captionsetup[subfigure]{labelformat=empty}
    \centering
    \subfloat[]{
        \centering
        \includegraphics[height=8cm,width=0.44\linewidth]{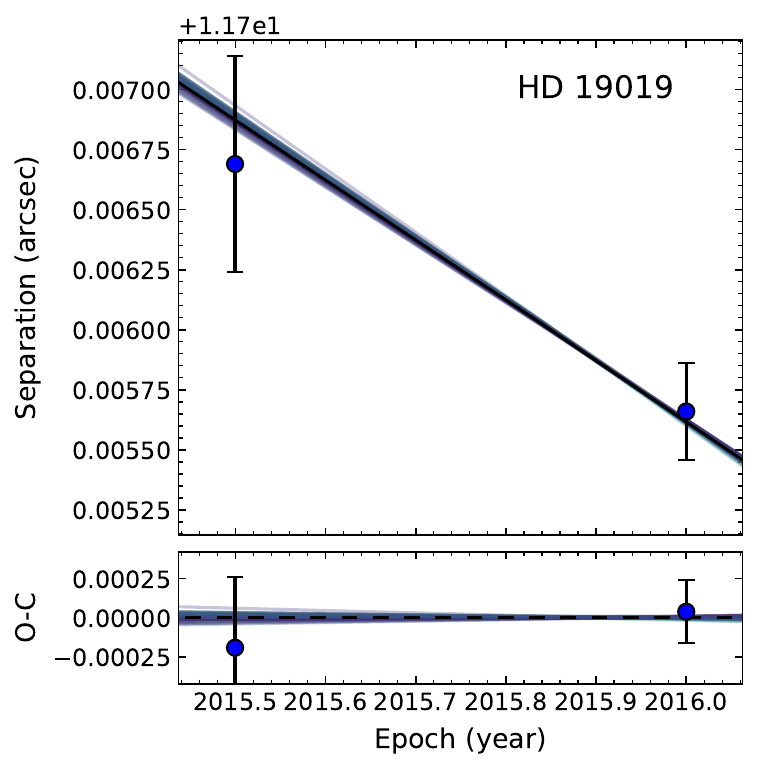}
        }
        \hfill
    \subfloat[]{
        \centering
        \includegraphics[height=8cm,width=0.52\linewidth]{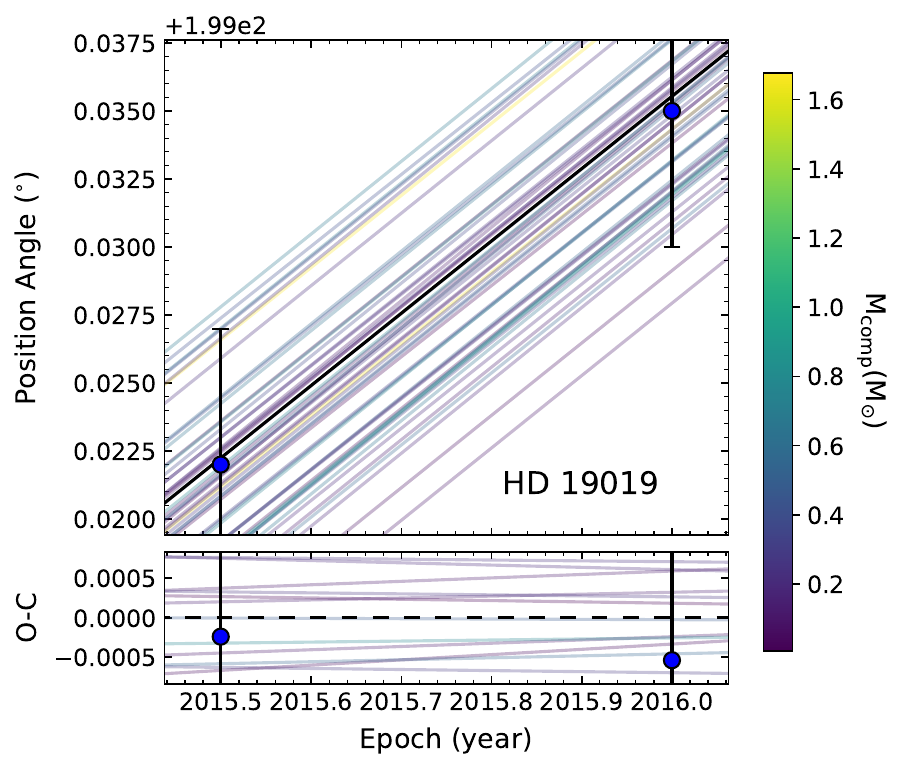}
    } \\
    \subfloat[]{
        \centering
        \includegraphics[height=8cm,width=0.985\linewidth]{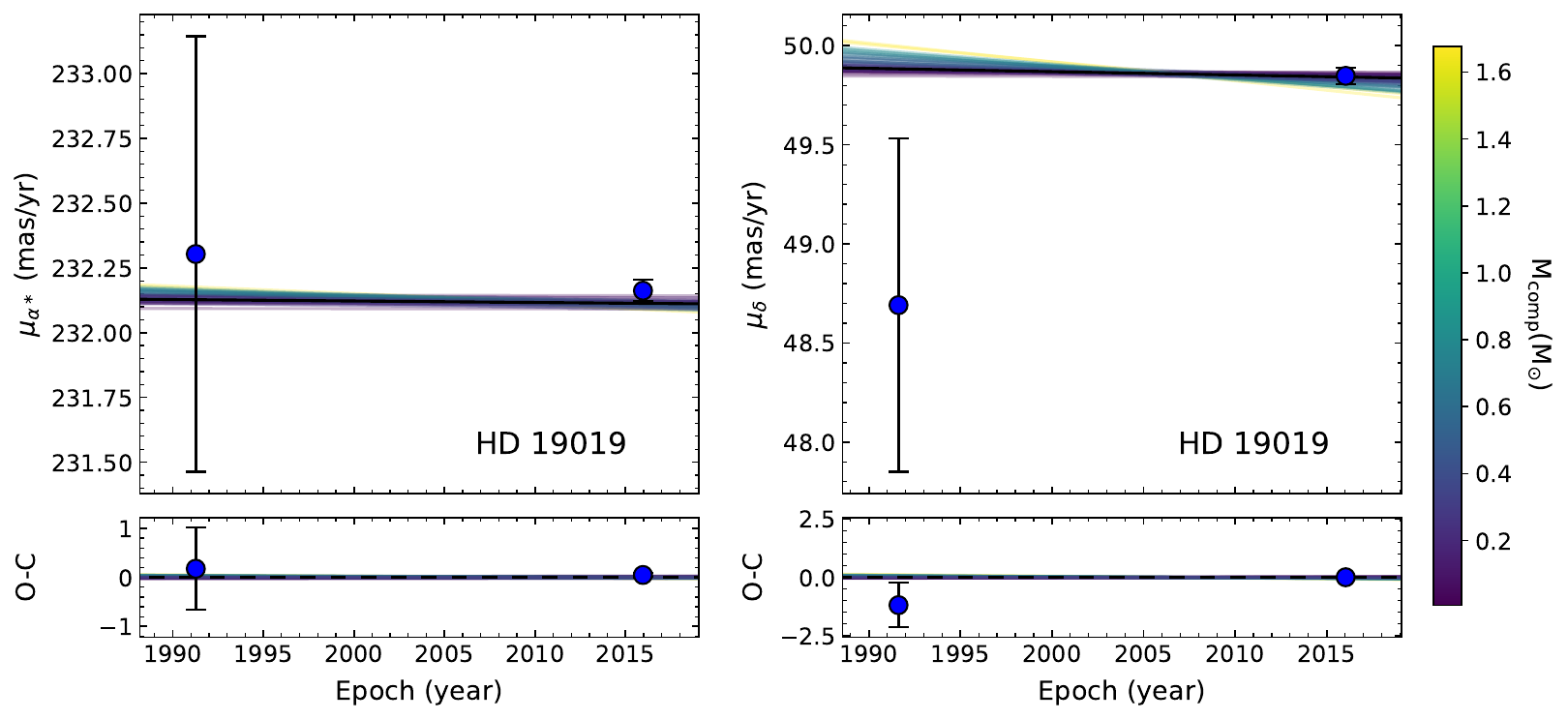}
    }
    \caption{Relative (top panel) and absolute (bottom panel) astrometry for the HD 19019 system. The observations are shown as blue dots with error bars. The thick black line is the maximum likelihood orbit, and the thin lines, color-coded by the companion mass, are 50 other orbits drawn randomly from the posterior distribution. The bottom part of each panel shows the residuals after subtracting the maximum likelihood orbit from the measurements.}
    \label{fig:relAst_19019}
\end{figure*}

\clearpage

\begin{table*} 
    \centering
    \begin{tabular}{c@{\hskip 4.1cm}c@{\hskip 4.1cm}c} \hline \hline
         Parameter & Prior Distribution & Posteriors $\pm 1\sigma$ \\ \hline
         \multicolumn{3}{c}{Fitted parameters} \\ \hline
         Primary mass ($M_{\odot}$) & $2.77 \pm 0.26$ & ${2.86}_{-0.22}^{+0.25}$ \\
         Companion mass ($M_{\odot}$) & $1/M$ (log-flat) &  ${0.798}_{-0.041}^{+0.10}$ \\
         Parallax (mas) & $21.094 \pm 0.032$ & $21.094 \pm 0.025$ \\
         Semimajor axis $a$ (AU) & $1/a$ (log-flat) &  ${49.8}_{-5.4}^{+12}$ \\
         Inclination $i$ ($\degree$) & $\sin{i}$ &  ${30}_{-15}^{+13}$ \\
         $\sqrt{e}\sin{\omega}$ & Uniform &  ${0.07}_{-0.49}^{+0.41}$ \\
         $\sqrt{e}\cos{\omega}$ & Uniform & ${0.01}_{-0.47}^{+0.47}$ \\
         Mean longitude at $t_{\rm ref}=2455197.50$ JD ($\degree$) & Uniform &  ${156}_{-126}^{+80}$ \\
         PA of the ascending node $\Omega$ ($\degree$) & Uniform &   ${174}_{-159}^{+121}$ \\ \hline
         \multicolumn{3}{c}{Derived parameters} \\ \hline
         Orbital period (yr) & ... &  ${184}_{-30}^{+65}$ \\
         Semimajor axis (mas) & ... & ${1050}_{-115}^{+250}$ \\
         Eccentricity $e$ & ... &   ${0.342}_{-0.19}^{+0.094}$\\
         Argument of periastron $\omega$ $(\degree)$ & ... &  ${163}_{-111}^{+141}$ \\
         Time of periastron $T_0$ (JD) & ... & ${2474412}_{-2469}^{+4137}$ \\
         Mass ratio & ... &   ${0.284}_{-0.024}^{+0.033}$ \\ \hline
    \end{tabular}
    \caption{Posteriors of the HD 27483 system.}
    \label{tab:post_27483}
\end{table*}

\begin{table*} 
    \centering
    \begin{tabular}{c@{\hskip 2.1cm}c@{\hskip 2.1cm}c@{\hskip 2.1cm}c} \hline \hline
         \multirow{2}{1.1cm}{Parameter} & Prior Distribution &  Posteriors $\pm 1\sigma$ & Posteriors $\pm 1\sigma$ \\ 
         & (Both companions) & (HD 27786 Ab) & (HD 27786 B) \\ \hline
         \multicolumn{4}{c}{Fitted parameters} \\ \hline
         Primary mass ($M_{\odot}$) & $1.540 \pm 0.077$ & \multicolumn{2}{c}{${1.714}_{-0.060}^{+0.061}$} \\
         Companion mass ($M_{\sun}$) & Uniform & $0.443 \pm 0.012$ & ${1.36}_{-0.41}^{+0.51}$ \\
         Parallax (mas) & $23.79 \pm 0.11$ & \multicolumn{2}{c}{$23.87 \pm 0.11$} \\
         Semimajor axis $a$ (AU) & $1/a$ (log-flat) & ${12.04}_{-0.14}^{+0.13}$ & $183 \pm 16$ \\
         Inclination $i$ ($\degree$) & $\sin{i}$ & ${172.8}_{-4.0}^{+3.2}$ & ${158.8}_{-9.2}^{+10}$ \\
         $\sqrt{e}\sin{\omega}$ & Uniform &  ${0.574}_{-0.16}^{+0.053}$ & ${0.07}_{-0.26}^{+0.21}$ \\
         $\sqrt{e}\cos{\omega}$ & Uniform &  ${0.15}_{-0.37}^{+0.31}$ & ${0.09}_{-0.21}^{+0.16}$ \\
         Mean longitude at $t_{\rm ref}=2455197.50$ JD ($\degree$) & Uniform &  ${48}_{-34}^{+294}$ & ${211}_{-140}^{+72}$ \\
         PA of the ascending node $\Omega$ ($\degree$) & Uniform &   ${199}_{-33}^{+35}$ &   ${221}_{-140}^{+67}$ \\ \hline
         \multicolumn{4}{c}{Derived parameters} \\ \hline
         Orbital period (yr) & ... & ${28.42}_{-0.28}^{+0.31}$ & ${1436}_{-251}^{+203}$ \\
         Semimajor axis (mas) & ... & ${287.4}_{-3.4}^{+3.3}$ &  ${4357}_{-367}^{+385}$ \\
         Eccentricity $e$ & ... &    ${0.4016}_{-0.0084}^{+0.0083}$ & ${0.074}_{-0.052}^{+0.085}$ \\
         Argument of periastron $\omega$ $(\degree)$ & ... &  ${77}_{-32}^{+34}$ & ${124}_{-82}^{+176}$ \\
         Time of periastron $T_0$ (JD) & ... &  ${2456909}_{-28}^{+29}$ & ${2648314}_{-89023}^{+202895}$ \\
         Mass ratio & ... &   ${0.2585}_{-0.0050}^{+0.0059}$ & ${0.80}_{-0.25}^{+0.31}$ \\ \hline
    \end{tabular}
    \caption{Posteriors of the HD 27786 system.}
    \label{tab:post_27786}
\end{table*}

\clearpage

\begin{table*} 
    \centering
    \begin{tabular}{c@{\hskip 3.5cm}c@{\hskip 3.5cm}c} \hline \hline
         Parameter & Prior Distribution & Posteriors $\pm 1\sigma$ \\ \hline
         \multicolumn{3}{c}{Fitted parameters} \\ \hline
         Primary mass ($M_{\odot}$) & $0.968 \pm 0.044$ &        $0.955 \pm 0.043$ \\
         Companion mass ($M_{\odot}$) & $1/M$ (log-flat) &  $0.591 \pm 0.011$ \\
         Parallax (mas) & $37.8677 \pm 0.0243$ & $37.867 \pm 0.016$ \\
         Semimajor axis $a$ (AU) & $1/a$ (log-flat) &  ${26.25}_{-0.66}^{+0.71}$ \\
         Inclination $i$ ($\degree$) & $\sin{i}$ &  $88.86 \pm 0.21$ \\
         $\sqrt{e}\sin{\omega}$ & Uniform &  ${0.8088}_{-0.0048}^{+0.0044}$ \\
         $\sqrt{e}\cos{\omega}$ & Uniform & ${0.190}_{-0.029}^{+0.027}$ \\
         Mean longitude at $t_{\rm ref}=2455197.50$ JD ($\degree$) & Uniform &  $21.1 \pm 2.1 $ \\
         PA of the ascending node $\Omega$ ($\degree$) & Uniform &   ${169.31}_{-0.52}^{+0.53}$ \\
         RV jitter $\sigma$ (\ms) & Log-flat over [0,1000 \ms] &  ${3.55}_{-0.34}^{+0.39}$ \\ \hline
         \multicolumn{3}{c}{Derived parameters} \\ \hline
         Orbital period (yr) & ... &  ${108.2}_{-4.8}^{+5.2}$ \\
         Semimajor axis (mas) & ... & ${994}_{-25}^{+27}$ \\
         Eccentricity $e$ & ... &   $0.690 \pm 0.017$ \\
         Argument of periastron $\omega$ $(\degree)$ & ... &  ${76.8}_{-1.7}^{+1.9}$ \\
         Time of periastron $T_0$ (JD) & ... & ${2481070}_{-1524}^{+1682}$ \\
         Mass ratio & ... &   ${0.620}_{-0.030}^{+0.033}$ \\ \hline
    \end{tabular}
    \caption{Posteriors of the HD 114174 system.}
    \label{tab:post_114174}
\end{table*}

\begin{table*} 
    \centering
    \begin{tabular}{c@{\hskip 1.63cm}c@{\hskip 1.63cm}c@{\hskip 1.63cm}c} \hline \hline
         Parameter & Prior Distribution & Posteriors $\pm 1\sigma$ & {\Gaia} solution $\pm 1\sigma$ \\ \hline
         \multicolumn{3}{c}{Fitted parameters} & ...\\ \hline
         Primary mass ($M_{\odot}$) & $1.158 \pm 0.058$ & ${1.154}_{- 0.055}^{+0.056} $ & ...\\
         Companion mass ($M_{\odot}$) & $1/M$ (log-flat) &  ... & ${0.557}^{+0.08}_{-0.072}$\\
         Parallax (mas) & $29.537 \pm 0.017$ & $29.537 \pm 0.017$ & $29.537 \pm 0.017$ \\
         Semimajor axis $a$ (AU) & $1/a$ (log-flat) &  ${3.758}_{-0.060}^ {+0.070}$ & $3.68 \pm 0.18$ \\
         Inclination $i$ ($\degree$) & $\sin{i}$ &  ... & ${53.1}^{+4.8}_{-5.5}$ \\
         $\sqrt{e}\sin{\omega}$ & Uniform &  ${-0.3024}_{-0.0006}^{+ 0.0007}$ & ${0.07}^{+0.06}_{-0.26}$ \\
         $\sqrt{e}\cos{\omega}$ & Uniform & ${-0.1907}_{-0.0016}^{+0.0017}$ & ${0.34}^{+0.06}_{-0.65}$\\
         Mean longitude at $t_{\rm ref}=2455197.50$ JD ($\degree$) & Uniform &  $44.52 \pm 0.05$ & ...\\
         PA of the ascending node $\Omega$ ($\degree$) & Uniform &  ... & ${212}^{+11}_{-191}$ \\
         RV jitter $\sigma$ (\ms) & Log-flat over [0,10 \ms] &  ${0.02}_{-0.02}^{+1.77}$  & ... \\ \hline
         \multicolumn{3}{c}{Derived parameters} & ... \\ \hline
         Companion minimum mass $M_{\rm sec}\sin{i}$ ($M_{\odot}$) & ... & ${0.461}_{-0.015}^{+0.017}$  & ${0.444}^{+0.081}_{-0.076}$\\
         Orbital period (d) & ... &  $2070.47 \pm 0.16 $ & $1968.6 \pm 154.0$ \\
         Semimajor axis (mas) & ... &  ${111.0}_{-1.8}^{+2.1}$ & $108.6 \pm 5.3$\\
         Eccentricity $e$ & ... &   $0.1278 \pm 0.0008$ & $0.143 \pm 0.036$\\
         RV semi-amplitude $K$ (\ms) & ... &  ${5560.3}_{-8.4}^{+8.1}$ & ... \\
         Argument of periastron $\omega$ $(\degree)$ & ... &  $237.76 \pm 0.22$ & ${19.5}^{+191.7}_{-9.7}$ \\
         Time of periastron $T_0$ (JD) & ... & ${2455344}_{-66}^{+77}$ & $2456380 \pm 160$ \\
         Mass ratio & ... &   $0.419_{-0.025}^{+0.066}$ & ${0.481}^{+0.070}_{-0.061}$ \\ \hline
    \end{tabular}
    \caption{Posteriors of the HD 118475 system.}
    \label{tab:post_118475}
\end{table*}

\clearpage

\begin{table*} 
    \centering
    \begin{tabular}{c@{\hskip 1.57cm}c@{\hskip 1.57cm}c@{\hskip 1.57cm}c} \hline \hline
         Parameter & Prior Distribution & Posteriors $\pm 1\sigma$ & {\Gaia} solution $\pm 1\sigma$ \\ \hline
         \multicolumn{3}{c}{Fitted parameters} & ...\\ \hline
         Primary mass ($M_{\odot}$) & $1.84 \pm 0.40$ & ${1.80}^{+0.37}_{-0.39}  $ & ...\\
         Companion mass ($M_{\odot}$) & $1/M$ (log-flat) & ... & ${0.543}^{+0.078}_{-0.080}$\\
         Parallax (mas) & $9.011 \pm 0.051$ & $9.011 \pm 0.051$ & $9.011 \pm 0.051$ \\
         Semimajor axis $a$ (AU) & $1/a$ (log-flat) &  ${1.65}^{+0.10}_{-0.12} $ & ${1.66}^{+0.10}_{-0.12}$ \\
         Inclination $i$ ($\degree$) & $\sin{i}$ &  ... & ${46.5}^{+5.6}_{-6.6}$ \\
         $\sqrt{e}\sin{\omega}$ & Uniform &  ${0.377}^{+0.025}_{-0.029}$ & ${0.557}^{+0.043}_{-0.044}$ \\
         $\sqrt{e}\cos{\omega}$ & Uniform & ${0.438}^{+0.028}_{-0.024}$ & ${0.325}^{+0.052}_{-0.057}$\\
         Mean longitude at $t_{\rm ref}=2455197.50$ JD ($\degree$) & Uniform &  $341.0^{+2.4}_{-2.6}$ & ...\\
         PA of the ascending node $\Omega$ ($\degree$) & Uniform &  ... & ${52.4}^{+5.1}_{-5.4}$ \\
         RV jitter $\sigma$ (\ms) & Log-flat over [0,1000 \ms] &  ${0.07}^{+9.94}_{-0.07}$  & ... \\ \hline
         \multicolumn{3}{c}{Derived parameters} & ... \\ \hline
         Companion minimum mass $M_{\rm sec}\sin{i}$ ($M_{\odot}$) & ... & ${0.385}^{+0.046}_{-0.056} $  & ${0.389}^{+0.081}_{-0.080}$\\
         Orbital period (d) & ... &  $509.6 \pm 1.2$ & $505.0 \pm 2.5 $ \\
         Semimajor axis (mas) & ... &  ${14.88}^{+0.89}_{-1.11} $ & ${14.94}^{+0.93}_{-1.06}$\\
         Eccentricity $e$ & ... &   ${0.336}^{+0.015}_{-0.014} $ & $0.419 \pm 0.046 $\\
         RV semi-amplitude $K$ (\ms) & ... &  ${6218}^{+135}_{-113}$ & ... \\
         Argument of periastron $\omega$ $(\degree)$ & ... &  ${40.5}^{+3.4}_{-3.5} $ & ${59.6}^{+5.5}_{-5.0}$ \\
         Time of periastron $T_0$ (JD) & ... & ${2455451}^{+184}_{-157}$ & $2455281 \pm 8$ \\
         Mass ratio & ... &   ${0.24}^{+0.15}_{-0.04}$ & ${0.297}^{+0.035} _{-0.028} $ \\ \hline
    \end{tabular}
    \caption{Posteriors of the HD 136138 system.}
    \label{tab:post_136138}
\end{table*}

\begin{table*} 
    \centering
    \begin{tabular}{c@{\hskip 3.35cm}c@{\hskip 3.35cm}c} \hline \hline
         Parameter & Prior Distribution & Posteriors $\pm 1\sigma$ \\ \hline
         \multicolumn{3}{c}{Fitted parameters} \\ \hline
         Primary mass ($M_{\odot}$) & $0.980 \pm 0.049$ & ${0.983}_{-0.049}^{+0.048}$ \\
         Companion mass ($M_{\odot}$) & $1/M$ (log-flat) &  ${0.526}_{-0.037}^{+0.039}$ \\
         Parallax (mas) & $28.2794 \pm 0.0259$ & $28.2795 \pm 0.0079$ \\
         Semimajor axis $a$ (AU) & $1/a$ (log-flat) &  ${41}_{-10}^{+15}$ \\
         Inclination $i$ ($\degree$) & $\sin{i}$ &  ${101.7}_{-8.8}^{+12}$ \\
         $\sqrt{e}\sin{\omega}$ & Uniform &  ${0.901}_{-0.050}^{+0.035}$ \\
         $\sqrt{e}\cos{\omega}$ & Uniform & ${-0.21}_{-0.18}^{+0.25}$ \\
         Mean longitude at $t_{\rm ref}=2455197.50$ JD ($\degree$) & Uniform &  ${52}_{-23}^{+19}$ \\
         PA of the ascending node $\Omega$ ($\degree$) & Uniform &   ${149}_{-20}^{+10}$ \\
         RV jitter $\sigma$ (\ms) & Log-flat over [0,1000 \ms] &  ${4.3}_{-1.5}^{+2.6}$ \\ \hline
         \multicolumn{3}{c}{Derived parameters} \\ \hline
         Orbital period (yr) & ... &  ${212}_{-75}^{+126}$ \\
         Semimajor axis (mas) & ... &  ${1154}_{-294}^{+419}$ \\
         Eccentricity $e$ & ... &   ${0.896}_{-0.088}^{+0.064}$ \\
         Argument of periastron $\omega$ $(\degree)$ & ... &  ${103}_{-16}^{+11}$ \\
         Time of periastron $T_0$ (JD) & ... & ${2466913}_{-819}^{+1550}$ \\
         Mass ratio & ... &   ${0.536}_{-0.047}^{+0.048}$ \\ \hline
    \end{tabular}
    \caption{Posteriors of the HD 169889 system.}
    \label{tab:post_169889}
\end{table*}

\end{document}